\documentclass[twoside,12pt]{article}
\usepackage{epsfig}
\usepackage{braket}
\usepackage{comment}

\def\Journal#1#2#3#4{{#1} {#2} (#4) #3 }

\def\NPA{{\em Nucl. Phys.} A}

\def\PLB{{\em Phys. Lett.} B}

\def\PRL{\em Phys. Rev. Lett.}

\def\PREP{\em Phys. Rep.}

\def\PRD{{\em Phys. Rev.} D}
\def\PRC{{\em Phys. Rev.} C}

\newcommand{\be}{\begin{equation}}
\newcommand{\ee}{\end{equation}}
\newcommand{\bea}{\begin{eqnarray}}
\newcommand{\eea}{\end{eqnarray}}

\makeatletter
\newcommand{\subsubsubsection}{\@startsection{paragraph}{4}{\z@}%
  {1.0\Cvs \@plus.5\Cdp \@minus.2\Cdp}%
  {.1\Cvs \@plus.3\Cdp}%
  {\reset@font\sffamily\normalsize}
}
\makeatother
\begin{document}

\title{ \vspace{1cm} Nuclei in Core-Collapse Supernovae Engine}
\author{S.\ Furusawa,$^{1,2}$ and H.\ Nagakura$^{3,4}$ 
\\
$^1$College of Science and Engineering, Kanto Gakuin University, Kanagawa, Japan\\
$^2$iTHEMS, RIKEN, Saitama, Japan\\
$^3$Department of Astrophysical Sciences, Princeton University, NJ, USA\\
$^4$Devision of Science, National Astronomical Observatory of Japan, Tokyo, Japan}
\maketitle
\begin{abstract} 
Herein,  we review the nuclear equations of state (EOSs) 
and the constituent nuclei of core-collapse supernovae (CCSNe) 
  and their roles in CCSN simulations. 
 Various nuclei such as deuterons, iron, and extremely neutron-rich nuclei compose in the central engines of CCSNe.
 The center of a collapsing core is dominated by neutron-rich heavy nuclei 
 prior to the occurrence of core bounce. Their weak interactions significantly affect the neutrino emission and
  the size of the produced proto-neutron star. 
 After a core bounce,  heavy nuclei are dissolved to protons, neutrons, and light nuclei 
 between the expanding shock wave and the newly formed neutron star (NS).
 Some of the key components in determining the  shock-wave dynamics  and supernova explosion of  outer envelopes 
 are neutrino interactions of nucleons and light nuclei such as deuterons. 
 An EOS provides the relations between thermodynamical properties and the nuclear composition, and is needed to simulate this explosion. Further investigations on uniform and non-uniform nuclear matter are needed to improve the understanding of the mechanism  of CCSNe and  the properties of supernova nuclei. 
 The knowledge of the EOS for uniform nuclear matter is being continually improved by a combination of microscopic calculations, terrestrial experiments, and NS observations.
 With reference to various  nuclear experiments and current theories, 
 the finite temperature effects on heavy nuclei,  formation of light nuclei in dilute nuclear matter, and  transition to uniform nuclear matter should be improved in the model of the EOS for non-uniform nuclear matter.
  \end{abstract}

\section{Introduction \label{intro}}
Core-collapse supernovae (CCSNe) are energetic events  that occur at the end of the evolution of massive stars with masses exceeding $\sim10 M_{\odot}$. 
Their typical explosion energy is  approximately $10^{51}$ $\rm{erg}$, and a black hole or  a neutron star (NS) is 
generated as a by-product.
Currently, the explosion mechanism  is  not  understood owing to its complexity.
CCSNe are also excellent accelerators of cosmic rays, major sites for nucleosynthesis,  and
excellent emitters of neutrinos and gravitational waves \cite{janka12,kotake12}.  

The central engines of CCSNe are iron-group cores, which are formed at the center of massive stars during their chemical evolution. 
The gravitational collapse of a core  is stopped  by repulsive nuclear interactions, and 
shock waves are generated  at the surface of a hot new-born prott-NS (PNS) on core bounce occurrence.
Numerous neutrinos with energies approximately $10^{53}$ $\rm{erg}$ are  emitted and 
facilitate the thermal expansion of the  shock matter. 

Neutrinos, nuclear matter, electrons, muons, and photons
form the supernova matter of the central engine. 
Nuclear matter is  composed  of nucleons (protons and neutrons) and nuclei.
The thermodynamic conditions of supernova matter vary over a wide range of density, temperature, and neutron (or proton) fraction.  
The equation of state (EOS) of   nuclear matter determines not only its thermodynamic properties
 but also its composition  in chemical equilibrium for a given condition. 

The theory of uniform nuclear matter 
based on a nuclear interaction model
is  a critical component in the construction of the nuclear EOS.
In most general-purpose EOSs  for CCSN and NS merger simulations,
Skyrme-type interactions \cite{lattimer91,schneider17, schneider19,raduta19}  or 
relativistic mean field (RMF) theories
\cite{shen98a,shen98b,shen11,hempel10,steiner13,sheng11a, sheng11b,furusawa13a,furusawa17a,typel18} are adopted.
Both  are based  on phenomenological medium-dependent interactions
and the model parameters  
are determined to reproduce some nuclear properties such as the nuclear saturation density. 
Some EOSs for supernova simulations have been formulated using 
 microscopic models considering
realistic interactions determined using nucleon--nucleon scattering data
 \cite{togashi17,furusawa17d, furusawa20a}.

For non-uniform nuclear matter calculations, 
classic EOSs \cite{lattimer91,shen98a,shen98b,shen11} adopt  single-nucleus approximation (SNA), 
in which  the full distribution of nuclei is represented by a single nucleus with optimized properties.
 Optimization of the nuclear structure such as nuclear decompression and nuclear pastas can be considered in such a calculation;
however,  a realistic nuclear composition cannot be obtained \cite{burrows84,furusawa17b}. 
An extended nuclear statistical equilibrium (NSE) model  is a modern approach to describe non-uniform nuclear matter in supernova EOSs,
 which yields  the number densities of all nuclei and nucleons in a thermodynamical state \cite{hempel10, blinnikov11, furusawa11}.
In addition, hybrid models of the NSE and the SNA are available for low densities and high densities, respectively \cite{schneider17,sheng11a, sheng11b}.


The nuclear composition determined by  the EOS, 
 weak interaction rates of non-uniform nuclear matter  at subsaturation  densities, 
and  the EOS of  uniform nuclear matter at supra-nuclear densities
affect the dynamics of supernova explosion and the  neutrino emission from CCSNe \cite{sumiyoshi05,nagakura19a}.
However, in  a hot, dense, and/or neutron-rich environment, 
the free energies of nuclei and uniform nuclear matter  are not constrained well.
Consequently,   
the models of nuclei and uniform nuclear matter for the EOSs currently used in supernova simulations
show a wide range of variations. 
The reader is referred to a review of the EOSs for CCSNe and NSs \cite{oertel17}
 as well as to a  detailed comparison of l  general-purpose EOSs \cite{raduta21}.
Moreover, an online service CompOSE summarizes data tables of various EOSs \cite{typel15, oertel17}.
Herein, we review the  EOSs for CCSN simulations, with a focus on nuclei. 
We discuss the nuclei present in a CCSN,  their roles in simulations, and their behavior. 

In Sec.~\ref{sec:sn}, we introduce the mechanism of CCSNe, related neutrino interactions,  and  thermodynamic conditions for supernova simulations. 
Theories of uniform nuclear matter are presented in Sec.~\ref{sec:um}.
Non-uniform nuclear matter and the EOS for supernova simulations 
 are described in Sec.~\ref{sec:num}.   
 The nuclei considered in supernova simulations are discussed in Sec.~\ref{sec:res}.
 Finally, a summary discussion is provided in Sec.~\ref{sec:conc}.

\section{Core-Collapse Supernovae \label{sec:sn}}
In this section,  we briefly review a standard scenario for a CCSN explosion and 
 the status of numerical simulations of the CCSN engine.
For more detailed  explanations, refer to the  reviews  in \cite{janka16,burrows21}.  

\subsection{Explosion scenario} \label{sec.intro.sc}
Main-sequence stars are supported by the thermal pressure generated by
 the nuclear burning reactions occurring around the center of a star. 
As the core shrinks and the temperature rises,
 $\alpha$ nuclei, which are even--even nuclei containing equal number of protons and neutrons,  are produced by nuclear fusion
and accumulate in the center as ash.
The ash-$\alpha$ nuclei form a new nuclear fuel, producing  $\alpha$ nuclei with larger mass numbers. 
By this cycle, massive stars in their final evolution stages  have an onion-like structure
 with an iron core surrounded by layers of  $^{28}_{14}$Si, ($^{16}_{8}$O, $^{20}_{10}$Ne, and $^{24}_{12}$Mg), ($^{12}_{6}$C and $^{16}_{8}$O), $^{4}_{2}$He and  $^{1}_{1}$H. 
 An iron core is formed  of  iron-group nuclei with  proton number $Z \sim 26$, and it does not  burn because these nuclei have larger binding energies per baryon than the other nuclei. 

 The mass of an iron core is approximately 1.5 $M_{\odot}$ \cite{woosely02}.
 The increase in the temperature and density of a core induce iron photodissociation and electron capture, respectively.
 The iron core becomes unstable and therefore, it starts to collapse. 
When the density increases and  the chemical energy of electrons exceeds the mass difference between parent and daughter nuclei, electron  capture occurs.
Reduction in the  electron degeneracy pressure  leads to gravitational collapse, inducing further electron capture as
$e^- + (N, Z) \longrightarrow \nu_e +  (N+1, Z-1)$,
where $(N,Z)$ denotes nucleus with  neutron number $N$ and proton number $Z$, 
 $\nu_e$ is  an electron-type neutrino, and $e$ is  an electron. 
In contrast, 
when the temperatures exceed 0.4 MeV, 
photodissociation of the iron-group nuclei into protons and neutrons occurs as follows:
 \begin{eqnarray}
 (N, Z)  & \longrightarrow  & \frac{Z}{2} \alpha+(N-Z) n  \ ,       \ \ \ \ \     N \geq Z   \  ,   \label{eqpd1}\\  
 (N, Z)  & \longrightarrow  & \frac{Z}{2} \alpha+(Z-N) p  \  ,    \ \ \ \ \     N < Z   \ , \\  
\alpha  & \longrightarrow  & 2 p+2 n ,  \label{eqpd3}  
 \end {eqnarray}
 where 
$\alpha$  is an $\alpha$ particle ($^4$He nucleus), $n$ is  a neutron, and $p$ is  a proton. 
This endothermic reaction suppresses the increase in  the thermal pressure caused by the core contraction, thereby accelerating the collapse. 
This photodissociation typically occurs before the electron capture, and gravitational collapse begins.
For relatively light stars of approximately 10~$M_\odot$, electron capture may occur first because of the low central temperature of the iron core.

 In the early stages of a core collapse,
the majority of neutrinos produced by the electron capture can escape from the core.
When the density exceeds 10$^{12}$~g/cm$^3$, the time scale for diffusion becomes shorter than the dynamical time scale, $\mathcal{O}$~(1 ms). 
and the neutrinos are  trapped and degenerated. 
 The core during collapse is divided into two parts: the inner core, which falls with a speed proportional to its radius, and the outer core,
  which is free falling from the outer layers. 
  When the density of the inner core exceeds the nuclear saturation density ($\rho_0 \sim 2.7 \times$ 10$^{14}$ g/cm$^3$  or $n_0 \sim 0.16$ fm$^{-3}$), the nuclear interaction becomes  repulsive
  and the EOS becomes stiff. 
  The gravitational collapse is halted and  shock waves are created. 
  This is  called a core bounce, and the core-collapse phase takes $\mathcal{O}$~(10 ms)  (from the beginning of the collapse to the core bounce).
 The region through which the shock waves pass is heated, 
 and numerous neutrinos are emitted via electron capture by protons, 
 $\nu_e + n \longleftrightarrow   e^- + p $,
 after the reactions expressed in Eqs.~(\ref{eqpd1}--\ref{eqpd3}).  
However, the neutrinos are almost trapped 
before the shock waves cross the neutrino sphere, which is the surface above which they can escape from the star.
A large neutrino flux is emitted outward after the shock waves cross the neutrino sphere,
which is called a neutronization burst, and it occurs  after the bounce on the time scale of   $\mathcal{O}$~(10 ms).

The shock waves travel to the outer boundary of the iron core in $\mathcal{O}$~(100 ms)  and
cause explosion of the outer layers in  $\mathcal{O}$~(10 h), resulting in a CCSN explosion. 
Parts of the outer layers
 are hit by the shock waves and  accrete on the surface of the produced PNS. 
The lepton number of PNS composed of the above-mentioned inner core and the accreting matter is still large. 
The neutrinos carry away many leptons and the internal energy of the PNS on the time scale of $\mathcal{O}$~(1 min), and the PNS becomes an NS.
A black hole may remain instead of  an NS, 
depending on the final stellar structure; however,  the details are unknown.

Neutrinos, unlike electrons, neutrons, protons, and nuclei,  are unaffected
 by electromagnetic  or strong interactions. 
 Consequently, the produced neutrinos have a long mean free path 
 and carry away  approximately  99$\%$ energy released by the gravitational collapse and the contraction of the PNS.
The remaining approximately 1$\%$ energy is considered to be used for the kinetic energy of the explosion. 
The neutrinos also provide information about the core, from which electromagnetic waves cannot escape.

\subsection{Supernova simulations}
Twice decades ago, many groups conducted one-dimensional (1D) numerical simulations assuming spherical symmetry, 
demonstrating  that the shock waves formed by the bounce of collapsing cores are decelerated and stalled
 by the energy losses due to  the nuclear dissociations  and  the neutrino emission \cite{sumiyoshi05}.
Since then, several two-dimensional (2D) and three-dimensional (3D) numerical simulations have been conducted
with the neutrinos emitted from a PNS reinvigorating stalled shocks and allowing it to propagate outward again.
This neutrino-heating  mechanism is the most promising scenario for a shock revival in a CCSN engine.

Multi-dimensional (multi-D) effects such as convection and 
 the standing accretion shock instability (SASI)
are essential in increasing the efficiency of neutrino heating in the central engine of an explosion \cite{iwakami08}.
Recently, 
multi-D numerical simulations have  successfully  modeled the
 relaunch of  a stalled shock wave, which may eventually produce supernova explosions \cite{takiwaki14, lentz15, roberts16, oconnor18,bernhard19, nagakura19b, robert19,vartanyan19,nagakura19c,burrows20,nagakura20,jade20,kuroda20,takiwaki21}.
However, most CCSN simulations are for less than $\mathcal{O}$~(1 s), making it impossible to determine whether they can reproduce 
a canonical explosion energy of  $10^{51}$~erg and $^{56}$Ni mass as well as
 how nuclear burning in accreting matter and ejecta occurs  \cite{yamamoto13, nakamura14, nakamura19}. 
The progenitor that provides the initial condition of supernova simulations
 and is one of the key components of the dynamics of CCSNe is also uncertain \cite{suwa15, suwa16, nagakura18b, yoshida19, yoshida21}.

The difficulties in the numerical simulations of CCSNe arise from  two factors: 
multiple physics involvements in the system and the difficulty in using approximations. 
The compactness of a PNS is governed by strong interactions, 
 weak interactions determine neutrino reactions, electromagnetic interactions 
 affect the size of the nuclei in supernova matter, and gravitational interactions influence the dynamics of explosions and the structure of a PNS.
 In numerical simulations, the hydrodynamics and neutrino transport calculations must be solved in 3D and 6D phase spaces, respectively.
Furthermore, relativistic effects should not be ignored \cite{nagakura14, akaho21}. 
However, various approximations cannot be used.
Supernova simulations  with spherical symmetry approximation (1D simulations) cannot reproduce explosions, as previously mentioned.
For neutrino transport, the diffusion approximation may be applied in the center, where scattering occurs frequently, 
and the free-streaming approximation may be adopted in the outer regions, where neutrinos are rarely scattered. 
In the middle region,  where the shock waves expand owing to neutrino energy depositions,
no approximations are available for neutrino transport.

%

\begin{figure}
\begin{center}
\includegraphics[width=10cm]{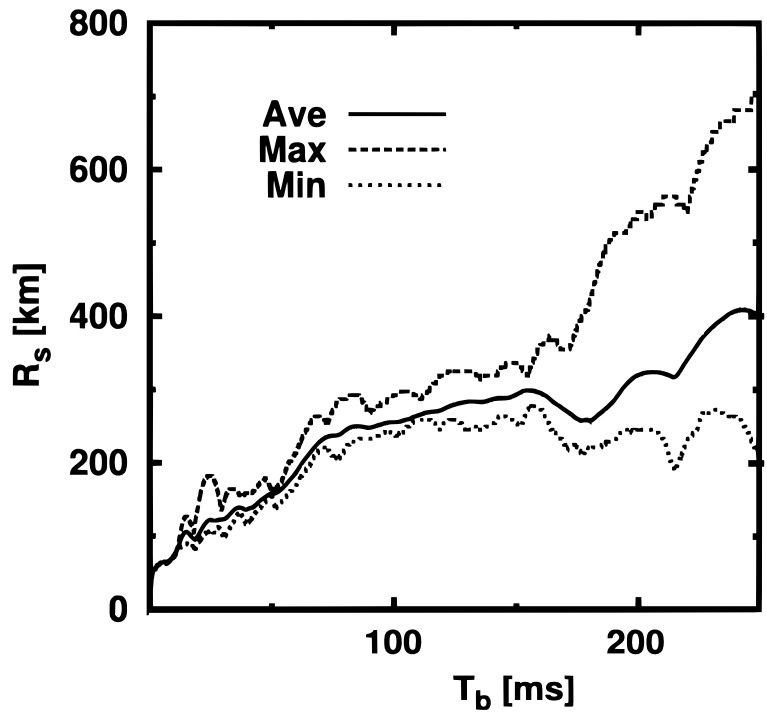}
  \caption{ Time evolutions of maximum (dashed line), minimum (dotted line) and average  (solid line) shock radii
 of the 2D numerical simulation \cite{nagakura19b}  of model of a 11.2 $M_{\odot}$ star  using  FYSS (VM)  EOS \cite{furusawa17d}.  } 
  \label{fig.sumisn}
\end{center}
\end{figure}

At present, there is no complete supernova simulations and different approximations are employed in the simulations. 
In 3D simulations, neutrino momentum distributions  in the phase space are not completely solved. 
For  some 2D simulations,  axial symmetry is assumed, 
whereas the full Boltzmann neutrino transport is solved in the  phase space \cite{nagakura19b}.
Based on Fig.~\ref{fig.sumisn}, the full Boltzmann neutrino transport 2D simulations of a massive star using 11.2 $M_{\odot}$ with the  FYSS (VM)  EOS \cite{furusawa17d}
show  the expansion of its shock waves. 
 The shock radii depend on the polar angle, as shown subsequently in Fig.~\ref{fig_rt},
 and their minimum, average, and maximum values are presented in Fig.~\ref{fig.sumisn}.
 The average and maximum radii of the shock waves are increased  owing to the neutrino heating.
The details of the EOS and the simulation are presented  in the following sections. 

\subsection{Matter in supernovae}
Supernova matter in the central engine of a CCSN consists of neutrinos, nuclear matter with electrons, muons, and photons. 
Neutrinos are not always in thermal or chemical equilibrium with nuclear matter,
 and thus, cannot be included in the EOS. 
 Their non-equilibrium distributions should be computed using transport equations.
Electrons  are treated as ideal Fermi gases, whereas photons as ideal Bose gases.
The presence of  muons in a PNS has been reported \cite{bollig17}.
However, they are located  at the center of a PNS because of  their larger masses  than electrons, and
thus, are  not considered in this review.
In addition, hyperons, quark matter,  and pion- and kaon-condensates  may drastically alter the dynamics of CCSNe \cite{fischer11}.
In this review, we focus on the nuclei  and nuclear matter present below and around the  nuclear saturation density, and  the details of such new degrees of freedom at high densities are not discussed.

\begin{figure}
\begin{center}
\includegraphics[width=10cm]{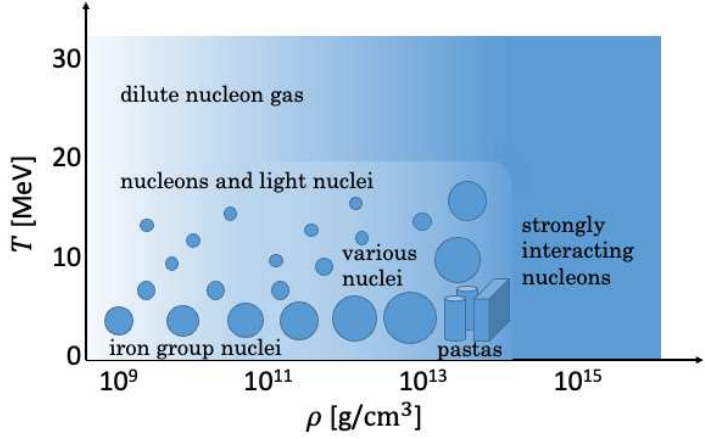}
  \caption{Nuclear phases 
in  typical environment of  CCSNe are schematically shown in the plane of number density-temperature. 
The details depend on the models of nuclear matter and nuclei and proton fraction.
\label{fig_phase}}  
\end{center}
\end{figure}

\subsubsection{Hadronic matter}
Depending on the density, temperature, and proton fraction, 
nuclear matter is formed of  various phases.
 As shown in Fig.~\ref{fig_phase}, 
matter with free nucleons  are available    at sub-nuclear densities  and at high temperatures.
At low temperatures,  a mixture of nuclei and free nucleons are formed. The mass numbers of nuclei are larger at higher densities and at lower temperatures.
Nuclear pastas may be seen at low temperatures (below about 3 MeV)  and just below nuclear saturation densities  \cite{newton09,watanabe11,roggero17}.
 Strongly interacting nucleons  appear at supra-nuclear densities. 
 The proton fraction, $Y_p$, is defined as the total proton number density, $n_p^{total}$, per baryon number density, $n_B$ as follows:  $Y_p=n_p^{total}/n_B$. 
The details of nuclear matter are discussed in subsequent sections (Secs.~\ref{sec:um}--\ref{sec:res}).

\subsubsection{Photon and lepton}
The Helmholtz free energy density of supernova matter without neutrinos, $f_t$, is expressed as
\begin{eqnarray}
\ f_t &=&f_\gamma + f_{e} +f_B   \ ,  \label{eq:all}  
\end{eqnarray}
where $f_{\gamma}$ is the free energy of photons and $f_{e}$ is that of electrons and positrons.
The free energy of nuclear matter, $f_B$, is explained subsequently in Sec.~\ref{sec:num}.

Photons are  ideal bosons and their free energy densities are calculated as 
\begin{eqnarray}
f_\gamma &=&- \left( \frac{\pi^2 }{45 \hbar^3 c^3 }    \right) T^4   ,
\end{eqnarray}
 where $\hbar$ is the reduced Planck's constant and $c$  is the speed of light.
Electrons and positrons are ideal fermions that have no interactions \cite{shapiro83}. 
The net electron number density is defined as the difference between the number density of electrons,   $n_{e^-}$,
and  that of positrons, $n_{e^+}$,  as 
\begin{eqnarray}
n_e & =& n_{e^-} - n_{e^+}   \label{eq_ne} \\
 n_{e^-} &=&\int_{0}^{+\infty} F_F(\mu_e, p) \frac{2 d^3p}{(2 \pi \hbar)^3}   , \\
 n_{e^+} &=& \int_{0}^{+\infty} F_F(-\mu_e, p) \frac{2 d^3p}{(2 \pi \hbar)^3}  , \\ 
F_F(\mu_e, p) &=& 1/ \left\{1+\exp \left( \frac{\sqrt{p^2+m_e} - \mu_e}{T} \right)  \right\}   , \label{eq_ff}   
\end{eqnarray}
where 
 $F_F$ is the Fermi distribution function. 
The factor of 2 in the denominator in Eq.~(\ref{eq_ne}) originates from the spin degree of freedom.
The total proton fraction and the baryon number density are related to the net electron number density by the charge neutrality of the system as 
\begin{equation}
n_e=Y_p n_B .
\end {equation} 
The chemical potential of electrons, $\mu_e$, is obtained by solving  
Eq.~(\ref{eq_ne}) for a given $n_e$.
The pressure, $p_e$, and free energy of electrons and positrons  are respectively expressed as
\begin{eqnarray}
p_e &=&  \int_{0}^{+\infty} F_F(\mu_e, p) \frac{2 \sqrt{p^2+m_e}   d^3p}{(2 \pi \hbar)^3} + \int_{0}^{+\infty} F_F(-\mu_e, p) \frac{2  \sqrt{p^2+m_e} d^3p}{(2 \pi \hbar)^3} , \\
f_e &=&  n_e \mu_e   - p_e ,
\end{eqnarray}
Note that the Coulomb interactions between electrons and  between electrons and  protons are considered  in the Coulomb energies of nuclei, $f(N, Z)$.
In uniform nuclear matter without nuclei, the local charge density is zero everywhere, because $n_e=n_p$, where $n_p$ is the proton number density.
Hence, the Coulomb energies do not have to be considered. 

\begin{figure}[h]
\vspace{-2cm}
\hspace{2.5cm}
\includegraphics[width=15cm]{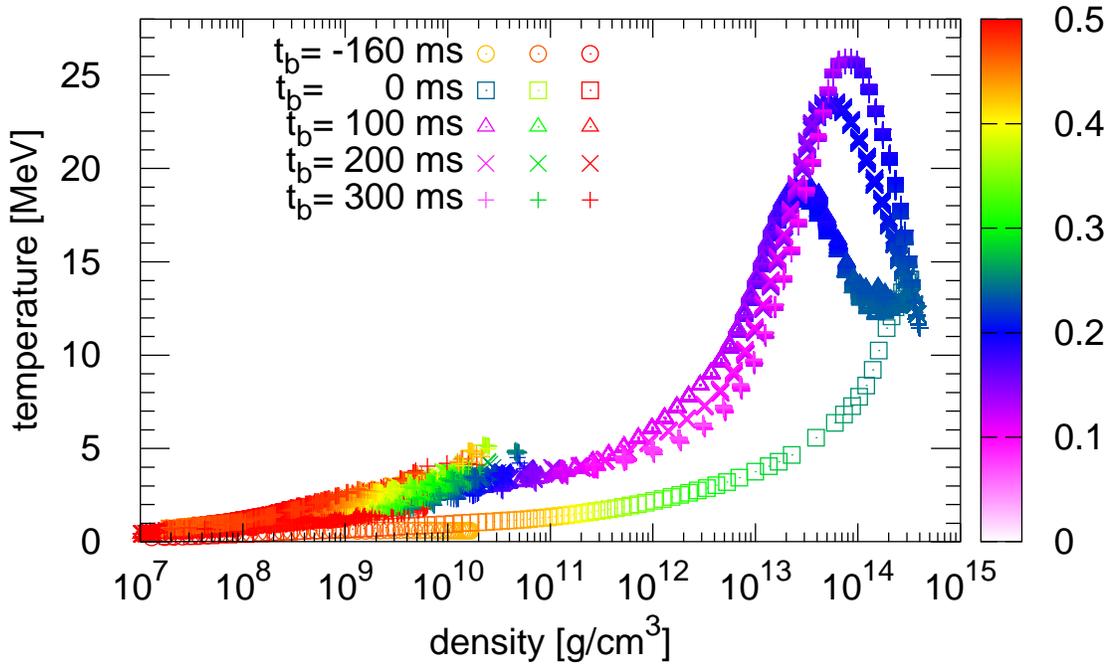}
\vspace{3cm}
\caption{Temperature versus density variation in the supernova simulation \cite{nagakura19b}. Color represents corresponding proton fraction. Values  of meshes at several times before or the core bounce are superposed:  $t=-$160 ms (circles, initial phase), 0 ms (squares,  core bounce), 100 ms (triangles), 200 ms (crosses), and 300 (pluses) ms.}
\label{fig_rtyp}
\end{figure}

\begin{figure}[h]
\hspace{2.5cm}
\includegraphics[width=13cm]{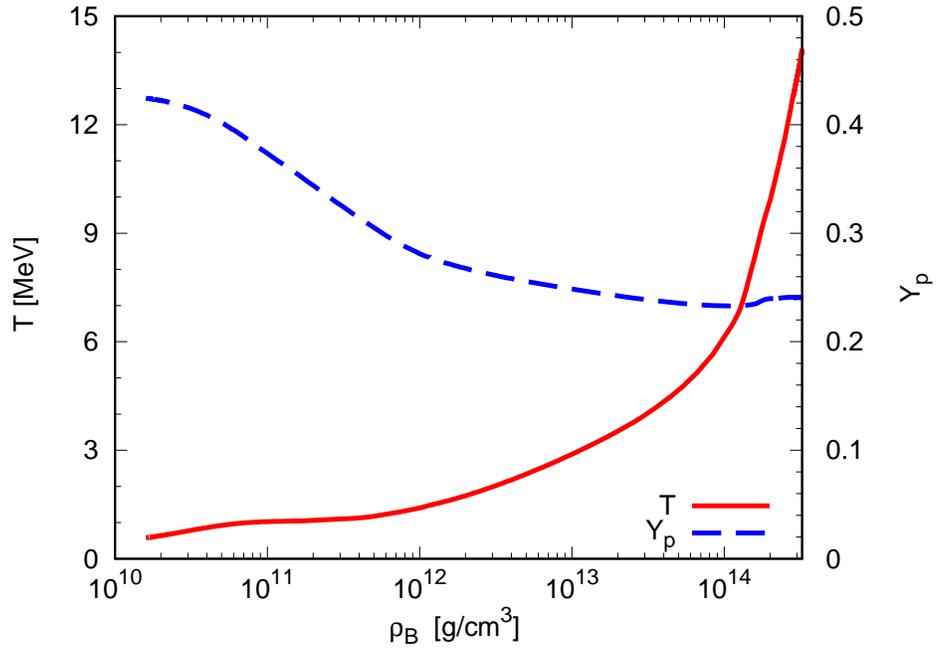}\\
\hspace{2.5cm}
\vspace{2cm}
\caption{
Temperature (red solid line) and proton fraction (blue dashed line) as  functions of central density in core-collapse phase  \cite{nagakura19b}.
}
\label{fig_pro}
\end{figure}

\subsubsection{Thermodynamic conditions}\label{sec.intro.sm}
Density $\rho_B$, temperature $T$, and proton fraction $Y_p$ 
 at each time step in the above-mentioned 2D supernova simulation  \cite{nagakura19b} are superposed in Fig.~\ref{fig_rtyp}.
 Here, the density is defined as $\rho_B=m_u n_B$, in the  where $m_u=931$ MeV  is the atomic mass unit \cite{shen98b}.
In the final phase of the massive star evolution or in the initial phase of the core collapse,
the thermodynamic conditions at the center of the core are  
($\rho_B$, $T$, $Y_p$) =($1.6 \times 10^{10}$ g/cm$^3$, 0.58 MeV, 0.42). 
During the core collapse, $\rho_B$ and $T$ increase almost adiabatically. 
The baryonic entropies are approximately 1~$k_B$,  as shown subsequently in Fig.~\ref{fig_pro5}.
Figure~\ref{fig_pro} presents the temperature and  the proton fraction as functions of the central density of
 a collapsing core and a PNS. 
The proton fraction decreases to 0.24 owing to electron capture by heavy nuclei, and the temperature rises up to 14  MeV 
at the core bounce ($\rho_B=3.2 \times 10^{14}$ g/cm$^3$ and $t=0$).
At densities exceeding  $\rho_B \sim 10^{11}$--$10^{12}$~g/cm$^3$,
neutrinos are almost trapped, and the proton fraction is barely changed, 
whereas the emission and absorption of the  trapped neutrinos slightly change $Y_p$.

\begin{figure}
\hspace{1cm}
\includegraphics[width=12cm]{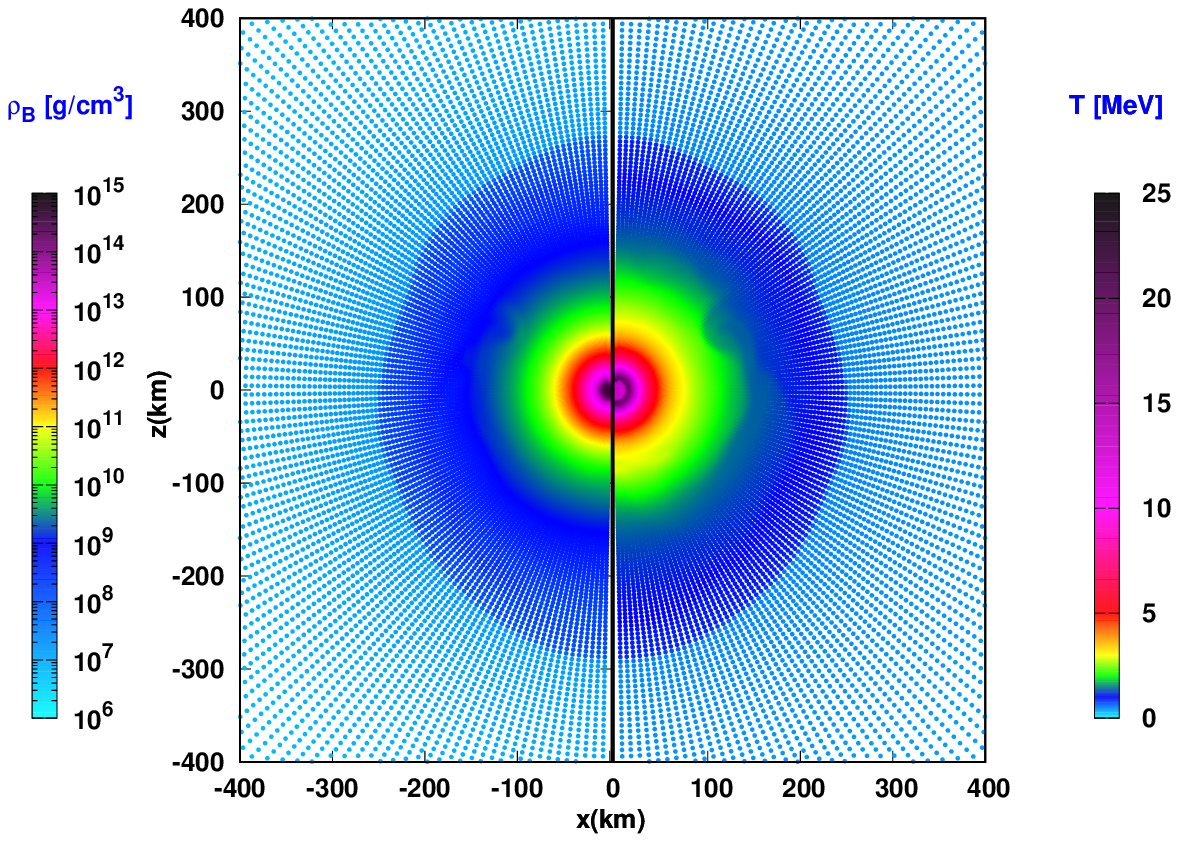}\\
\hspace{1cm}
\includegraphics[width=12cm]{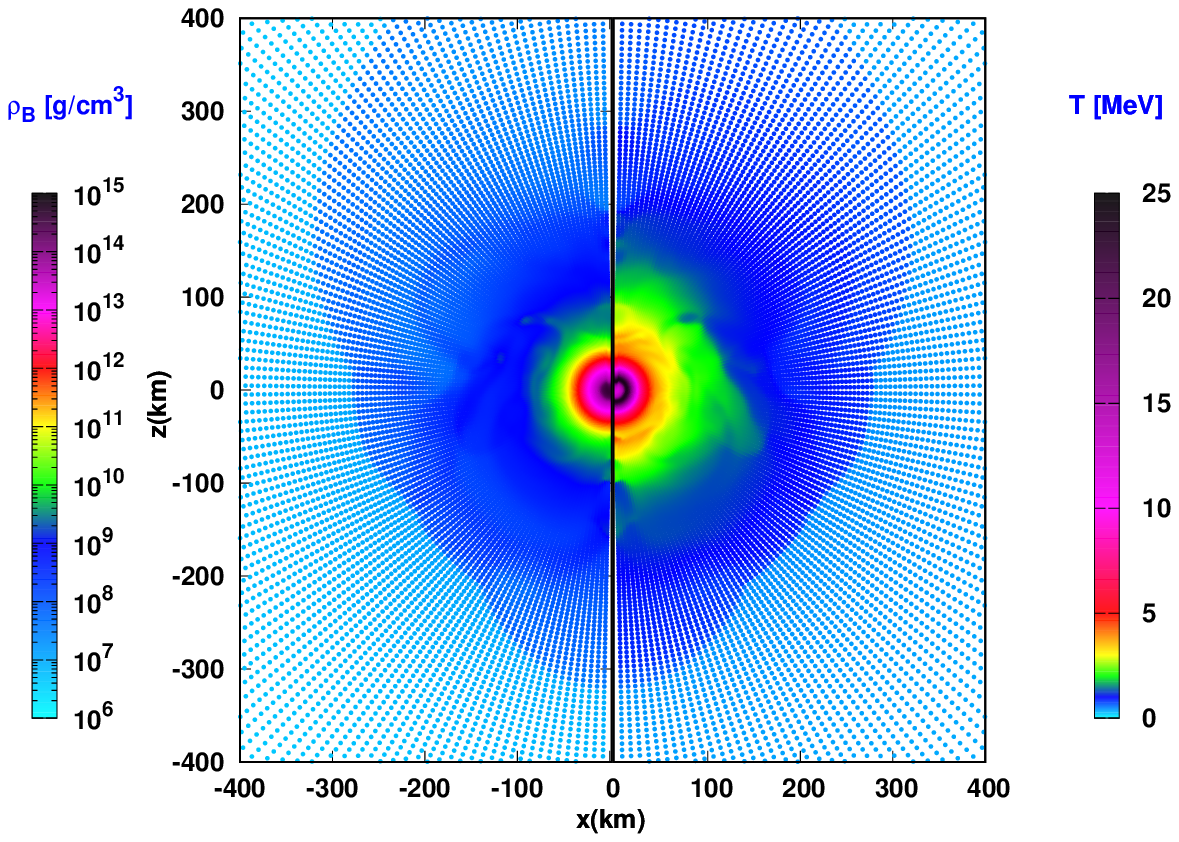}
\vspace{2cm}
\caption{Density (left) and temperature (right) distributions  at 100 ms (top panel) and 200 ms (bottom panel)   after core bounce
in the supernova simulation \cite{nagakura19b}   using  FYSS (VM)  EOS \cite{furusawa17d}. }
\label{fig_rt}
\end{figure}

\begin{figure}
\hspace{1cm}
\includegraphics[width=12cm]{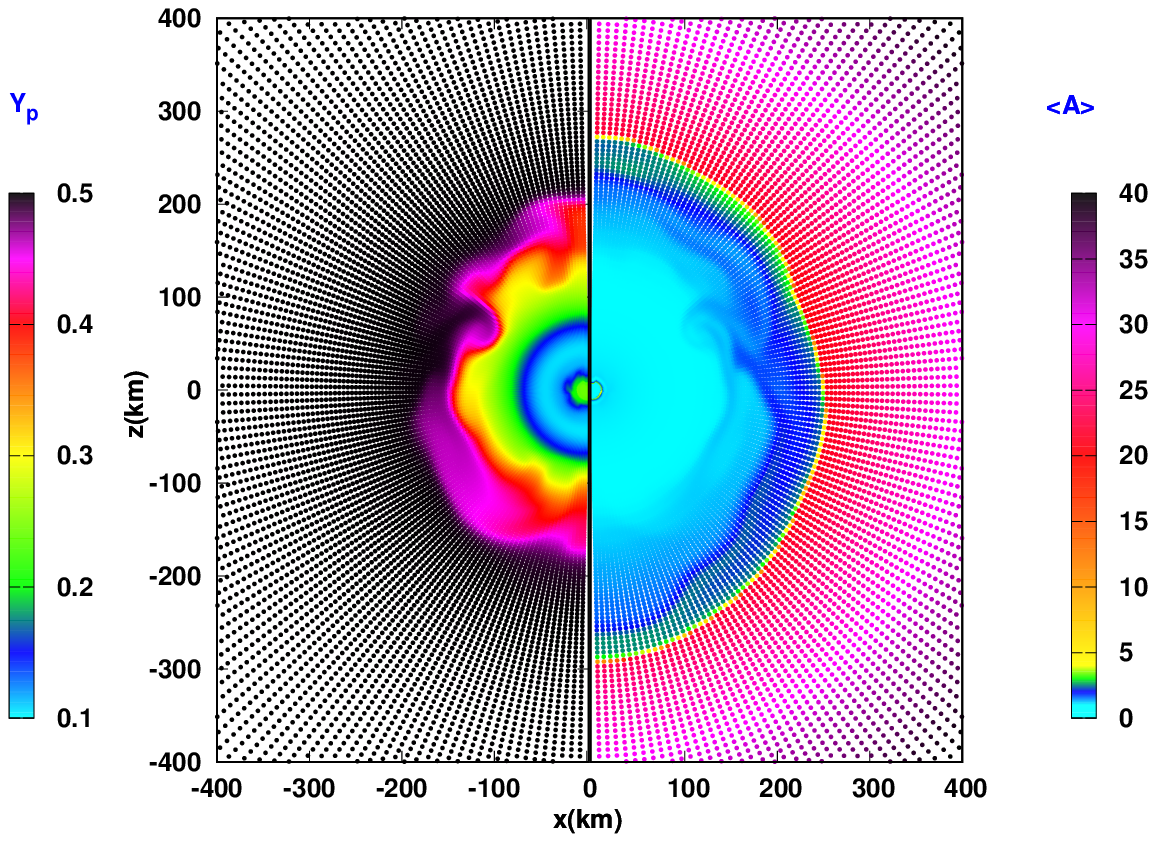}\\
\hspace{1cm}
\includegraphics[width=12cm]{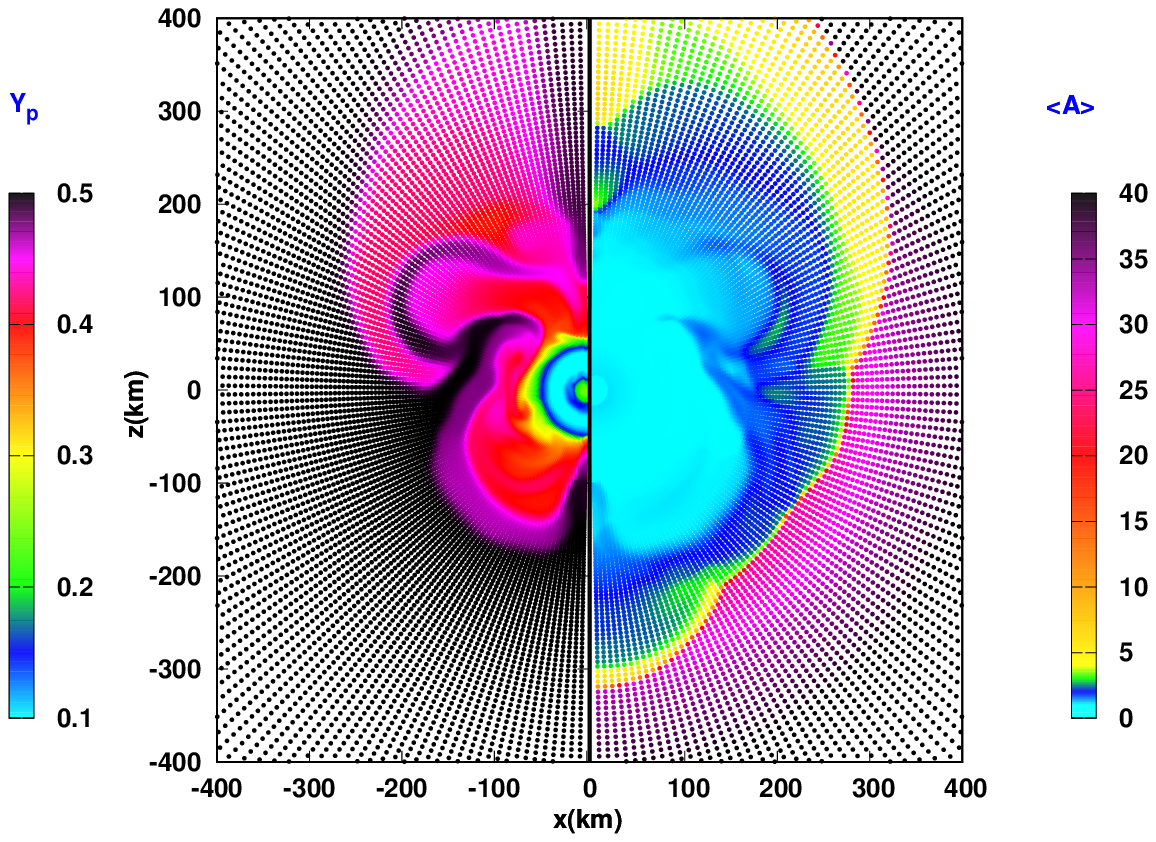}
\vspace{2cm}
\caption{Distributions of  total proton fractions  (left) and average nucleon numbers of nucleons and nuclei  (right)  at 100 ms (top panel) and 200 ms (bottom panel)   after core bounce
in the supernova simulation \cite{nagakura19b}   using  FYSS (VM)  EOS \cite{furusawa17d}. }
\label{fig_fye}
\end{figure}

After the core bounce, shock heating increases the entropy and temperature of the matter above the surface of the produced PNS. 
The large neutrino emission via electron capture by protons significantly reduces $Y_p$. 
Around the neutrino sphere at  $\rho_B \sim 10^{11}$--$10^{12}$~g/cm$^3$,  the temperature is approximately $5.0$~MeV and the proton fraction is approximately 0.09.
The distributions of $\rho_B$, $T$, and $Y_p$  in the star at $t=100$~ms and 200 ms after core bounce are  shown in Figs.~\ref{fig_rt} and~\ref{fig_fye}, respectively.
The values around the shock waves are $T\sim1$~MeV, $\rho_B\sim10^9$~g/cm$^3$, and $Y_p=0.5$. 
During the shock-revival phase,
the temperature around the surface of the PNS reach close to 30 MeV and the central density increases up to approximately $5.0 \times 10^{14}$ g/cm$^3$.  


\newpage

\subsection{Weak interactions}\label{sec.intro.inter}
Consideration of weak interactions in CCSN simulations 
plays an important role  in both pre- and post-bounce stages, as noted in Sec. \ref{sec.intro.sc}.
 In some simulations, 
an open-source neutrino interaction library  Nulib  is utilized, which is  based on some studies \cite{burrows06, sullivan16}.
Many-body corrections may have to be considered in the calculations of weak rates  \cite{burrows98,burrows99,horowitz17}.
In this section, we introduce  the neutrino reactions included in the supernova simulation \cite{nagakura19b}; they  are listed in Tab.~\ref{tab_wr}.  

\begin{table}[t]
\begin{tabular}{cccc}
\hline
(i) & $\nu_e + n \longleftrightarrow   e^- + p   $ & (viii) & $\nu + (N, Z) \longrightarrow \nu  + (N, Z)$  \\
(ii) &  $\bar{\nu}_e + p \longleftrightarrow  e^+ + n$  & (ix) &  $\nu_e + d  \longleftrightarrow  e^- + p + p$\\
(iii) & $\nu + n/p \longrightarrow  \nu + n/p  $  &  (x) &  $ \bar{\nu_e} + d  \longleftrightarrow  e^+ + n + n $  \\
(iv) &  $\nu + e^{\pm} \longrightarrow  \nu + e^{\pm} $  & (xi) & $e^- + d   \longleftrightarrow  \nu_e + n+n $ \\
(v) & $e^-  + e^+ \longleftrightarrow  \nu_e +\bar{\nu}_e $  & (xii) & $e^+ + d \longleftrightarrow \bar{\nu_e} + p+p$ \\
(vi) & $n/p+n'/p' \longleftrightarrow n/p+n'/p'+ \nu_e +\bar{\nu}_e$  & (xiii) & $\nu_e + ^{3}{\rm H}  \longleftrightarrow  e^- + ^{3}{\rm He}$ \\
(vii) & $e^- + (N, Z) \longrightarrow \nu_e +  (N+1, Z-1)$  & (xiv) & $\bar{\nu_e} + ^{3}{\rm He}  \longleftrightarrow  e^+ + ^{3}{\rm H}$ \\
\hline
\end{tabular}
\caption{\label{tab_wr}%
Weak interactions in supernova simulations.}
\end{table}

Referring to Tab.~\ref{tab_wr}, the electron capture by free protons (reaction (i)) has a small  impact on the evolution of $Y_p$  before the  bounce. 
Free nucleons are the  main sources of the neutrino emission and absorption in the post-bounce phase,
which affect the cooling and heating rates via reactions (i) and (ii).
The neutrino scattering by nucleons (reaction (iii)) is important  for the mean free path of neutrinos,  
 whereas electrons  play a minor role in the total neutrino opacity. 
However, they are important in decreasing the neutrino energies through an inelastic reaction in reaction (iv). 

Heavy nuclei have a major influence on the evolution of $Y_p$ in collapsing cores in the pre-bounce phase
via reactions (vii) and (viii). 
The former determines the number of  many neutrinos  that are emitted.  
The latter affects the deleptonization by the neutrino trapping, although it does not directly change $Y_p$.
For some nuclei, 
the calculated electron capture rates are presented in data tables \cite{fuller82,oda94,langanke00,langanke03}.
  These tables are based on shell model calculations or  the Monte Carlo approach 
with a random phase approximation.
For the nuclei where no data are available,
 the following approximation formula  as a  function of the $Q$ value \cite{fuller85,langanke03} is frequently adopted.
\begin{eqnarray}
 \lambda_i &=&\frac{({\rm ln}2)B}{K} \left( \frac{T}{m_e c^2} \right)^{5} \left[ F_4(\eta(N,Z)) - 2 \chi(N,Z)F_3(\eta(N,Z)) + \chi_i^2 F_2(\eta(N,Z))  \right] , \label{eq:appro}  \\
 F_k(\eta)&=&\int_0^{\infty}\frac{u^kdu}{1+\exp(u-\eta)}  \ , \label{eq_gf}  
\end{eqnarray}
where  $K=6146$ s, $\chi(N,Z)=(Q(N,Z) - \Delta E)/T$, and  $\eta(N,Z)=(\mu_e +  Q(N,Z) -\Delta E)/T$, where the electron chemical potential is denoted as $\mu_e$.
The parameters of a typical matrix element  ($B=4.6$) and 
 the transition energy from an excited state in a parent nucleus to a daughter state ($\Delta E=2.5$ MeV) are fitted to the results of shell-model calculations for
the pf-shell nuclei  \cite{langanke03}.
The $Q$ value of each nucleus, 
 $Q(N,Z)$,  is calculated as $Q(N,Z)=m(N,Z)-m(N+1,Z-1)$,
where $m(N, Z)$ is the nuclear mass considering the Coulomb energy shifts \cite{furusawa17b, kato21}. 

We note that the calculation of electron capture rates of the heavy nuclei such as $^{78}_{28}$Ni and  $^{80}_{30}$Zn, 
is still controversial \cite{langanke21}, which are important nuclei for deleptonization of core-collapse as discussed later.
In actual, shell gaps are not taken into consideration the fitting formula of Eq.~(\ref{eq:appro}), which may overestimate the reaction rates  \cite{titus18, raduta17}.
On the other hand, the nuclear shell structures at zero temperature may smear out by thermalization and the reaction rates increase by thermal unblocking effects at finite temperatures   \cite{dzhioev20,litvinova21, giraud22}. 

A cross-section of the neutrino-nucleus scattering rate (reaction (viii))  is evaluated as
\begin{eqnarray}
 \sigma_i (E_{\nu_e}) =  \frac{G_W^2}{8 \pi ( \hbar c)^4} E_{\nu_e}^2 A^2 \left\{ 1- \frac{2 Z}{A}(1-{\rm sin}^2 \theta_W)  \right\}^2 \frac{2 y  + {\rm exp}(2 y) -1}{y^2} ,
 \label{eq:sca}
\end{eqnarray}
where $A=N+Z$, $y= 1.92 \times 10^{-5} A^{2/3} E_{\nu_e}^2$, 
$G_W$ and $\theta_W$ are  the weak coupling constant and  the Weinberg angle, respectively.
A non-degenerated nucleus and  an isoenergetic  zero-momentum transfer  are assumed \cite{bruenn85}.
Inelastic  weak interactions of light nuclei are considered in some simulations \cite{fischer16, nagakura19a},
For neutrino absorptions on deuterons, (ix)~and~(x), the data of vacuum cross-section
are used  \cite{nakamura01}. 
The neutrino absorption rate (ix) is expressed as
\begin{equation}
1/\lambda (E_\nu)= n_d \int {\rm d} p_e \left[ \frac{{\rm d}\sigma_{\nu d}}{{\rm d} p_e}(E^*_{\nu}) \right] (1-F_F(E_e)] ,
\end{equation} 
where $ n_{d}$ is the deuteron number density and $F_F$ denotes the Fermi-Dirac distribution of electrons (see Eq.~(\ref{eq_ff})). 
Electron capture by deuterons (reactions (xi) and  (xii)) is estimated 
by assuming that the matrix elements of electron and positron capture
 are equivalent to those of neutrino absorption for reactions~(ix)~and~(x) as  
\begin{equation}
 \frac{{\rm d} \sigma_{e ^{2}{\rm H}}} {{\rm d} p_{\nu}}  \sim \frac{1}{2} \frac{{\rm d} \sigma_{\nu ^{2}{\rm H}}} {{\rm d} p_e}.
\end{equation}
Factor $1/2$ originates from the difference between the spin degrees of freedom of  neutrinos and electrons. 
The energy deposition by the relative motion of two nucleons is negligible because of the small injection energies of leptons. 

Nuclei with three bound nucleons, $^3 \rm H$ and $^3 \rm He$, interact with neutrinos via breakup or charge exchange, the latter of which is the dominant neutrino opacity source and is reaction~(xiii) or (xiv). 
 The rate of charge exchange in  reaction~(xiii) is  calculated as 
\begin{equation}
1/\lambda (E_\nu)= n_{^{3}{\rm H}} \left[\frac{G^2_W V^2_{ud}}{\pi (\hbar c)^4}\right] p_e E_e [ 1-f_e(E_e)] B(GT),
\end{equation} 
where  $ n_{^{3}{\rm H}}$ is  the triton number density, $B(GT)=5.87$, and $V_{ud}=0.967$.
All reverse reactions can be evaluated by a detailed balance with the rate of absorption.
%

The inelastic interactions between neutrinos and other nuclei with $A>3$
 have been neglected in current  supernova simulations of the neutrino heating.
Haxton \cite{Haxton88}  pointed out the importance of these reactions as  
\begin{equation}
\nu + (N, Z) \longrightarrow \nu' + (N, Z)^*,
\end{equation}
where $(N,Z)^*$ denotes an excited nucleus.
Langanke et al. \cite{langanke08} revealed that 
the reactions of heavy nuclei with$ Z=24$--$28$ \cite{juodagalvis05} reduce the high energy tail of the neutrino energy spectrum. 
The $\alpha$ particles are available around shock waves, as shown subsequently in Fig.~\ref{fig_xdal}.
Although the effects of the reactions  for the $\alpha$ particles on  the dynamics of shock waves and neutrino observations
 are expected to be minor, they should  be further investigated \cite{furusawa13b,ohnishi07}.

\section{Uniform Nuclear Matter \label{sec:um}}
The free energies of 
the nucleons dripping out of nuclei at a sub-nuclear density or strongly interacting nucleons at a supra-nuclear density
are  calculated using theories of uniform nuclear matter.
In some EOSs, nuclear bulk energies of non-uniform nuclear matter are also provided theoretically. 

\subsection{Models}
In the following sections,
we briefly outline some theoretical approaches. 
We refer the reader to a more comprehensive review \cite{oertel17} of nuclear matter theories. 
In this review,  a general-purpose EOS  is named as XX(YY).  XX represents the 
initials of the authors in the original publication of the EOS model, which is discussed in Sec.~\ref{sec:num}.  
YY denotes the name of the uniform nuclear matter, which is introduced in this section (i.e., the name of  the parameter set  such as TM1 and SLy4) or  the initials of the theory, such as VM for the variational method.

\subsubsection{Skyrme-type interactions}
Skyrme-type interactions, which are represented as expansions of the effective interaction
 in powers of momenta and density-dependent three-body contributions, 
 are the most well-known choice for calculating the effective interaction energy.
There are many Skyrme interaction parameter sets,  and 240 were compared  by Dutra et al. \cite{dutra12}.
The first general-purpose EOS  with  Skyrme interactions was built by Lattimer and Swesty (LS) \cite{lattimer91}.  
The values of incompressibility, the definition of which is introduced in section~\ref{sec_pro},  are set to $K=180, 220$, and $375$,
which  are represented as LS (180, 220, 375) . 
Recently, Skyrme SLy4 parameters \cite{chabant98}  have  been commonly used in studies on NSs and CCSNe, 
 and have been fitted to produce nuclear binding energies and radii (summarized subsequently in Sec.~\ref{sec_pro}) as well as pure neutron matter calculated from  nucleon-nucleon interaction data  \cite{akmal98}.
Raduta et al. used these parameters to construct a general-purpose EOS, called RG EOS  \cite{raduta19}.
Schneider et al. \cite{schneider17, schneider19} also used SLy4  and other parameter sets for some  general-purpose EOSs: SRO (SLy4, APR, NRAPR, SkAPR, LS220, KDE0v1, LNS).

In the Skyrme-type interactions of the SRO EOS, 
the internal energy density, $\epsilon_B(n_B, x, T)$,  with $n_B$,
proton fraction $x$, and temperature $T$ is expressed as
\begin{eqnarray} 
\epsilon_B(n_B,x,T)&=&\frac{\hbar^2\tau_n}{2m_n^*}+
     \frac{\hbar^2\tau_p}{2m_p^*}   +\left(a+4bx(1-x)\right) n_B^2 \nonumber  \\
& &  +\sum_{j=0,1,2} \left( c_j+4 d_j x(1-x)\right) n_B^{1+\delta_j} +(1- x) n_B m_n+ x n_B  m_p\,,    \label{eq:sky}
\end{eqnarray}
where $a$, $b$, $c_j$, $d_j$, and $\delta_j$ are parameters of the
Skyrme forces, $\tau_n$ and $\tau_p$ are the kinetic energy
densities of neutrons and protons, respectively, and $m_n$ and $m_p$ are the masses of neutrons and protons, respectively.
 The first and second terms correspond to the non-relativistic kinetic energy
density of neutrons $n$ and protons $p$, respectively.  The third term
 represents two-body nucleon interactions,
 and  the summation over $j$  approximates the effects of many-body or density-dependent interactions.
The last two terms express the rest masses of neutrons and protons, respectively.
%

The effective masses,  $m^*_p$ and $m^*_n$, are given by
\begin{eqnarray}  
 \frac{\hbar^2}{2m_{n}^*} &=&\frac{\hbar^2}{2m_{n}}+\alpha_1 n_{n}+\alpha_2n_{p} \ ,  \label{eq_efm1} \\
 \frac{\hbar^2}{2m_{p}^*}&=&\frac{\hbar^2}{2m_{p}}+\alpha_1 n_{p}+\alpha_2n_{n} \  . \label{eq_efm2} 
\end{eqnarray}
Parameters $\alpha_1$ and $\alpha_2$ are chosen to reproduce observables of uniform nuclear matter together with
$a$, $b$, $c_j$, $d_j$, and $\delta_j$ \cite{dutra12}.
For example, for the SLy4 parameter set,   $\alpha_1=81.8$
 MeV fm$^5$ and $\alpha_2=32.5$ MeV fm$^5$. 
The number density of nucleons, $n_i$  ($i$ denotes neutron $n$ or proton $p$),  is expressed as $n_p=x n_B$ and $n_n =(1-x)n_B$.
It determines the degeneracy parameters, $\eta_i$, leading to the kinetic terms in Eq.~(\ref{eq:sky}).
They are expressed as
\begin{eqnarray}
 n_i&=&\frac{1}{2\pi^2}\left(\frac{2m_i^*T}{\hbar^2}\right)^{\frac{3}{2}} F_{1/2}(\eta_i) \ ,  \\
 \tau_i &=&\frac{1}{2\pi^2}\left(\frac{2m_i^*T}{\hbar^2}\right)^{\frac{5}{2}} F_{3/2}(\eta_i) \ , 
 \end{eqnarray}
where  $F_k$ is the relativistic Fermi integral of order $k$, as expressed in Eq.~(\ref{eq_gf}).
 The entropy density and the free energy density are obtained by
 \begin{eqnarray}
 s_B&=& \frac{1}{n} \sum_{i=n,p} \left(\frac{5\hbar^2\tau_i}{6m_i^*T}-n_i\eta_i \right) ,\\  
 f_B &= & \epsilon_B - s_ B T  \ .
\end{eqnarray}
In the LS EOS,  the summation over many-body interactions in Eq.~(\ref{eq:sky}) is replaced by  a simple term $cn_B^\delta$ and the effective masses are set as the rest masses;  $\alpha_1=\alpha_2=0$.

\subsubsection{Relativistic mean-field models} \label{sec_rmf}
The second general-purpose EOS, STOS (TM1), is based on the RMF  for the free energies of nucleons \cite{shen98a, shen98b} 
 with the TM1 parameter set \cite{sugahara94}. 
In the RMF, nuclear interactions are described by the exchange of mesons. 
The employed Lagrangian is as follows:
\begin{eqnarray}
{\cal L}_{RMF} & = & \bar{\psi}\left[i\gamma_{\mu}\partial^{\mu} -M
-g_{\sigma}\sigma-g_{\omega}\gamma_{\mu}\omega^{\mu}
-g_{\rho}\gamma_{\mu}\tau_a\rho^{a\mu}
\right]\psi  \\ \nonumber
 && +\frac{1}{2}\partial_{\mu}\sigma\partial^{\mu}\sigma
-\frac{1}{2}m^2_{\sigma}\sigma^2-\frac{1}{3}g_{2}\sigma^{3}
-\frac{1}{4}g_{3}\sigma^{4} \\ \nonumber
 && -\frac{1}{4}W_{\mu\nu}W^{\mu\nu}
+\frac{1}{2}m^2_{\omega}\omega_{\mu}\omega^{\mu}
+\frac{1}{4}c_{3}\left(\omega_{\mu}\omega^{\mu}\right)^2   \\ \nonumber
 && -\frac{1}{4}R^a_{\mu\nu}R^{a\mu\nu}
+\frac{1}{2}m^2_{\rho}\rho^a_{\mu}\rho^{a\mu},
\end{eqnarray}
where $\psi$, $\sigma$, $\omega$, and $\rho$ denote nucleons, scalar-isoscalar mesons, vector-isoscalar mesons, and vector-isovector mesons, respectively, and 
$W_{\mu\nu} =\partial^{\mu}\omega^{\nu}- \partial^{\nu}\omega^{\mu} $ and 
$R^{a}_{\mu\nu}= \partial^{\mu}\rho^{a\nu}- \partial^{\nu}\rho^{a\mu} +g_{\rho}\epsilon^{abc}\rho^{b\mu}\rho^{c\nu} $. 
Nucleon-meson interactions are expressed as Yukawa couplings, and isoscalar mesons ($\sigma$ and 
$\omega$) interact with themselves. 
$M$ is the mass of nucleons and is assumed to be $938$ MeV.
In the TM1 parameter set, the masses of mesons---$m_{\sigma}$, $m_{\omega}$, and $m_{\rho}$---and 
the coupling constants---$g_{\sigma}$, $g_{\omega}$, $g_{\rho}$, $g_2$, $g_3$, and $c_3$---are determined;
therefore not only the saturation of 
uniform nuclear matter but also the properties of finite nuclei can be best reproduced \cite{sugahara94}. 

In the mean field theory, mesons are assumed 
to be classical and replaced by their ensemble averages. 
The Dirac equation for nucleons is quantized, and the free energies are 
evaluated based on their energy spectrum. 
The  nucleon number density is expressed as
\begin{eqnarray}
 n_{i}&=&\frac{1}{\pi^2}    \int_0^{\infty} dk\,k^2\,\left( F_F(\nu_i,k) -F_F(-\nu_i,k)  \right)  \label{eq_rmfn} \ ,  
\end{eqnarray}
where $F_F$ is the Fermi distribution function of Eq.~(\ref{eq_ff}) and 
$\nu_i$ is  the kinetic
part of the chemical potential.
The equations of motion of meson fields are written as
\begin{eqnarray}
\sigma_0 &=& -\frac{g_{\sigma}}{m_{\sigma}^2} \displaystyle{\sum_i 
   \frac{\gamma}{2\pi^2} \int_0^{\infty} dk\,k^2\, 
   \frac{M^{*}}{\sqrt{k^2+{M^*}^2}}
   \left(F_F(\nu_i,k) + F_F(-\nu_i,k) \right) }
  -\frac{1}{m_{\sigma}^2}\left(g_{2}\sigma_0^{2}+g_{3}\sigma_0^{3}\right)
,  \\  
\omega_0 &=&
   \frac{g_{\omega}}{m_{\omega}^2} 
   \left(n_p+n_n\right)
  -\frac{c_3}{m_{\omega}^2}\omega_0^3
,  \\  
\rho_0   &=&
   \frac{g_{\rho}}{m_{\rho}^2} 
   \left(n_p-n_n\right) .
   \end{eqnarray}
The effective mass is defined as $M^*=M+ g_\sigma \sigma_0$.
In RMF calculations, $M^*$ and $\sigma_0$ are solved self-consistently for a  given $n_i$.
 Note that the effective mass in the relativistic framework---Dirac mass---differs
  from the non-relativistic ones (see  Eqs.~(\ref{eq_efm1}) and (\ref{eq_efm2})).
  The former  is  defined using the scalar part of the nucleon self-energy in the
Dirac field equation and can only be determined from relativistic approaches, 
whereas the latter parameterizes the momentum dependence of a single-particle potential \cite{dalen10,li18}. 

The chemical potentials including nuclear interactions are expressed as
\begin{eqnarray}
 \mu_{p}       &=& \nu_p      +g_{\omega}\omega +g_{\rho}\rho, \\
 \mu_{n}       &=& \nu_n      +g_{\omega}\omega -g_{\rho}\rho.
 \end{eqnarray}
 where $\nu_p$ and $\nu_n$ are given by Eq.~(\ref{eq_rmfn}).
The energy density and the entropy density  are obtained by
\begin{eqnarray}
\epsilon_B &=& \displaystyle{\sum_i \frac{\gamma}{2\pi^2} 
   \int_0^{\infty} dk\,k^2\, 
   \sqrt{k^2+{M^*}^2}  \left(F_F(\nu_i,k)+F_F(-\nu_i,k)\right) } 
  +\frac{1}{2}m_{\sigma}^2\sigma_0^2+\frac{1}{3}g_{2}\sigma_0^{3}
  +\frac{1}{4}g_{3}\sigma_0^{4}  \\ \nonumber
 & & +g_{\omega}\omega_0 \left(n_p+n_n\right)
  -\frac{1}{2}m_{\omega}^2\omega_0^2-\frac{1}{4}c_{3}\omega_0^{4}
  \\ \nonumber
 & & +g_{\rho}\rho_0 \left(n_p-n_n\right)
  -\frac{1}{2}m_{\rho}^2\rho_0^2,   \\
  s_B & =&  \displaystyle{\sum_{i=p,n,\Lambda} \frac{1}{\pi^2} \int_0^{\infty} dk\,k^2 }
    \left[ -F_F(\nu_i,k)\ln F_F(\nu_i,k)
            -\left(1-F_F(\nu_i,k)\right)\ln \left(1-F_F(\nu_i,k)\right) \right. \nonumber\\
 & & \left. -F_F(-\nu_i,k)\ln F_F(-\nu_i,k)
            -\left(1-F_F(-\nu_i,k) \right)\ln \left(1- F_F(-\nu_i,k) \right) \right] .
\end{eqnarray}

The HS (TM1) EOS \cite{hempel10} and the FYSS (TM1) EOSs  \cite{furusawa11} are also constructed using the TM1 parameter set. 
The RMF with the TM1 parameter set exhibits stiff properties of neutron-rich nuclear matter as listed in Tab.~\ref{tab1_bulk} and discussed in Sec.~\ref{sec_pro}.  

Therefore, Shen et al. \cite{shen20} updated the parameter set and constructed the STOS (TM1e) EOS
by introducing an additional $\omega$-$\rho$ coupling term in  the Lagrangian.
Hempel et al. adopted many parameter sets  in the RMF to construct EOSs: HS (NL3, TMA,DD2, FSUgold, IUF, SFHx, SFHo) \cite{hempel10,steiner13}.  
 Typel  et al. also formulated the GRDF1 and GRDF2 EOSs using parameter DD2 by introducing density-dependent parameters \cite{typel18}.
 Shen et al. developed EOSs with two parameter sets: SHO with FSUgold  and SHT with NL3 \cite{sheng11a,sheng11b}.

\subsubsection{Variational method}
Skyrme-type interactions  and  RMF are phenomenological medium-dependent interactions.
Their model parameters are 
determined to reproduce some nuclear properties such as the nuclear saturation density. 
Togashi et al.  \cite{togashi17}  constructed a general purpose EOS, 
TNTYST (VM) EOS, based on a realistic two-body nuclear potential,  Argonne v18  \cite{wiringa95}, and a three-body potential, UIX \cite{carlson83,pudliner95}, 
 employing the VM \cite{togashi13}.
 The VM is also used for the FYSS (VM) EOS  \cite{furusawa17d}.

Uniform nuclear matter is derived from  nuclear potentials supplemented by three-body forces, 
which reproduce the saturation properties of nuclear matter \cite{togashi13}.
The variational calculations of the free energies of free nucleons  are based on Refs. \cite{kanzawa07,kanzawa09}.
The nuclear Hamiltonian is composed of two-body potentials $V_{ij}$ and three-body potentials $V_{ijk}$, as in the Fermi hypernetted chain  variational calculations \cite{akmal98}
expressed as follows:
\begin{eqnarray}
\label{VM}
H=-\sum^N_{i=1}  \frac{\hbar^2}{2 m} \nabla^2 +\sum^N_{i<j}V_{ij} + \sum_{i<j<k}^{N} V_{ijk},
\end{eqnarray} 
where $m$ is set as the mass of  a neutron. 
The free energies derived from two-body interactions are obtained by 
combining 
 the VM extension of Schmidt and Pandharipande \cite{schmidt79, mukherjee07} with the AV18 two-body potential \cite{wiringa95} and
 the healing distance condition.
 The latter reproduces the internal energy per baryon of symmetric
nuclear matter and neutron matter at zero temperature of Akmal et al. \cite{akmal98}.
The internal energies of three-body interactions are
based on  the UIX three-body potential  \cite{carlson83,pudliner95}. 
The total free energy per baryon is calculated by minimizing it  in relation to the effective masses of nucleons. 
The optimized free energies agree reasonably with those of  Akmal et al. \cite{akmal98} at zero temperature
and those of Mukherjee  \cite{mukherjee09} at finite temperature. 

\subsubsection{Dirac Br\"{u}ckner Hartree Fock}
The Dirac Br\"{u}ckner Hartree--Fock (DBHF) theory also employs
 the bare nuclear interaction adjusted to account for nucleon--nucleon scattering data. 
In contrast to non-relativistic many-body theories  with  a three-body potential  such as the VM EOS, 
the DBHF theory reproduces nuclear saturation properties starting from two-body forces.
In the calculations of the DBHF theory, 
  three integral equations are solved:  Bethe--Salpeter equation,  single-particle self-energy, and Dyson's equation.
In the DBHF calculation \cite{katayama13},  
the Bonn A potential for two-body interactions \cite{brockmann90} and the subtracted T-matrix representation \cite{gross99} are utilized.



In the FYSS (DBHF) EOS,
analytical formulas of
the interaction energy of homogeneous nuclear matter and the effective masses of nucleons are used,
which are fitted to the DBHF calculation 
in the zero-temperature limit \cite{katayama13}. 
The free energy of a baryon of homogeneous matter consists of  kinetic and interaction parts expressed as  $ F_B(n_B,x,T) =  F_{kin}(n_B,x,T)  +  E_{int}(n_B,x)$.
The kinetic part of the energy and  entropy per baryon at a finite temperature \cite{horowitz87, sumiyoshi94} are expressed as 
\begin{eqnarray}
 \label{eq:parakine}
F_{kin}(n_B,x,T) &=& \frac{2}{2 \pi^2 n_B}  \sum_{i=p,n} 
\int_0^{\infty} {\rm d}k k^2  (F_F(\nu_i,k) +F_F(-\nu_i,k))  \nonumber \\
&& \times [ E_i^*(k)  + (M_i-M_i^*) \frac{M_i^*}{E_i^*(k)} ]  - \frac{s T}{n_B},  
\end{eqnarray} 
where $M^*_{i}$  and $E^*_{i}(k)=\sqrt{k^2+M_{i}^{*2}}$ are the effective masses and  the effective energies, respectively.
Entropy densities $s$ and kinetic chemical potentials $\nu_{p}$ and $\nu_{n}$  are the same in the RMF presented in Sec.~\ref{sec_rmf}.
The effective masses are evaluated using the fitting formulas for scalar and vector potentials  in Ref.  \cite{katayama13})---$\Sigma_i^S (k_{Fn},k_{Fp},k)$ and $\Sigma_i^V (k_{Fn},k_{Fp},k)$---as functions of
 the kinetic momentum and the Fermi momentums of protons and neutrons as follows:
\begin{eqnarray}
M_i^*=\frac{M_i+\Sigma^S_i(k_{Fn},k_{Fp},k_{Fi})}{1+\Sigma^V_i(k_{Fn},k_{Fp},k_{Fi})} \ \  (i=p \ {\rm or} \ n).
\end{eqnarray}
The data from DBHF calculations covering a wide range of $x$ are 
 not provided, 
and only the energy densities of symmetric nuclear matter, $E_{snm}$, and neutron matter, $E_{nm}$ 
are obtained using the fitting formulas  as functions of $n_B$ \cite{katayama13}.
The interaction energy in the FYSS (DBHF) EOS is obtained by subtracting the kinetic term from the energy
per baryon at zero temperature as
\begin{eqnarray}
E_{int}(n_B,x)= \{ 4 x(1-x) E_{snm}(n_B) +  (1-2x)^2 E_{nm}(n_B) \}/n_B - F_{kin}(n_B,x,T=0). \label{eq:int}
\end{eqnarray}
Although this quadratic expression is an approximation, 
the obtained results reproduce well
the exact calculations for asymmetric nuclear matter \cite{katayama13}.

\subsubsection{Chiral effective field theory}  \label{sec:chiral}
The chiral effective field theory ($\chi$EFT)  also describes the nuclear interaction microscopically.
It is  based on
the effective chiral Lagrangian, which respects the required symmetries and
 is expanded in powers of  $q/\Lambda_\chi$. 
Here, $q$  denotes a (small) momentum or pion mass,
and $\Lambda_\chi \sim 1$~GeV corresponds to the scale of chiral symmetry breaking.  
Pions appear naturally as degrees of
freedom at low energies, because they are Goldstone bosons related to
the spontaneous breaking of the chiral symmetry in quantum chromodynamics (QCD).
The classification of different contributions to the interaction and connection \cite{aoki12}
are advantages of the $\chi$EFT.  
In addition, 
at each order,  nucleonic contact operators are obtained, which  correspond to short-range interactions,
 whose strength is controlled by  low-energy constants (LECs) fitted to experimental
data.
For example, in  a recent calculation  \cite{holt17},
  LECs  were fitted to the  binding energy and lifetime of a triton. 
The energies of symmetric and neutron matter  were compared based on
 nucleon-nucleon potentials at different orders---$(q/\Lambda_\chi)^2$, $(q/\Lambda_\chi)^3$, and $(q/\Lambda_\chi)^4$---in  the chiral expansion,  corresponding to next-to-leading order, next-to-next-to-leading order, and next-to-next-to-next-to-leading order (N3LO). The cutoff parameter, $\Lambda_\chi$, is set as 415, 450, or 500 MeV.
It is  currently difficult to build  a general-purpose EOS based on the $\chi$EFT that covers a wide range of densities owing to the high computational cost.
However,  the results would be helpful to improve other general-purpose EOSs.
Ab-initio calculations using the $\chi$EFT can be used to constrain the EOS  in neutron-rich conditions \cite{hebeler10, hebeler14}. 
 We refer the readers to the recent review of the $\chi$EFT \cite{drischler21}.

\subsection{Bulk properties} \label{sec_pro}
 Some parameters defined  in a few expansions of the energy per nucleon at $T=0$~MeV,
can be used to describe the  properties of nuclear matter.
 The energy per baryon with the rest mass subtracted, $\omega(n_B,x)$,  is approximately  divided
  between the symmetric nuclear matter energy  and the symmetry energy as 
\begin{equation}
\omega(n_B,x) \approx \omega(n_B, 0.5)  +   \{ \omega(n_B, 0)-\omega(n_B, 0.5) \}(1-2x)^2   \label{eq:int0}.
\end{equation}
Their leading terms or derivatives of the wave number at  $n_B=n_0$
are the following parameters: energy per baryon of symmetric nuclear matter $E_0$,  incompressibility  $K_0$, 
 nuclear symmetry energy $J_0$, and  slope of the symmetry energy $L_0$.
\begin{eqnarray}
 \omega(n_B, 0.5)  &=&  E_0 + \frac{K_0}{2} \left( \frac{n_B-n_0}{3n_0} \right)^2  + ...  \ , \\
  \omega(n_B, 0)-\omega(n_B, 0.5)& = &  J_0 +  L_0 \left( \frac{n_B-n_0}{3n_0} \right)  + ...  \ . 
\end{eqnarray}

Measurements of density distributions and nuclear masses \cite{vries87, audi03,audi12, wang14}
yield  $n_0=0.15-0.16$~fm$^3$ and $E_0=-16\pm 1\,$MeV;
however, their extractions are ambiguous.
$K_0$ is  estimated  using  experimental data from 
 isoscalar giant monopole resonances in heavy nuclei:
$K_0=240\pm 10\,$MeV~\cite{colo04},   
and $K=248 \pm 8\,$MeV~\cite{piekarewicz04}.
Isospin diffusion measurements,
 giant and pygmy resonances,
isoscaling, isobaric analog states,
pion  and kaon production, and measurements of the neutron skin thickness in heavy nuclei,
can constrain the values of $S_0$ and $L_0$.
This can also be achieved by astrophysical observations of the radius, mass, and tidal deformability of NSs
(see  a recent review of NSs and  EOSs   \cite{burgio21}). 
These data led to some constraints: $29.0 < J_0 < 32.7$~MeV and $40.5<L<61.9\,$MeV~\cite{lattimer13}, $30.2<J_0<33.7\,$MeV and $35<L<70\,$MeV \cite{danielewicz14},  $25<L<50$~MeV \cite{zenihiro21},  $42<L<117$ MeV \cite{estee21}, 
and $J_0=38.1 \pm 4.7$MeV and $L =106 \pm 37$~MeV \cite{reed21}.
The last constraints are based on  a recent experiment on the neutron skin thickness of nuclei \cite{adhikari21}.
The isovector part of  nuclear matter is still uncertain, as evidenced by the diversity of the constraints. 

The uniform nuclear matter theory  in $\beta$-equilibrium is also characterized by NS properties.
A set of  the masses and radii  of NSs corresponds to a $\beta$-equilibrium EOS.
The maximum mass of NSs, $M_{max}$,  has to be greater than $2.14^{+1.1}_{-1.0}\ M_{\odot}$,
 which is  the mass of one of the heaviest observed pulsars  \cite{cromartie19}.
A gravitational wave observation of the NS merger,  GW170817 \cite{abbott17,shibata17},
also estimated 
that the radius of an NS with $1.4 M_{\odot}$,  $R_{1.4}$ 
would be smaller than 12.9 km \cite{abbott18}  or 13.4~km \cite{annala18},
based on observed tidal deformability.

The nuclear saturation properties and NS properties of some EOSs of  uniform nuclear matter are summarized  in Tab.~\ref{tab1_bulk}.
Most nuclear and NS properties are satisfied using
the  Skyrme interaction with the SLy4 parameter set and the VM model.
Subsequently, they are utilized to construct  general-purpose EOSs, such as the SRO, RG, TNTYST, and FYSS EOSs.  
The symmetric matter of the DBHF EOS is soft, and 
it yields the highest nuclear saturation density of symmetric nuclear matter, $n_0$.
This deep effective potential for $x=0.5$ (low $E_0$ and large $n_0$) is  one of the characteristics  of the DBHF theory.
For a long  time, the TM1 parameter set was used; however, in this decade, it has been discovered that its NS matter  is extremely stiff (large values of $J_0$, $L_0$, and $R_{1.4}$).
However, a recent experiment may revive it \cite{reed21, adhikari21}, and  we cannot conclude which EOS for uniform nuclear matter is the best at present.


\begin{table}[t]
\begin{tabular}{c|cccccc|cc}
\hline
   Model  &   type &  $n_0$ & $E_0$  & $K_0$ & $J_0$  & $L_0$   & $R_{1.4}$  & $M_{\rm max}$\\
   &   & (fm$^{-3}$)   & (MeV)    & (MeV)  & (MeV)  & (MeV)  & (km) & ($M_{\odot}$)     \\
 \hline
LS220 \cite{lattimer91} & Skyrme  &  0.155& -16.6  &  220   & 28.6 & 73.8 & 12.7 & 2.07 \\
SLy4  \cite{chabant98}  & Skyrme  & 0.159  &  -16.0  & 230  & 32.0  & 45.9 &  11.7 &  2.05  \\
\hline
TM1  \cite{sugahara94} & RMF &  0.145& -16.3  &  281   & 36.9 & 111 & 14.5 & 2.23 \\
TM1e  \cite{shen20}  & RMF & 0.145& -16.3  &  281   & 31.4 & 40.0 & 12.5 & 2.12 \\
DD2 \cite{typel10}  &  RMF & 0.149& -16.0  &  243   & 31.7 & 55.0 & 13.1 & 2.42 \\
SFHo \cite{steiner13} &  RMF & 0.158& -16.2  &  245   & 31.6 & 47.1 & 11.9 & 2.06 \\
SFHx \cite{steiner13} &  RMF & 0.160& -16.2  &  239   & 28.7 & 23.2 & 12.0 & 2.13 \\
\hline
 VM  \cite{togashi13}& micro & 0.160 &  -16.0  &  245   & 30.0 & 35.0 & 11.5  & 2.21 \\ 
DBHF \cite{katayama13} & micro & 0.179& -16.6  &  232   & 34.5 & 66.8 & 12.9  & 2.34 \\    
$\chi$EFT \cite{holt17} & micro & 0.166& -15.9  &  250   & 31.3 &  41.9 & -  & - \\ 
\hline
\hline
\end{tabular}
\caption{\label{tab1_bulk}%
Bulk properties of EOSs for uniform nuclear matter.}
\end{table}

KOKO
\begin{figure}
\begin{center}
\includegraphics[width=10cm]{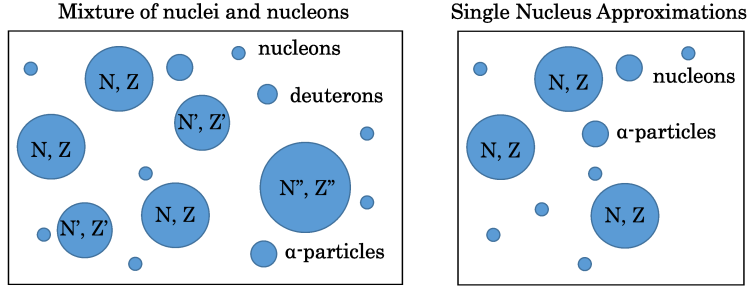}
  \caption{Schematic pictures of nuclear matter at a thermodynamical state for NSE EOSs (left) and for SNA EOSs (right).
\label{fig_sna}}  
\end{center}
\end{figure}

\section{Non-uniform Nuclear Matter  \label{sec:num}}
This section  reviews  non-uniform nuclear matter, which is a mixture of nucleons and nuclei as illustrated in Fig.~\ref{fig_sna}.
The total Helmholtz free energy density for all ingredients  is  expressed as:
\begin{eqnarray}
\ f_B &=& f_{pn}+\sum_{N,Z}  f(N,Z),  \label{eq:total}
\end{eqnarray}
where $f_{pn}$ is the free energy  density of the nucleons outside the nuclei
 and $f(N,Z)$ is that of individual nuclei
 with  neutron number $N$ and proton number $Z$. 
 For uniform nuclear matter, 
 there are no contributions of nuclei, and  $f_B=f_{pn}$.

\subsection{Single nucleus approximation}
In the first two general EOSs, LS and STOS EOSs, the SNA  is assumed; 
the entire ensemble of nuclei is replaced by a single representative nucleus as illustrated in Fig.~\ref{fig_sna}.
The free energy density is expressed as:
\begin{eqnarray}
\label{eq:total2}
\ f_B =f_{pn} + f_{\alpha} + f_{rep}, 
\end{eqnarray}
where $f_\alpha$  and $f_{rep}$ are free energy densities of $\alpha$ particles and  representative nuclei, respectively.
In SNA EOSs, the ensemble of light nuclei is frequently represented by $\alpha$ particles. 

\subsubsection{Compressible liquid drop model} \label{sec_cldm}
In the LS EOS,  the compressible liquid drop model (CLDM) is adopted \cite{lattimer91}.
The free energy of the representative nucleus is expressed by
\begin{eqnarray}
\label{eq:cldm}
\ f_{rep}=A n_{rep} \{ \omega(n_B^{in}, x^{in}, T) + F_{C} (n_B^{in}, x^{in}, u) + F_{S} (n_B^{in}, x^{in}, u)  + F_{t} (n_B^{in}, u) \}
\end{eqnarray}
where  $n_B^{in}$ and $x_p^{in}$ are the density and proton fraction of the nucleons inside nuclei, respectively.
The uniform nuclear matter theory, Skyrme interactions, is used to calculate the bulk energy, $\omega$.
The other Coulomb, surface, and translational energies---$F_C$, $F_S$, and $F_t$---and the volume fraction in the cell, $u$, are described in Sec. \ref{sec.intro.nse}.
In the calculation,  $n_B^{in}$, $x_p^{in}$, $u$,  $\alpha$-particle density $n_\alpha$ for $f_\alpha$,  density and charge fraction of dripped nucleons  $n_B^{out}$ and $x_p^{out}$ for $f_{pn}$, as well as radius of the cell  $r$ (a total of seven quantities) are optimized to reduce the total free energy density under charge and baryon conservations.
The optimization of $n_B^{in}$ corresponds to compression or decompression of nuclei.

\subsubsection{Thomas-Fermi approximation}
The STOS and TNTYST EOSs  \cite{shen98a, togashi17} use 
the Thomas-Fermi (TF) approximation for finite nuclei with dripped nucleons and $\alpha$ particles \cite{oyamatsu93}.
Thus, the TNTYST EOS can be represented as STOS (VM).
The Wigner-Seitz cell is assumed to be a sphere whose volume is the same as the unit cell in a body centered cubic lattice.
The nucleon distribution in the cell, $n_{n/p}(r)$, where  $n$ and $p$ denote neutrons and protons, respectively, 
is defined  as
\begin{equation}
n_{n/p}\left(r\right)=\left\{
\begin{array}{ll}
\left(n_{n/p}^{in}-n_{n/p}^{out}\right) \left[1-\left(\frac{r}{R_{n/p}}\right)^{t_{n/p}}
\right]^3 +n_{n/p}^{out},  & 0 \leq r \leq R_{n/p} \ ,  \\
n_{n/p}^{out},  & R_{n/p} \leq r \leq R_{cell}  \ , \\
\end{array} \right. \label{eqshen}
\end{equation}
where $r$ is the distance from the center of the nucleus and
$R_{cell}$ denotes the radius of the Wigner-Seitz (WS) cell;
$n_{n/p}^{in}$ and $n_{n/p}^{out}$ are the densities at $r=0$ and $ r \geq R_{n/p} $;
and $R_{n/p}$ and $t_{n/p}$ are the boundary and the relative surface thickness
of the representative nucleus.
The distribution function of $\alpha$ particles, $n_{\alpha}(r)$,
is assumed as
\begin{equation}
n_{\alpha}\left(r\right)=\left\{
\begin{array}{ll}
-n_{\alpha}^{out} \left[1-\left(\frac{r}{R_p}\right)^{t_p}
\right]^3 +n_{\alpha}^{out},  & 0 \leq r \leq R_p \ , \\
n_{\alpha}^{out},  & R_{p} \leq r \leq R_{cell}  \ . \\
\end{array} \right. 
\end{equation}
The local free  energy density is calculated using  the uniform nuclear matter theory 
(the RMF theory  with the TM1 or TM1e parameter set  in STOS EOSs or the VM in the TNTYST EOS).

\begin{figure}[h]
\begin{center}
\includegraphics[scale=0.2]{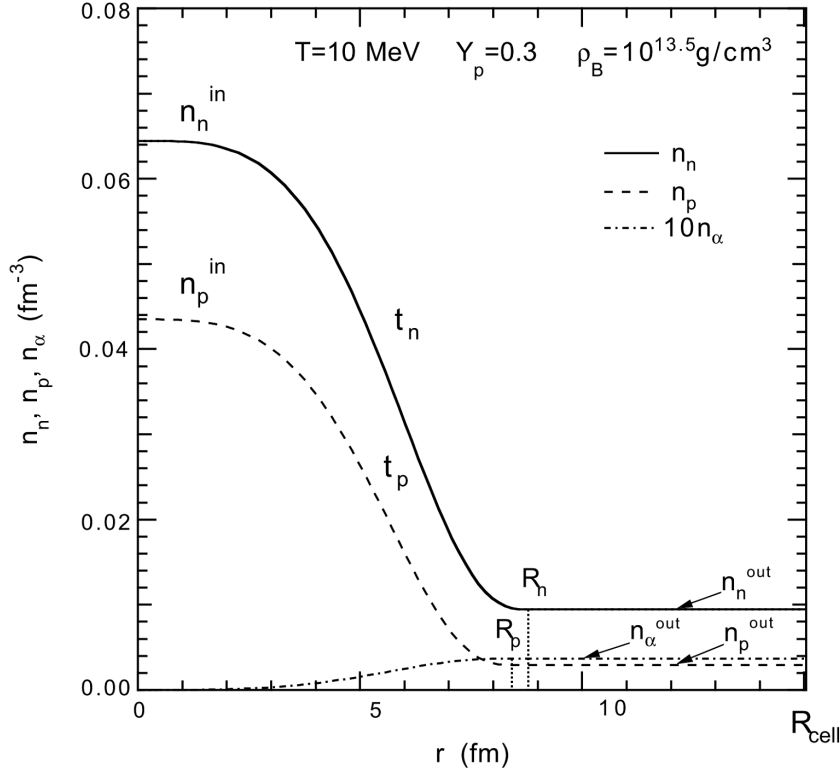}
  \caption{
Distributions of neutrons, protons, and $\alpha$ particles  in WS cell of STOS EOS \cite{shen98b} at
$T = 10$ MeV, $Y_p = 0.3$, and $\rho_B = 10^{13.5} \rm{g/cm^3}$ are plotted as solid, dashed, and
dot-dashed curves, respectively. To show $\alpha$ particle distribution clearly, it is
enlarged by factor of 10.}
 \label{fig.shen}
\end{center}
\end{figure}

Figure~\ref{fig.shen} (Fig.~1 in \cite{shen98b}) shows the distributions of nucleons and $\alpha$ particles at
$T=10$ MeV, $Y_p=0.3$, and $\rho_B=10^{13.5} \rm{g/cm^3}$.
There are two constraints for the given  $Y_p$ and $n_B$, which are expressed as follows: 
\begin{eqnarray}
n_B &=& \frac{3}{4 \pi R_{cell}^3}  \int_{cell} \left[\, n_n\left(r\right) +
                         n_p\left(r\right) + 4n_{\alpha}\left(r\right) 
                      \,\right] d^3r   \ ,
 \\ 
 Y_p n_B &=& \frac{3}{4 \pi R_{cell}^3}  \int_{cell} \left[\, n_p\left(r\right) + 2n_{\alpha}\left(r\right) 
                      \,\right] d^3r \ . 
\end{eqnarray}
The equilibrium state is obtained by minimizing the free energy density with respect to  
the ten variables under the two constraints: $R_{cell}, n_n^{in}, n_n^{out}, R_n, t_n,$ $n_p^{in}, n_p^{out},$ $R_p, t_p,
and  n_{\alpha}^{out}$.

 Thus, in the SNA, the structure of a representative nucleus in the WS cell is solved to reproduce the minimum free energy density under a given condition.   
In the other SNA EOS (LS EOS \cite{lattimer91}), the thickness of the nuclear surface
 between $n_{p/n}^{in}$ and $n_{p/n}^{out}$ in the cell  is not considered, which corresponds to the limit of $t_{n/p}=\infty$
 in Eq.~(\ref{eqshen}).

\subsection{Nuclear statistical equilibrium and extensions} \label{sec.intro.nse}
Nuclear EOSs present some relations among thermodynamical properties, such as the pressure as a function of $\rho_B$,
 $T$, and $Y_p$. In addition, the composition of nuclear matter (number densities of  nucleons and all nuclei) is determined by the 
NSE EOS, in which
the free energy of a model is minimized using the parameters of the free energy. 
In the following,  first the standard NSE is introduced and subsequently several extended NSE EOSs are discussed.
The free energy of  nuclei, $f(N,Z)$, is expressed as 
\begin{equation}
f(N,Z)=n(N,Z)  \{ F_g(N,Z) + F_t(N,Z) \}
\end{equation}
where $n(N, Z)$  represents the nuclear number density, $F_g(N, Z)$ represents the nuclear gross energy 
corresponding to the nuclear mass  (Secs.~\ref{sec_hnfree} and \ref{sec_lnfree}),
and $F_t(N,Z)$ represents the translational energy (Sec.~\ref{sec_trans}).

\subsubsection{Standard NSE}   \label{sec_snse}
The standard NSE is used in various calculations: the final phase of stellar evolution,
which is the initial condition of CCSN simulations, 
 initial condition for nucleosynthesis of the cooling ejecta from supernova explosions or NS mergers,  
low-density part  in hybrid EOSs \cite{schneider17}  (see Sec. \ref{sec_hybrid}), and  weak rate data table \cite{juodagalvis10}.
 As already noted, in the NSE at temperatures above $T \sim 0.4$ MeV,  nuclear reactions via strong and electromagnetic interactions reach  equilibrium as follows:
\begin{eqnarray}
(N, Z) \longleftrightarrow (N, Z-1) + p \ ,  \\
(N, Z) \longleftrightarrow (N-1, Z) + n \ .
\end{eqnarray}
  
The abundances of nuclei under NSE for a given $\rho_B$, $T$, and $Y_p$ 
are obtained by minimizing the model free energy with respect to the many parameters under 
the two  constraints of  mass and charge conservations as
\begin{equation}
 n_p+n_n+\sum_{N,Z}{(N+Z) n(N, Z)}=n_B=\rho_B/m_u  ,  \label{eq_const1}
\end{equation}
\begin{equation}
 n_p+\sum_{N,Z}{Z n(N, Z)}=Y_p n_B ,  \label{eq_const2}
\end{equation}
where $m_u$ is the atomic mass unit.
The parameters are $n(N, Z)$ and number densities of free protons $n_p$ and neutrons $n_n$. 
By introducing Lagrange multipliers $\alpha$ and $\beta$ for these constraints, the minimization
conditions are expressed as 
\begin{eqnarray}
 \frac{\partial}{\partial {n_p}} \{f_B - \alpha(n_p+n_n+\sum_{N,Z}{(N+Z) n(N, Z)} - n_B   )  - \beta(n_p+\sum_{N,Z}{Z n(N, Z)}-Y_p n_B) \} &=&0,  \\
 \frac{\partial}{\partial {n_n}} \{f_B - \alpha(n_p+n_n+\sum_{N,Z}{(N+Z) n(N, Z)} - n_B   )  - \beta(n_p+\sum_{N,Z}{Z n(N, Z)}-Y_p n_B) \} &=&0.
\end{eqnarray}
The differentials yield the relations of Lagrange multipliers $\alpha$ and $\beta$ and the chemical potentials of
protons  and neutrons  as follows: 
\begin{eqnarray}
 \alpha & = &\frac{\partial f_B}{\partial n_n}= \mu_n, \\
 \beta & = & \frac{\partial f_B}{\partial n_p} -  \frac{\partial f}{\partial n_n}= \mu_p -\mu_n.
\end{eqnarray}
In contrast, differentiation with respect to the number densities of nuclei yields the relations between the chemical 
potentials of nuclei, $\mu(N, Z)=\partial f_B /\partial n(N ,Z)$, and those of protons and neutrons:
\begin{equation}
\label{eq:nse}
 \mu(N,Z)=N \mu_n + Z \mu_p.
\end{equation}

Here, we assume nucleons and nuclei as ideal Boltzmann gases with constant masses as follows:
\begin{eqnarray}
f(N, Z) &=& n(N, Z)  \left[  m(N, Z)  +T \ln \left\{ \frac{n(N,Z)}{g(N, Z) n_{Q}(N,Z)} \right\}  - T \right]  \label{eq_bg} \\
 n_{Q}(N,Z) & = &  \left(\frac{m(N,Z) k_B T}{2\pi \hbar ^2}   \right)^{3/2} \ , 
\end{eqnarray}
where  $n_{Q}(N,Z)=\lambda_T(N,Z)^{-3}$  with thermal wavelength 
$\lambda_T=\left( \frac{ 2\pi \hbar ^2}{m k_B T}   \right)^{1/2}   $. When $n(N, Z) > n_{Q}(N, Z)$, 
quantum statistical properties such as degeneracy of Fermi particles  or condensation of Bose particles emerge,
although such states at high densities  and low temperatures do not appear in supernova matter \cite{furusawa20c}.
The partial differentials of Eq.~(\ref{eq_bg}) with respect to $n(N, Z)$ and Eq.~(\ref{eq:nse})
lead to  $n(N, Z)$ as follows:
\begin{eqnarray}
n(N, Z) = g(N, Z)\, n_{Q}(N,Z) \exp \left( \frac{Z \mu_p+ N \mu_n - m(N, Z)}{T} \right), \label{eq:nin} 
\end{eqnarray}
where   $g(N, Z)$ is the degeneracy factor of the nucleus. 
In the standard NSE EOS,  only the spin degree of freedom is considered as $g(N, Z)=g_0(N,Z)$,
because the nuclei are assumed to be unexcited. 
 For $m(N,Z)$, some experimentally known masses in vacuum  \cite{audi12,huang21,wang21}  are utilized.
Eq.~(\ref{eq:nin}) is a transformed expression of the Saha equation. 


\subsubsection{Extended NSE}
\paragraph{Free energies of dripped nucleons} \label{sec_nucleon} 
The extended NSE EOS is complex and $f_{pn}$ is not simply the free energies of Boltzmann gases, different from  the standard NSE EOS. 
In  some extended NSE EOSs such as the HS and FYSS EOSs,  
the excluded volume is considered in the free energy density of 
free nucleons, which is expressed as
\begin{equation}
\label{fpn}
f_{pn} =  \xi (n_p'+n'_n)  \omega(n'_p+n'_n,\frac{n'_p}{n_p'+n'_n},T),
\end{equation}
where $\xi=1-V_{ex}/V$, $V$ is the total volume, $V_{ex}$ is the excluded volume that is occupied by nuclei and expressed as
 $V_{ex}=\sum_{N,Z} V_{N}(N,Z)$, $V_N$ is the nuclear volume  as defined as $V_N=A/n_s(N,Z, 0)$ with  saturation density $n_s(N, Z, T=0)$ (explained in Sec.~\ref{sec_hnfree}),
 $n'_{p/n}$ is the local number density of protons
 and neutrons in the unoccupied volume ($V-V_{ex}$) for nucleons and is defined as $n'_{p/n}=(N_{p/n})/(V-V_{ex})$, 
$N_{p/n}$ is the number of free protons and neutrons, and $\omega$ is the free energy per baryon of uniform matter of nucleons.
$\omega$ as a function of $n_B$,  $Y_p$, and $T$ is calculated using a model for uniform nuclear matter. 
For instance, the models in the FYSS (TM1, VM, DBHF) EOSs  \cite{furusawa17a, furusawa17d,furusawa20a} 
are the RMF theory with the TM1 parameter set \cite{sugahara94},  VM \cite{togashi13}, and  DBHF approach \cite{katayama13}, respectively.
Combinations of known uniform and non-uniform matter  models are summarized in  Table~\ref{tab1_eosdata}.

\paragraph{Mass of heavy nuclei} \label{sec_hnfree}

In most EOSs such as the HS and RG EOSs,  
the nuclear gross energy is not explicitly dependent on
 the temperature and expressed as an experimentally determined or a theoretically-predicted nuclear mass isolated in vacuum 
with Coulomb energy shifts.
\begin{eqnarray}  
 F_g(N, Z)  &= &m(N, Z) + \Delta_{Coul}(N, Z) , \\
 \Delta_{Coul} (N, Z) &=& \frac{3}{5} \left( \frac{4 \pi}{3} \right)^{-1/3}  e^2 n_{0}^2 \left(\frac{Z }{A}\right)^2 {V_{N}}^{5/3} 
 \left( -\frac{3}{2}u^{1/3}+\frac{1}{2}u \right) \ ,   \label{eq_cshift}
\end{eqnarray}
where $e$ is the elementary charge and  $u=V_{N}/V_{C}$ is the filling factor defined using nuclear volume  $V_{N}=A/n_0$  
and cell volume $V_{C}=Z/n_e$.
 Temperature dependence of nuclear energy is included in the internal degree of freedom, $g(N, Z, T)$, which is explained subsequently in Sec.~\ref{sec_trans}.

In the GRDF1 and GRDF2 EOSs \cite{typel18}, the  mass shift of heavy nuclei by  Pauli blocking is introduced as
\begin{eqnarray}  
 F_g(N, Z) =m(N, Z) + \Delta_{Coul}(N, Z)  +\Delta_{Pauli}(N, Z) .
\end{eqnarray} 
In the EOSs, the excluded volume effects introduced in Sec.~\ref{sec_trans} are not considered
 and the dissolution density at which nuclei disappear  is determined   by a simplified function of a  mass number.
The dissolution density for $\Delta_{Pauli}$ is expressed as $n^{diss}(A)=\frac{n_0}{3+28/A}$   in  GRDF1 EOS and $n^{diss}(A)=\frac{n_0}{2+32/A}$ in GRDF2 EOS,
and in  both,  $n^{diss}(4)=n_0 /10$  is obtained.

In the FYSS EOS \cite{furusawa20a}, the nuclear gross energy  is dependent on the temperature, and 
that  of heavy nuclei with $6 \leq Z$ consist of the bulk, Coulomb, surface, and shell energies as follows: 
\begin{equation}
F_g(N,Z)=F_{bulk}(N,Z) +F_{Coul}(N,Z)  +F_{surf}(N,Z)+F_{shell}(N,Z) \ .
\end {equation}
The nuclear bulk energy is  evaluated using the same model for dripped nucleons and uniform nuclear matter as 
$F_{bulk}(N,Z)=  A \{\omega(n_s,Z/A,T) \}$, where  $n_{s}(N, Z, T)$ is 
defined as the density at which the free energy per baryon, $\omega (n_B,Z/A,T)$, reaches its local minimum value around $n_0$.
For nuclei,  
when experimental or theoretical mass data are available (\cite{audi03, koura05, audi12,wang14},
the shell energies are defined as positive deviation s of the mass data from the gross part of the liquid-drop mass model, $F_{0shell}(N,Z)=M_{data}(N,Z)- [F_{bulk}(N,Z)+F_{surf}(N,Z) +F_{Coul}(N,Z)]_{n_B=0,T=0}$. 
 The temperature dependence is introduced as
$ F_{shell}(T)=F_{0shell} \ \tau /{\rm sinh}\tau$, 
 where $\tau =2 \pi^2 T/(41 A^{-1/3})$ \cite{nishimura14}.

The Coulomb and surface energies in the FYSS EOS
are calculated using the liquid-drop model in a WS cell that contains dripped nucleons and uniformly distributed electrons. 
The sum of cell volumes $V_C$ for all nuclei is the total volume of the system, $V$. 
The nuclear shape is assumed to change from a droplet to a bubble, as it is dependent on $u(N,Z)=V_N/V_C$. 
The nuclear volume is set as $V_{N}=A/n_{s}(N, Z)$, and a cell volume is obtained as $V_{C}=(Z - n'_p V_N)/(n_e-n'_p)$, 
under the assumption of charge neutrality in the cell including dripped protons.
In Eq.~(\ref{eq_cshift}), dripped protons are neglected.
A smooth function of $u$ \cite{ravenhall83,lattimer91} is  used for the Coulomb energy.
The surface energy is expressed as product of the nuclear surface area and the surface tension.
They are represented as
\begin{eqnarray}
F_{Coul}(N,Z)&=&\displaystyle{\frac{(36 \pi)^{1/3}}{5}
e^2 n_{s}^2 \left(Z/A - n'_p/n_{s}  \right)^2 V_{CAZ}^{5/3}
\mathcal{D}_C(u)} ,  \\
\mathcal{D}_C(u)&=&\frac{u^{5/3}(1-u)^2D(u)+u^2(1-u)^{5/3}D(1-u)}{u^2+(1-u)^2+C_{cp}u^2(1-u)^2}, \\
    D(u) &=&1-\frac{3}{2}u^{1/3}+\frac{1}{2}u  \ , \\
F_{surf}(N,Z)&=& 4 \pi \left(\frac{3 V_N}{4 \pi}\right)^{2/3} \sigma (T,n'_n,n'_p)  \mathcal{D}_S(u) , \\
\mathcal{D}_S(u) & = & \frac{ u^{2}(1-u)^{2/3}+u^{2/3} (1-u)^{2}} {u^2+(1-u)^2+C_{sp}  u^2 (1-u)^2} , \\
 \sigma (T,n'_n,n'_p) & = & 
\sigma_0  \left\{ \frac{16 + C_{st}} {(1-Z/A)^{-3} + (Z/A)^{-3}  +C_{st}} \right\} \left( \frac{T_c^2-T^2}{T_c^2+T^2}\right)^{5/4}
\left(1-\displaystyle{\frac{n'_p+n'_n}{n_{s}}} \right)^2 . \nonumber
\end{eqnarray}
The expressions for $\mathcal{D}_C$ and $\mathcal{D}_S$ asymptotically approach the factors for nuclear droplet and bubble phases.  
Coefficients $C_{cp}=-0.863$ and $C_{sp}=4.19$ are set to reproduce the Coulomb and surface energies of a nuclear slab phase at $u=0.5$.
The values of $\sigma_0$ and $C_{st}$
are optimized to minimize the sum of the positive shell energies per baryon in vacuum.
For instance, $(\sigma_0,C_{st})=(1.01$~MeV/fm$^2,$ 42.5~MeV)  for the FYSS (VM) EOS.
The critical temperature, $T_c(N,Z)$, is defined as the temperature
 at which $(\partial P_{bulk}/\partial n_B)|_{x=Z/A}=0$ and $(\partial^2 P_{bulk}/\partial n_B^2)|_{x=Z/A)}=0$, 
 where $P_{bulk}=n_B^2 \partial \omega(n_B,x,T)/\partial n_B$ \cite{furusawa18a}.
 

Thus, the modeling of the free energy density of nuclei differs from model  to model.
However, in all above EOSs, the change in the Coulomb energies  of nuclei 
at high densities  is included.
The  in-medium effects during a core collapse  \cite{furusawa20b} are introduced  subsequently in Sec.~\ref{sec_cc}.

\paragraph{Mass  of light nuclei} \label{sec_lnfree}
The treatment of the gross energy of light nuclei also differs with the EOS. 
In the HS and RG EOSs, their free energy models are the same as those of heavy nuclei (mass data with Coulomb energy shifts).
In the GRDF1 and GRDF2 EOSs \cite{typel18}, 
Pauli blocking and coupling of nucleons inside the nuclei to  meson fields are calculated 
 based on predictions from microscopic quantum statistical calculation \cite{roepke09}.
The Pauli energy shifts  \cite{typel10}
for deuterons,  $^3 \rm H$,  $^3 \rm He$, and $\alpha$ particles,
which are fitted to the results of quantum statistical calculations \cite{roepke09},
are expressed as
\begin{eqnarray}
\label{DEP}
 \Delta_{Pauli}(n_{pl},n_{nl},T)  & =&  - \tilde{n} 
 \left[1 + \frac{\tilde{n}}{2\tilde{n}^{0}(T)} \right]
 \delta B (T), \\
 \delta B(T) &=&\left\{ \begin{array}{ll}
  a_{1} / T^{3/2}  \left[ 1/\sqrt{y}  -  \sqrt{\pi}a_{3} \exp  \left( a_{3}^{2} y \right) 
 {\rm erfc} \left(a_{3} \sqrt{y} \right) \right]   &  {\rm{for}}   \  d, \\
  a_{1} /\left( T  y \right)^{3/2}  \ \   &  {\rm{for}} \   t,h,\alpha, 
\end{array} \right.
\end{eqnarray}
where  $\tilde{n} = 2(Z \ n_{pl} +N \ n_{nl}) /A$, $y = 1+a_{2}/T$, 
 $n_{pl/nl}$ represents the local proton and neutron number densities that include light nuclei as well as free nucleons, and $\tilde{n}^{0}(T) = B^{0}/\delta B(T)$, where $B^{0}$ is the  binding energy in vacuum.
 
In the FYSS EOSs, 
the free energy of light nuclei 
is based on the Pauli energy shift and the self-energy shift, expressed as  
  $ F(Z,N)=M_{data}(Z,N) + \Delta_{Pauli}  + \Delta_{self} + \Delta_{Coul}$. 
 \label{eqquasi}
The self-energy shift,  $\Delta_{Self}$, is the sum of
 the self-energy shifts of the individual nucleons composing the light nuclei, expressed as  
 $\Delta E_{n/p}=\Sigma^0_{n/p}(T,n'_p,n'_n)-\Sigma_{n/p}(T,n'_p,n'_n)$, where
  $\Sigma^0$ and $\Sigma$ are the vector and scalar potentials of nucleons and 
the contribution from their effective masses $\Delta_{\rm eff.mass} = s\left(1-m^{\ast}/m_B \right)$,  where $m^{\ast}=m_B-\Sigma_{n/p}(T,n'_p,n'_n)$ and $m_B=939$~MeV:
\begin{equation}
\label{SE}
\Delta_{self}(n'_{p},n'_{n},T)= (A-Z) \Delta E_{n}+ Z \Delta E_{p}+\Delta_{\rm eff.mass}\ . 
\end{equation}
For potentials $\Sigma^0$ and $\Sigma$, the  parametric formula
 for the RMF with the DD2 parameter set is utilized (see  Eqs.~(A1) and (A2) in the Ref. \cite{typel10}). 
The self-energy shift for the other light nuclei  ($Z\leq5$)  are set as zero. 
The Pauli energy shifts for the other light nuclei  are calculated 
similar as that  for  $\alpha$ particles.    
 Parameters $a_{1}$, $a_{2}$, $a_{3}$, and $s$ are listed in Table~I of Ref. \cite{typel10}.

\paragraph{Translational energies of nuclei}\label{sec_trans}  
In the HS, FYSS, and RG EOSs,  the translational energies of nuclei are expressed as Boltzmann gases with excluded volume effects
as follows:
\begin{eqnarray}%
 F_{t}(N,Z) &=& T \left\{ \log \left(\frac{ n(N,Z)/\kappa}{g(N,Z,T) n_Q(N,Z)  }\right)- 1 \right\}  \ ,   \label{eq:tra} \\
 n_{Q}(Z.N) &=& \left( F_g T/2\pi \hbar ^2  \right)^{3/2}  \ ,
\end{eqnarray}
where $\kappa=1-n_B/n_0$ and $g(T,N,Z)$  is a factor for internal degrees of freedom.
At $n_B=n_0$, the number density of nuclei, $n(N,Z)$, becomes zero, as expressed subsequently in  Eq.~(\ref{eq:nio}).

In the HS and RG, SRO EOSs, the temperature dependence of nuclear free energies is considered in the internal degree of freedom, $g(Z,N,T)$.
In the HS EOSs \cite{hempel10,steiner13},
the following formula \cite{fai82} is utilized:
\begin{equation}
\label{eq:ex}
g(N, Z, T)=g_{A}^0 +\frac{c_1}{A^{5/3}}\int_0^{16.2 A} dE e^{-E/T}\exp\left(\sqrt{2 a(A) E}\right) , \label{eq:fai} 
\end{equation}
where $a(A)=(A/8)(1-c_2 A^{-1/3})$MeV$^{-1}$, $c_1=0.2$MeV$^{-1}$, $c_2=0.8$,
and $g_{A}^0=1$ for even nuclei and $g_{A}^0=3$ for odd nuclei, and
the upper bound of the integral is set as a typical bulk energy, $16.2 A$.
In the RG EOS, the data table of level densities is employed \cite{egidy05}. 
Partition-function data  \cite{rauscher00}  are used in the SRO EOS.
Some of these temperature dependences were compared  in a systematical study  \cite{furusawa18b}.

In the FYSS EOS,   $g(T,N,Z)= (g_0(N,Z) -1)\displaystyle{\frac{\tau}{{\rm sinh}\tau}} +1$, where $\tau =2 \pi^2 T/(41 A^{-1/3})$
for heavy nuclei ($Z>5$). 
The temperature dependence of introduced in line with the  washout of the shell energy, $F_{shell}$, introduced in subsection~\ref{sec_hnfree}   \cite{furusawa17a}.
For light nuclei ($Z<6$), the excluded volume effects and the shell washout are not included because $\kappa=1$ and $g(T,N,Z)=g_0(N,Z)$. 

%

\subsubsection{Minimization of free energy and thermodynamical properties}\label{sec_min} 
The abundances of nuclei as a function of $\rho_B$, $T$, and $Y_p$ is 
obtained by minimizing the derived model free energy with 
respect  to the number densities of nuclei and nucleons under the constraints of baryon and charge conservations, 
as expressed in Eqs.~(\ref{eq_const1}) and (\ref{eq_const2}).
Differentiating the free energy for nuclei with respective to the number densities of nuclei, $n(Z,N)$, 
they can be expressed as follows:
\begin{eqnarray}
n(Z,N)  &=&  \kappa g(Z,N,T) \, n_{Q}(Z,N) \exp \left( \frac{\mu (Z,N)- F_g(Z,N)}{k_B T} \right). \label{eq:nio}   
\end{eqnarray} 
The chemical potential of nuclei are calculated using those of nucleons, $\mu(Z,N)=Z \mu_p+N \mu_n$.

In most general-purpose EOSs, the number densities of nuclei depend only on  $\mu_p$ and $\mu_n$.
Two conservation equations, i.e., Eqs.~(\ref{eq_const1}) and (\ref{eq_const2}), can be solved  for   $\mu_p$ and $\mu_n$ 
similar to the standard NSE  discussed in  Sec.~\ref{sec_snse}.
In the FYSS EOS, the gross energies of nuclei, $F_g$, depend on the number densities of dripped nucleons, $n'_p$ and $n'_n$, and, hence, 
the number densities of nuclei depend on $\mu_p$, $\mu_n$, $n'_p$ and $n'_n$.
Their relations are as follows:
\begin{eqnarray}
\mu_{p} &=& \frac{\partial f_B}{ \partial n_{p}}= \mu'_{p} (n'_p,n'_n,T)+ \sum_{N,Z} n(Z,N) {\frac{\partial F_g(Z,A)}{ \partial n_{p}}}   \ ,  \label{eq:chem1} \\ 
\mu_{n}  &=&\frac{\partial f_B}{ \partial n_{n}}= \mu'_{n} (n'_p,n'_n,T)+ \sum_{N,Z} n(Z,N) {\frac{\partial F_g(Z,A)}{ \partial n_{n}}} \ , 
\label{eq:chem2}
\end{eqnarray}
where  $\mu'_{p}$  and  $\mu'_{n}$ are the chemical potentials of nucleons in a vapor of volume $V'$, 
which are obtained from uniform matter calculation, $\omega$.
 The number densities of the nucleons in the total volume, $n_{p}$ and $n_n$,
are different from local number densities $n'_{p}$ and $n'_n$ as defined in Sec.~\ref{sec_nucleon}.
The second term originates from the dependences of $F_g(N, Z)$ on $n'_p$ and $n'_n$ and is not included in the standard NSE EOS and other general-purpose EOSs.
In the FYSS EOS,  not only the two conservation equations, i.e., Eqs.~(\ref{eq_const1}) and (\ref{eq_const2}), but also
 the two chemical potential in Eqs.~(\ref{eq:chem1}) and~(\ref{eq:chem2})  must be solved for the four variables: $\mu_p$, $\mu_n$, $n'_p$ and $n'_n$.

After minimization, the free energy density is obtained with the abundances of various nuclei 
and free nucleons as a function of $\rho_B$, $T$, and $Y_p$. 
Once the free energy density, $f_B$, is obtained, other physical properties are derived by its partial differentiation.
The baryon pressure and the baryon entropy density are calculated using the following thermodynamic relations:
\begin{eqnarray}
p_B & =&\left[ n_B^2 \frac{\partial}{\partial n_B } \left(\frac{f_B}{n_B}\right) \right]_{T,Y_p}  \ ,  \label{eq:pre} \\
 s_B& =& -\left(\frac{\partial{f_B}}{\partial {T}} \right)_{\rho_B,Y_p}  \   \label{eq:ent} \ .
\end{eqnarray}
The baryon internal energy density is calculated using the following  thermodynamical relation:
\begin{eqnarray}
 \epsilon_B & =& f_B -T s_B  \  .  \label{eq:ene}
\end{eqnarray}
In some EOSs \cite{shen98b}, the entropy and the internal energy are calculated  and Eq.~(\ref{eq:ene})  yields the free energy.
The consistency of the thermodynamic quantities is used to evaluate the numerical  calculations. 
\begin{equation}
        \frac{f_B}{n_B}=\mu_n (1-Y_p)+\mu_p Y_p-\frac{p}{n_B}  \ .
\end{equation}
The sound velocity of supernova matter is obtained using 
\begin{equation}
        c_s= \sqrt{\left( \frac{\partial{p}}{\partial {\rho}} \right)_{s,Y_p}}=\sqrt{ \Gamma \frac{p}{\rho}}  , 
\end{equation}
where $p$ is the total pressure,  $\rho$ is the total density, and $s$ is the total entropy.
The included contributions of electrons are comparable to the baryonic part, particularly in the core-collapse phase. 
Although the sound speed must  be lower  than the light speed,
some EOSs violate this condition at extremely high densities because of strong repulsion.
The adiabatic index, $\Gamma$, which characterizes the stiffness of supernova matter is calculated as
\begin{equation}
        \Gamma=\left( \frac{\partial{ \ln p}}{\partial { \ln \rho}} \right)_{s,Y_p} .
\end{equation}

\subsection{Hybrid models} \label{sec_hybrid}
In the SHO and SHT EOSs  \cite{sheng11a, sheng11b}, 
two different EOSs---NSE at low densities and SNA with the Hartree approximation at high densities---are employed.
The  critical density at which the two calculations are switched is determined by the free energy densities as
\begin{equation}
f_B= \mathrm{min}(f_{\mathrm{SNA}}, f_{\mathrm{NSE}}) \ ,
\end{equation}
where  $f_{\mathrm{SNA}}$ and $f_{\mathrm{NSE}}$ are the free energy densities of the EOSs based on the SNA and the NSE, respectively.
The NSE shows the lower free energy  densities  at low densities and vice versa at high densities.
In the SNA  approach, the density distribution in a spherical cell is optimized without any assumptions about the nuclear shape.
Thus, the description of the transition from non-uniform nuclear matter  to uniform nuclear matter is the most complex part in general-purpose EOSs.
In a spherical symmetry, some nuclear structures such as slab and rod phases cannot be obtained.
However, a spherical shell structure is observed  at intermediate densities between spherical and bubble phases,
 which is similar to the slab phase of nuclear pastas. 

In the SRO EOSs, the CLDM model  of the LS EOS (Sec.~\ref{sec_cldm}) is utilized at high densities.
The free energy densities of the SNA  and  EOSs at low densities are combined using 
 a density-dependent function as
\begin{eqnarray}
f_B & =& \chi(n_B) f_{\mathrm{SNA}}  + [1-\chi(n_B)] f_{\mathrm{NSE}}  \ , \\ 
\chi(n_B) &=& \frac{1}{2}\left[1 + \tanh \left(\frac{\log_{10}(n_B)-\log_{10}(n_t)}{n_\delta}\right)\right],
\end{eqnarray}
where the center of the  transition is set 
$n_t=10^{-4}$~fm$^{-3}$, 
 and its dimensionless width is  defined as $n_\delta=0.33$.
This approach ensure a more smooth connection between the NSE and SNA EOSs than that between the SHO and SRO EOSs.

\begin{table}[t] 
\begin{tabular}{c|ccc}
  Model  &  interaction &  nuclear model  & characteristics \\
 \hline
LS \cite{lattimer91} & Skyrme & SNA  (CLDM) &  discrete  distribution\\
 & (180, 220, 300) & &  \\
STOS \cite{shen98a, shen20} &  RMF (TM1, TM1e)  & SNA (TF) &  smooth distribution \\
TNTYST \cite{togashi17} &  VM  & SNA (TF)  & smooth distribution\\
\hline
SHO, SHT \cite{sheng11a, sheng11b} &  RMF (FSUgold, NL3)   & NSE or  & Virial expansion \\
 &   & SNA (Hartree)  & arbitrary distribution  \\
SRO \cite{schneider17, schneider19}  &  Skyrme (LS220, SLy4, APR, LNS  &  NSE  and & excitation data \cite{rauscher00}  \\
 &   NRAPR, SkAPR, KDE0v1) &  SNA (CLDM)    & discrete  distribution \\
\hline
HS\cite{hempel10, steiner13} &   RMF (TM1, DD2, NL3,  & extended NSE &   excitation formula \cite{fai82}  \\
 &    TMA, FSG,IUF, SFHo, SFHx)   & &  \\
FYSS\cite{furusawa17a,furusawa17d, furusawa20a} &  RMF(TM1), VM, DBFH & extended NSE &  excitation formula \cite{nishimura14}  
\\GRDF1, GRDF2 \cite{typel18} & RMF (DD2) & extended NSE &  energy shift models \cite{pais16} \\
RG \cite{raduta19} &  Skyrme (SLy4) & extended NSE &  excitation data \cite{egidy05}.
\end{tabular}
\caption{\label{tab1_eosdata}%
Bulk properties of nuclear matter obtained using  EOSs for uniform nuclear matter
 based on RMF theory using TM1 parameter set
 \cite{sugahara94, shen11}, VM \cite{togashi13, togashi17}, and  DBHF  \cite{katayama13}.}  
\end{table}

\subsection{Ambiguities in  supernova EOSs}
In the last two decades, various EOSs for supernova simulations have been formulated,
which are listed  in Tab.~\ref{tab1_eosdata}.
Uniform nuclear matter theories, which are compared in Sec.~\ref{sec:um}, are one of their characteristics.
The presence of exotic hadrons such as hyperons  and quark phase is also an open question, 
which is discussed subsequently in Sec.~\ref{exo}.

For non-uniform nuclear matter,  there is a major difference between  the SNA and NSE EOSs.
In  terms of thermodynamical properties,
this difference is insignificant; however, the nuclear composition,
 which determines weak interaction rates, differs \cite{burrows84, furusawa17c}. 
Hence, in the CCSN simulations,  the weak interactions
 of nuclei  should be calculated using an NSE EOS.
At least, some weak rate data  based on an NSE \cite{juodagalvis10} should be employed,
when the  SNA EOS is used in CCSN simulations.

Coulomb energy  density dependences are almost universal in  NSE EOSs. 
Other energy shifts caused by  the dense environment may affect
 the EOS only in  the final stages of core collapses of massive stars  with  densities exceeding $\rho_B \sim 10^{13}$~g/cm$^3$ \cite{furusawa20a}.
However,  temperature dependence of nuclear energy such as the shell washout and the internal degree of freedom, 
 exhibit different nuclear compositions and entropies even in the initial phase of the core-collapse at $T=1$~MeV \cite{furusawa18b}.

Transition to uniform nuclear matter is also a major task for the construction of  the complete supernova EOS;
however, its details  may have little impact on the dynamics of CCSNe,
owing to the narrow density region of the transition. 
In the FYSS and LS EOSs, nuclear bubble phases are based on liquid drop models, 
in which a nuclear structure is determined by  simple functions of volume fraction $u$ in spherical cells.
In  the SHO and SHT EOSs, the structures of nuclear pastas
 with a spherical symmetry (droplet, spherical shell, or bubble phase)
  are obtained by optimization of the density distribution in a spherical cell.
Although nuclear pastas are not considered,
 nuclei are diminished by  the excluded volume effects in the HS and RG EOSs 
and by the Pauli energy shifts in  the GRDF1 and GRDF2 EOSs \cite{typel18}.
More detailed  studies such as a molecular dynamics simulations \cite{schneider13} and 
 3D finite-temperature Skyrme-Hartree-Fock studies 
\cite{pais14},
will provide more insight into the issue.

Regarding combinations of the SNA and NSE EOSs,
 some self-consistent approaches can describe both the ensemble of various nuclei and individual nuclear properties \cite{gulminelli15,furusawa17c,furusawa18a}.
However, they necessitate significant computing power and are unsuitable
 for building a general-purpose EOS that can deal with a wide range of thermodynamic conditions.
Readers are referred  to some papers comparing these EOSs   \cite{furusawa11, buyukcizmeci13, furusawa18b, raduta21, hempel15}.

\begin{figure}[h]
\vspace{-3cm}
\hspace{2cm}
\includegraphics[width=13cm]{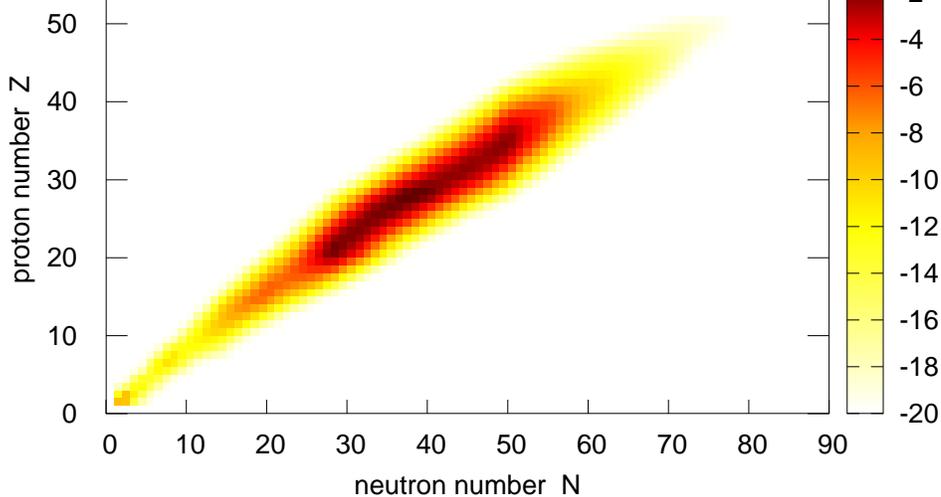}
\vspace{3cm}
\caption{
 Mass fractions of nuclei in log$_{10}$ on  $(N, Z)$ plane for $(\rho_B, T, Y_p) = (2.0 \times 10^{
10}$~g/cm$^{ 3}$, 0.64~MeV,  0.42. 
The thermodynamical condition is available at the center of a collapsing core 
in the supernova simulation \cite{nagakura19b} using FYSS (VM) EOS. }
\label{fig_zn1}
\end{figure}

\begin{figure}[h]
\vspace{-3cm}
\hspace{2cm}
\includegraphics[width=13cm]{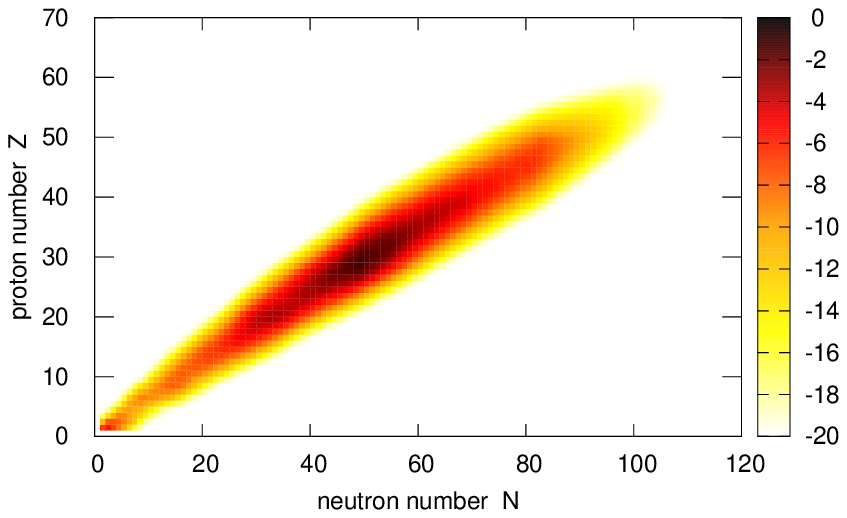}
\vspace{3cm}
\caption{
Mass fractions of nuclei in log$_{10}$ on  $(N, Z)$ plane for 
$(\rho_B, T, Y_p) = (2.0 \times 10^{
11}$~g/cm$^{ 3}$, 1.1~MeV, $0.34$). 
The thermodynamical condition is available at the center of a collapsing core 
in the supernova simulation \cite{nagakura19b} using FYSS (VM) EOS.  }
\label{fig_zn2}
\end{figure}

\begin{figure}[h]
\vspace{-3cm}
\hspace{2cm}
\includegraphics[width=13cm]{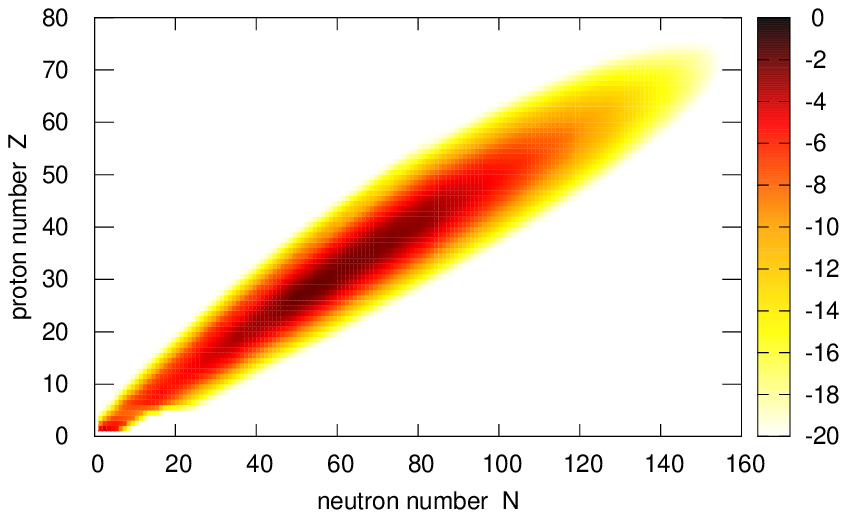}
\vspace{3cm}
\caption{
Mass fractions of nuclei in log$_{10}$ on  $(N, Z)$ plane for 
 $(\rho_B, T, Y_p) =(2.0 \times 10^{12}$~g/cm$^{ 3}$, 1.7~MeV, $0.27$).
The thermodynamical condition is available at the center of a collapsing core 
in the supernova simulation \cite{nagakura19b} using FYSS (VM) EOS.  }
\label{fig_zn3}
\end{figure}

\section{Nuclei in supernova simulations \label{sec:res}}
In this section, we discuss the distribution of the nuclei that were included in the supernova simulation \cite{nagakura19b} of an 11.2 $M_\odot$ progenitor using the FYSS (VM) EOS  \cite{furusawa17d}.

\begin{figure}[htbp]
  \begin{minipage}[b]{0.99\linewidth}
    \centering
\vspace{-1cm}
\includegraphics[width=13cm]{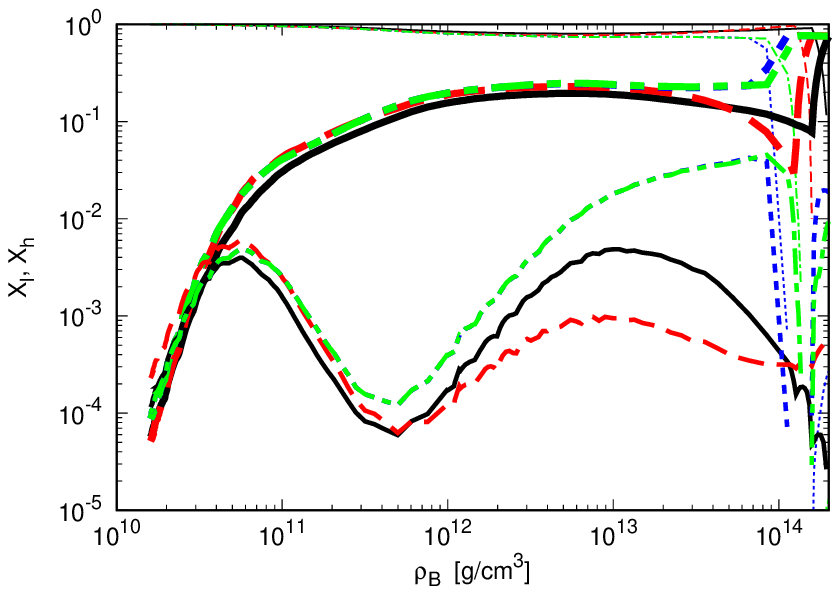}
\vspace{2cm}
\caption{
Mass fractions of heavy nuclei  (thin lines), light nuclei (middle lines), and neutrons (thin lines) for FYSS (VM) (black solid lines),   TNTYST (VM) (red dashed lines), 
HS (DD2) (blue dotted lines), and HS (SHFx) (green dashed-dotted lines) EOSs
at $(\rho_B, T, Y_p)$ of Fig.~\ref{fig_pro}
 in  center of the collapsing core of the supernova simulation \cite{nagakura19b}  using  FYSS (VM)  EOS \cite{furusawa17d}.
}
\label{fig_pro2}
  \end{minipage}\\
  \begin{minipage}[b]{0.99\linewidth}
    \centering
\vspace{-1cm}
\includegraphics[width=13cm]{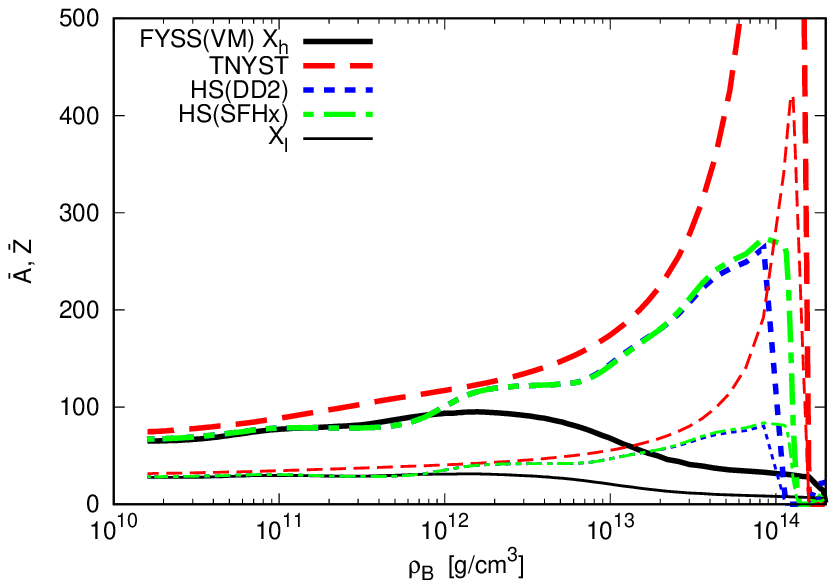}
\vspace{2cm}
\caption{
Average mass numbers (thick lines) and average proton numbers (thin lines)  of heavy nuclei 
 for FYSS (VM) (black solid lines),   TNTYST (VM) (red dashed lines), 
HS (DD2) (blue dotted lines), and HS (SHFx) (green dashed-dotted lines) EOSs
at $(\rho_B, T, Y_p)$  of Fig.~\ref{fig_pro}
 in  center of the collapsing core of the supernova simulation \cite{nagakura19b}  using  FYSS (VM)  EOS \cite{furusawa17d}.
}
\label{fig_pro3}
\end{minipage}
\end{figure}

\begin{figure}[h]
\hspace{2cm}
\includegraphics[width=13cm]{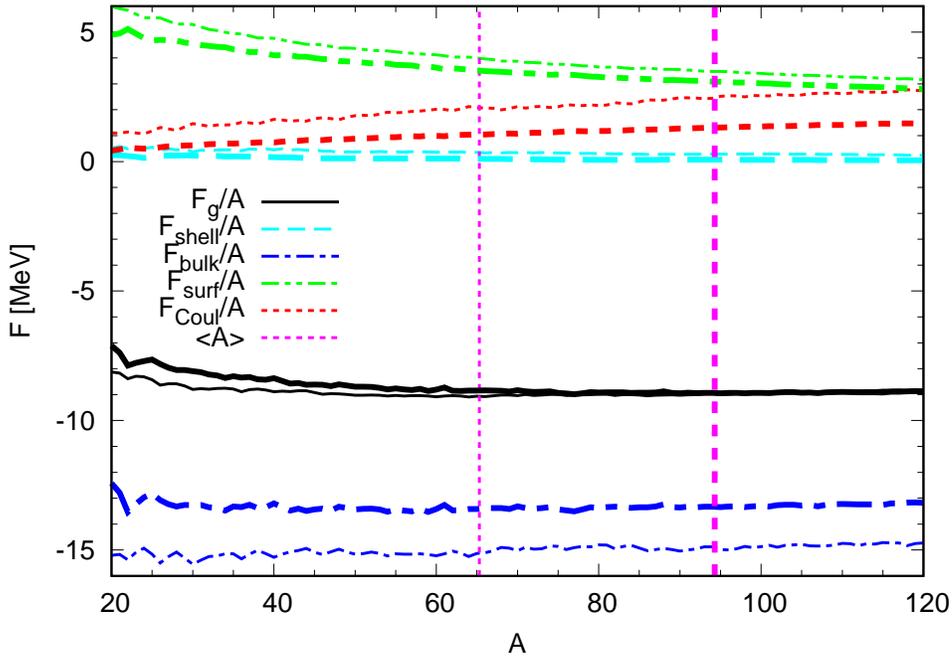}
\vspace{3cm}
\caption{Nuclear gross energy per baryon (black solid lines) and  shell (cyan dashed lines), bulk (blue dotted dashed lines), surface (green double-dotted dashed lines), and Coulomb (red dotted lines) energies  at   $(\rho_B, T, Y_p) = (2.0 \times 10^{10}$~g/cm$^{ 3}$, 0.64~MeV,  0.42)  (thin lines) and $(2.0 \times 10^{12}$~g/cm$^{ 3}$, 1.7~MeV, $0.27$)  (thick lines). 
The thermodynamical condition is available at the center of a collapsing core 
in the supernova simulation \cite{nagakura19b} using FYSS (VM) EOS. 
Rest masses are subtracted from gross and bulk energies. 
 They are averaged over the nuclides with  same mass numbers. Magenta dotted vertical lines display mass numbers of heavy nuclei averaged over entire ensemble of nuclei. }
\label{fig_dense}
\end{figure}

\begin{figure}[htbp]
  \begin{minipage}[b]{0.99\linewidth}
    \centering
\vspace{-1cm}
\includegraphics[width=11cm]{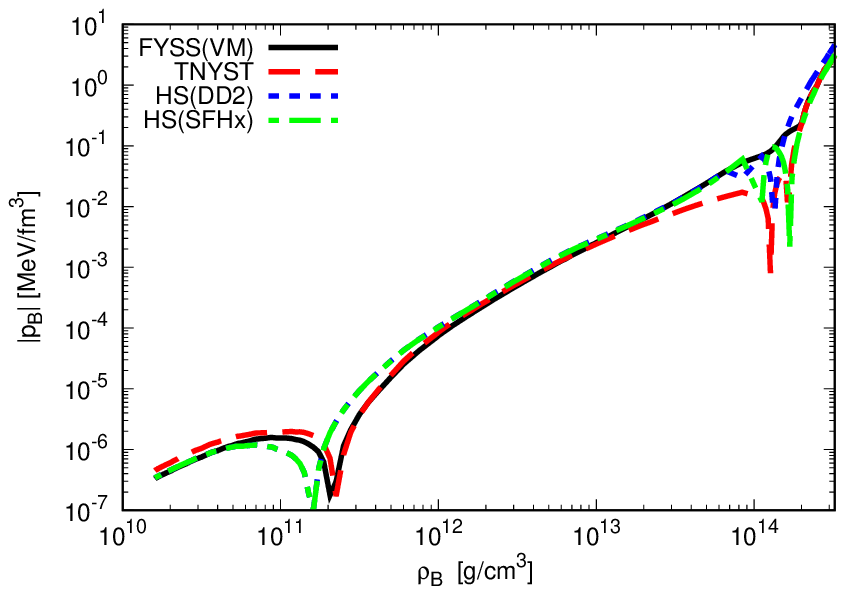}
\vspace{2cm}
\caption{
Absolute value of baryonic pressure  using FYSS (VM) (black solid lines),   TNTYST (VM) (red dashed lines), 
HS (DD2) (blue dotted lines), and HS (SHFx) (green dashed-dotted lines) EOSs
at $(\rho_B, T, Y_p)$ of Fig.~\ref{fig_pro}
 in  center of the collapsing core of the supernova simulation\cite{nagakura19b}  using  FYSS (VM)  EOS \cite{furusawa17d}.
Sign changes occur  from positive to negative at  $\rho_B \sim 2.0 \times 10^{11}$~g/cm$^{ 3}$
and  from negative to positive  at  approximately  $\rho_B \sim 10^{14}$~g/cm$^{ 3}$.
}
\label{fig_pro4}
  \end{minipage}\\
  \begin{minipage}[b]{0.99\linewidth}
    \centering
\vspace{-1cm}
\includegraphics[width=11cm]{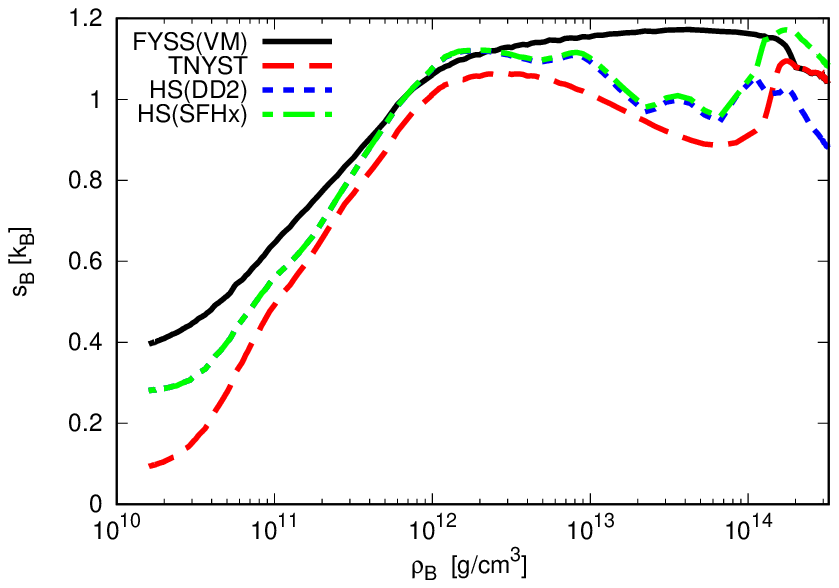}
\vspace{2cm}
\caption{
Entropies per baryon from FYSS (VM) (black solid lines),   TNTYST (VM) (red dashed lines), 
HS (DD2) (blue dotted lines), and HS (SHFx) (green dashed-dotted lines) EOSs
at $(\rho_B, T, Y_p)$  of Fig.~\ref{fig_pro}
 in  center of the collapsing core of the supernova simulation \cite{nagakura19b}  using  FYSS (VM)  EOS \cite{furusawa17d}.}
\label{fig_pro5}  \end{minipage}
\end{figure}

\subsection{Core-collapse phase} \label{sec_cc}
%
%
%
Snapshots of the nuclear mass fractions in the $(N, Z)$ plane at the center of the core-collapse simulation \cite{nagakura19b} 
at $(\rho_B, T, Y_p) = (2.0 \times 10^{10}$~g/cm$^{ 3}$, 0.64~MeV,  0.42),  at $(2.0 \times 10^{11}$~g/cm$^{ 3}$, 1.1~MeV, $0.34$) and at $(2.0 \times 10^{
12}$~g/cm$^{ 3}$, 1.7~MeV, $0.27$)
are shown in Figs.~\ref{fig_zn1}--\ref{fig_zn3}. 
The mass fraction, $X(N, Z)$, is defined as
 \begin{equation}
  X(N, Z) =A  n(N, Z) /n_B, 
 \end{equation} 
The initial state of the core collapse corresponds to the first one,
in which the iron-group nuclei are most abundant.
As the supernova matter becomes denser, electrons weaken the Coulomb repulsion between
 the protons inside the heavy nuclei, in turn, producing nuclei with larger mass numbers.
In addition, the reduction in $Y_p$ increases the chemical potential difference, $\mu_n-\mu_p$, and, therefore, neutron-rich nuclei become abundant.  
At $\rho_B=10^{11}$--$10^{12}$~g/cm$^3$, 
large electron-type neutrinos are emitted  during the core collapse.
Nuclei are abundant near the neutron magic numbers ($N=28, 50, 82$) such as $^{78}_{28}$Ni, $^{80}_{30}$Zn,  and $^{122}_{40}$Zr.
At $T=1.7$~MeV,  the nuclear shell effects are  partially reduced and non-magic nuclei  become 
as abundant as magic ones \cite{furusawa18b}.
These nuclei with $Z=$20--50 and $N =$40--90 that emerge at densities  $\rho_B=10^{11}$--$10^{12}$g/cm$^3$ 
are critical for determining the weak interaction rate and the subsequent core dynamics  \cite{sullivan16,furusawa18b}. 

Figures~\ref{fig_pro2} show the neutron mass fraction, $n_n/n_B$, and the total  mass fractions of
 heavy ($Z \geq 6$) and light nuclei ($Z\leq5$)  as  functions of the central density, 
  which  are calculated as $\sum_{Z\geq 6} X(N,Z)$ and $\sum_{Z\leq 5}X(N,Z)$, respectively.
At the beginning of the core collapse,   heavy nuclei are dominant in abundance.
When $\rho_B$ increases and $Y_p$ decreases, neutrons start to drip.   
For comparison, those for  the TNTYST (VM), HS (DD2), and HS (SFHx) EOSs  under the same thermodynamical
conditions $(\rho_B, T, Y_p)$ are displayed. 
However,  this comparison is only perturbative. 
In the supernova simulations using  the TNTYST (VM) or HS EOS, 
the thermodynamical conditions deviate from those using the  FYSS (VM) EOS.
 Comparisons of different EOSs in supernova simulations are discussed in Sec.~\ref{sec:eoscomp}. 
In the entire core-collapse phase, heavy nuclei practically dominate. 
The  TNTYST (VM) EOS  yields more dripped neutrons than the FYSS (VM) EOS, owing to the SNA.
In  the TNTYST (VM) EOS, there are fewer light nuclei ($\alpha$ particles) than in other EOSs above $\rho_B\sim10^{12}$~g/cm$^3$,
  because of the lack of deuterons and  $^3 \rm H$.
The differences in modeling the free energies of light and heavy nuclei causes a gap between the FYSS and HS EOSs.

The average mass numbers and proton numbers of heavy nuclei at the center of the core are 
shown in Fig.~\ref{fig_pro3}, which are expressed as %
\begin{eqnarray}
\bar{A}=\frac{\sum_{Z\geq 6} A n(N,Z) }{ \sum_{Z\geq 6} n(N,Z)} \ ,  \\
\bar{Z}=\frac{\sum_{Z\geq 6} Z n(N,Z) }{ \sum_{Z\geq 6} n(N,Z) } \ .
\end{eqnarray} 
The  shell effect causes a stepwise  growth,  because  nuclei with  neutron magic numbers $N=$~28, 50, and 82 ($A\sim$~60, 80, and 120) are stable in the HS EOSs. 
the FYSS (VM) EOS is essentially identical to the HS EOS models at low densities,  
because both are similar to  the standard NSE EOS, which is  introduced in Sec.~\ref{sec_snse}. 
The mass and proton numbers from the  FYSS (VM) EOS  reduce above $\rho_b\sim 10^{12}$~g/cm$^3$, 
because the reductions in shell and surface energies are more critical for nuclei with smaller mass numbers, as explained  subsequently. 
The FYSS and TNTYST EOSs show significant differences in the nuclear composition, even with the same nuclear interaction for uniform nuclear matter (VM).
By  contrast, the  HS (DD2 and SHFx) EOSs  yield similar results owing  to the same nuclear model.

The nuclear abundance during a core collapse is sensitive to the gross energy, $F_g(N, Z)$.
Figure~\ref{fig_dense} shows the average energies of nuclei and each contributions of  the energy shifts  as functions of the nuclear mass number
at  
$(\rho_B, T, Y_p) = (2.0 \times 10^{10}$~g/cm$^{ 3}$, 0.64~MeV,  0.42) and  $(2.0 \times 10^{
12}$~g/cm$^{ 3}$, 1.7~MeV, $0.27$). 
For example, the average  gross and Coulomb energies are defined as 
\begin{eqnarray}
\bar{F}_{g}(A)=\frac{\sum_{Z+N=A} F_{g}(Z,N)n(Z,N)}{\sum_{Z+N=A} n(Z,N)}\  , \\ 
\bar{F}_{Coul}(A)=\frac{\sum_{Z+N=A} F_{Coul}(Z,N)n(Z,N)}{\sum_{Z+N=A} n(Z,N)} \ .
\end {eqnarray}
The mass number at which  the average gross energy is the lowest increases at a high density owing to the Coulomb energy reductions. 
Because the density increases during a core collapse, large nuclei can form. 
In contrast, the shell and surface energies are reduced as the temperature rises, making 
 nuclei with small mass numbers more stable \cite{furusawa17a, furusawa18b}.
As $Y_p$ decreases,  the average bulk and gross energies 
increases owing to the high neutron richness in the nuclei (the low value of $Z/A$). 

Fig. \ref{fig_pro4} shows the absolute values of the baryonic pressure. 
In the initial phases, the pressure is negative, because the Coulomb energies of heavy nuclei decrease under  the compression \cite{furusawa11}. 
In addition, near narrow density region $\rho_B\sim 10^{14}$~g/cm$^{ 3}$,  
the EOSs other than the FYSS (VM) EOS yield the negative pressures. 
Note that a positive leptonic pressure is much larger than a baryonic one,
 below the nuclear saturation density.
Hence, the total pressure never becomes negative.
At  $\rho_B\sim 10^{13}$--$10^{14}$~g/cm$^{ 3}$,
in the TNTYST (VM) EOS, 
the thermal pressures of nuclei and the  baryonic pressures may be underestimated owing to the SNA,
in which translational motions of various nuclei are not included.
Above  $\rho_B\sim 10^{14}$~g/cm$^{ 3}$,  the HS (DD2) EOS  yields larger baryon pressures owing to its stiffer bulk properties. 

Baryonic entropies  are sensitive to the temperature dependence of nuclear free energies \cite{furusawa18b}. 
In actual, there are difference between the entropies obtained with different EOSs, as shown in Fig. \ref{fig_pro5}.
In the FYSS (VM) EOS, heavy nuclei with small mass numbers  are  more abundant than in the other EOSs, because of the shell and surface energy washouts.
Hence, the thermal entropy is larger in the former  than in the other EOSs and, 
in addition, the entropy  due to the temperature dependence of the nuclear gross energies contributes. 
 The TNTYST (VM) EOS yields the lowest baryonic entropy, which may also be owing to the SNA.


Nuclear excited states are considered in various formulations in  the EOSs \cite{hempel10d}.
The effect of excited states can be expressed as a partition function as
\begin{eqnarray}
  g_{exp}(T)  =  \sum_i g_i  \left(1+ \frac{M_0}{ \Delta E_i} \right)^{3/2} \exp \left( - \frac{\Delta E_i}{T} \right) ,
   \end{eqnarray}
where  $i$ denotes the sum over all known states.
The average excitation energies, $\Delta E$, in the degeneracy factor,
 $g(T)$, similar to Eq.~(\ref{eq:fai}), is expressed as 
\begin{eqnarray}
  \Delta E  =  \frac{\partial g}{\partial T} \frac{T^2} {g}  .
   \end{eqnarray}
 The nuclear excitation energies per baryon, $\Delta E /A$, approximately 0.03--0.06 MeV at $T=0.86$ MeV and  around 2--6 MeV at $T=8.6$ MeV,  depending on the nucleus and on the NSE EOS model \cite{hempel10d}.
In the supernova simulations, the nuclei may be around 0.03--6 MeV per baryon excited  during a core collapse.
Nevertheless, the excitation of each nucleus in supernova matter, which is related to the temperature dependence in $f(N, Z)$, 
is inadequately investigated.

\begin{figure}
\hspace{1cm}
\includegraphics[width=14cm]{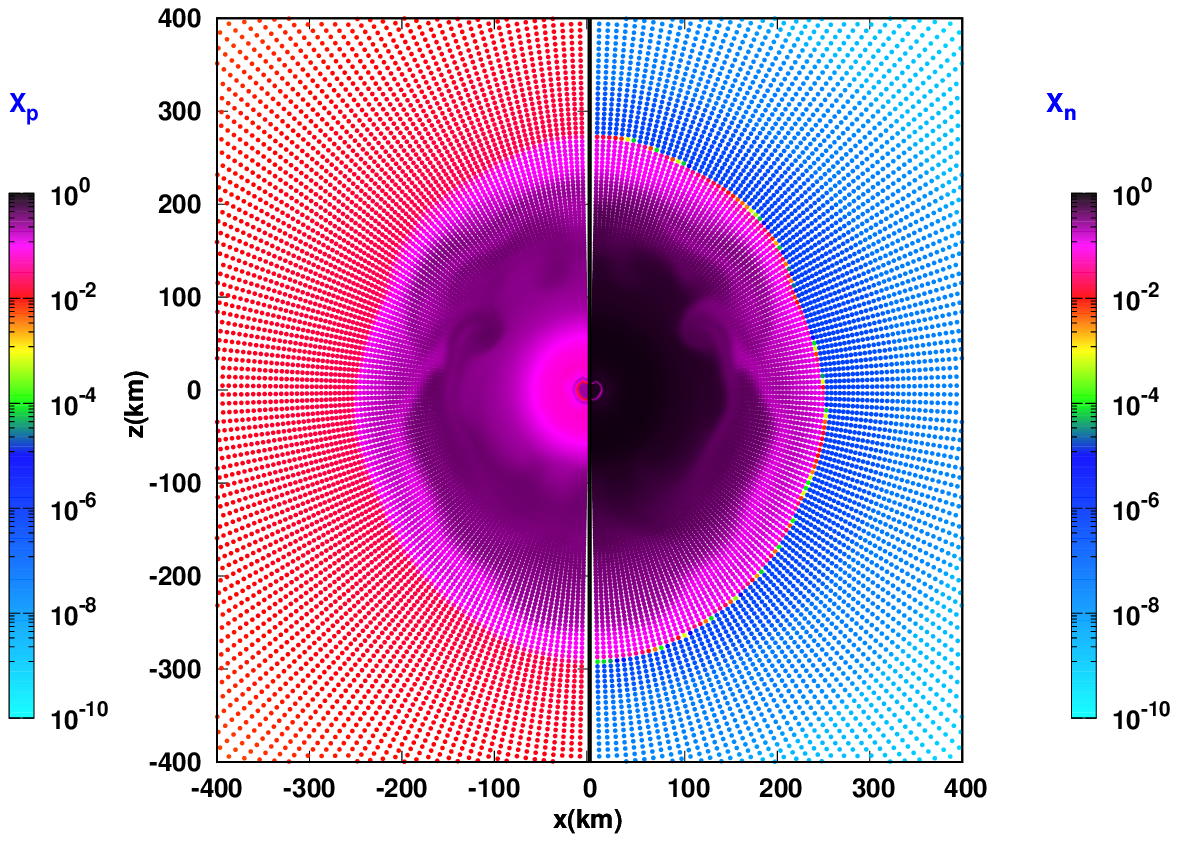}\\
\hspace{1cm}
\includegraphics[width=14cm]{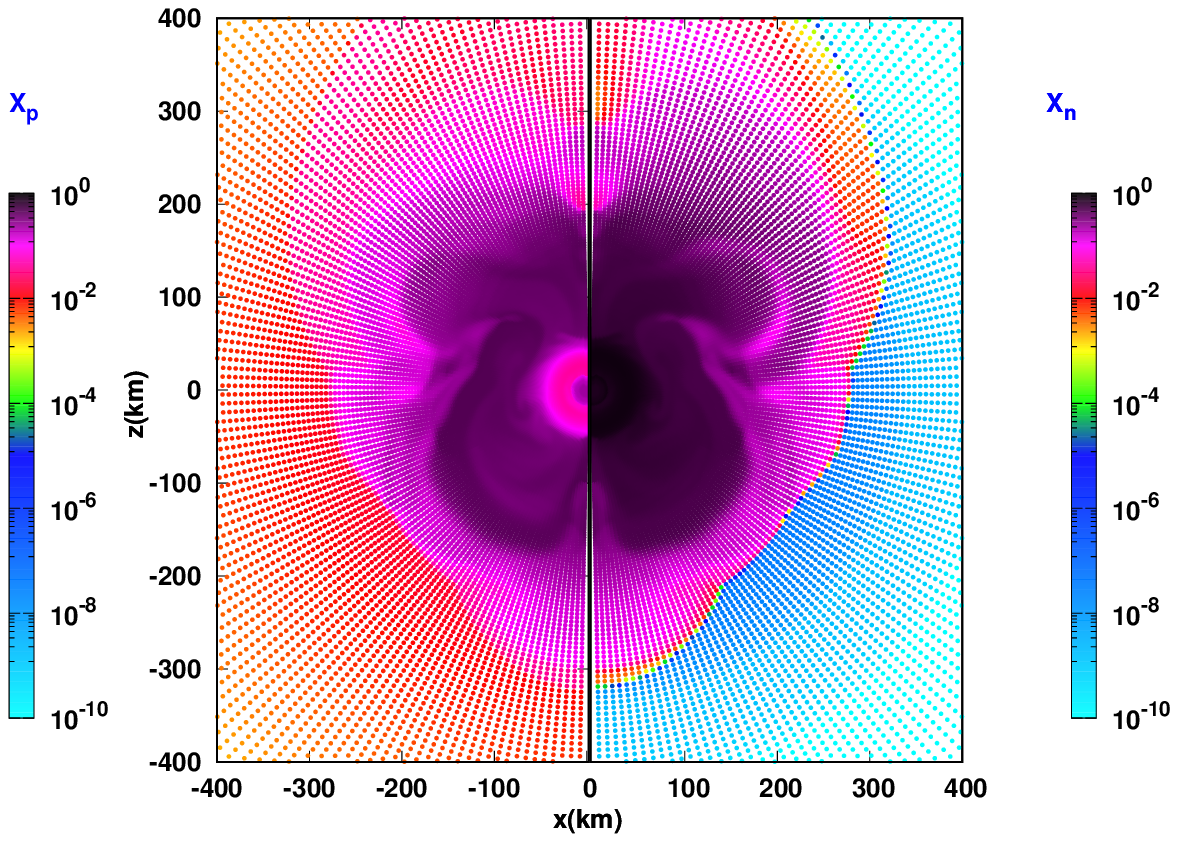}
\vspace{2cm}
\caption{Distributions of  free proton fractions  (left) and free neutron fractions  (right)  at 100 ms (top panel) and 200 ms (bottom panel)   after core bounce
 in   the supernova simulation \cite{nagakura19b}  using  FYSS (VM)  EOS \cite{furusawa17d}. }
\label{fig_xpn}
\end{figure}

\begin{figure}
\hspace{1cm}
\includegraphics[width=14cm]{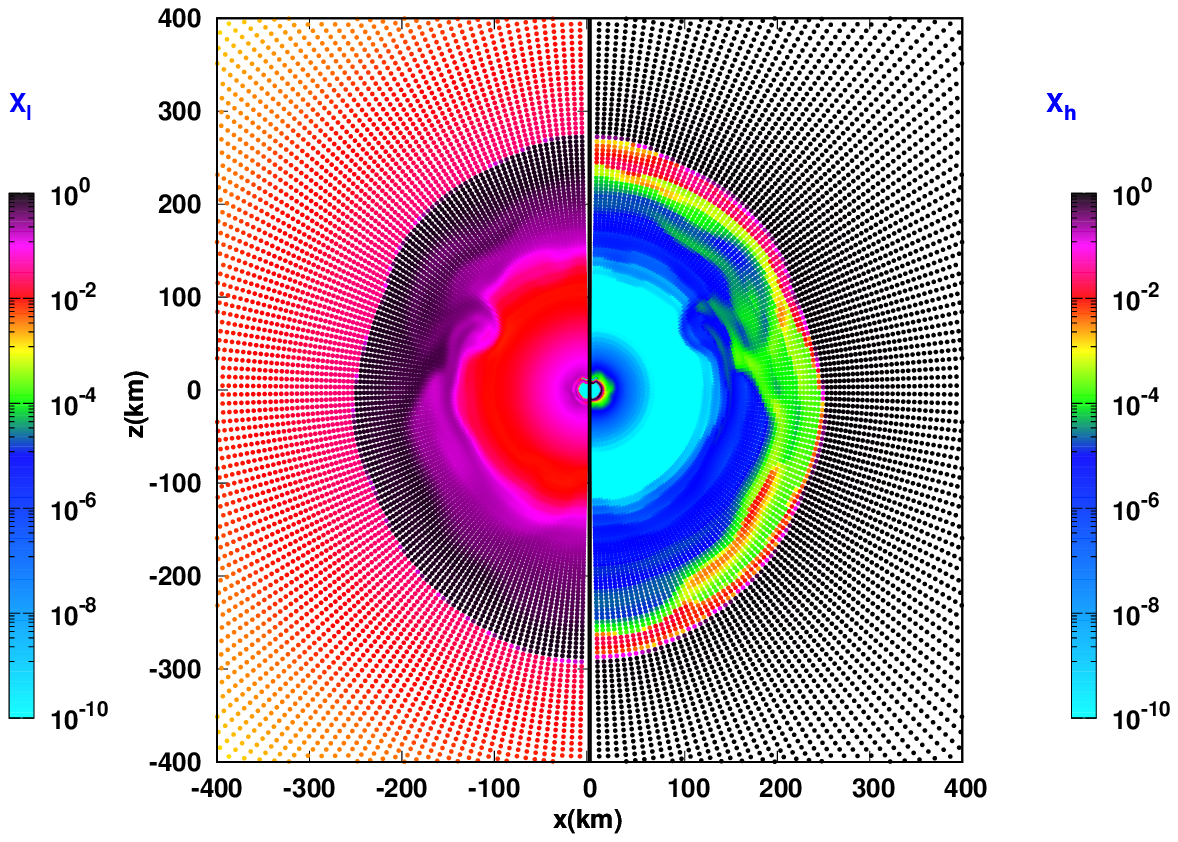}\\
\hspace{1cm}
\includegraphics[width=14cm]{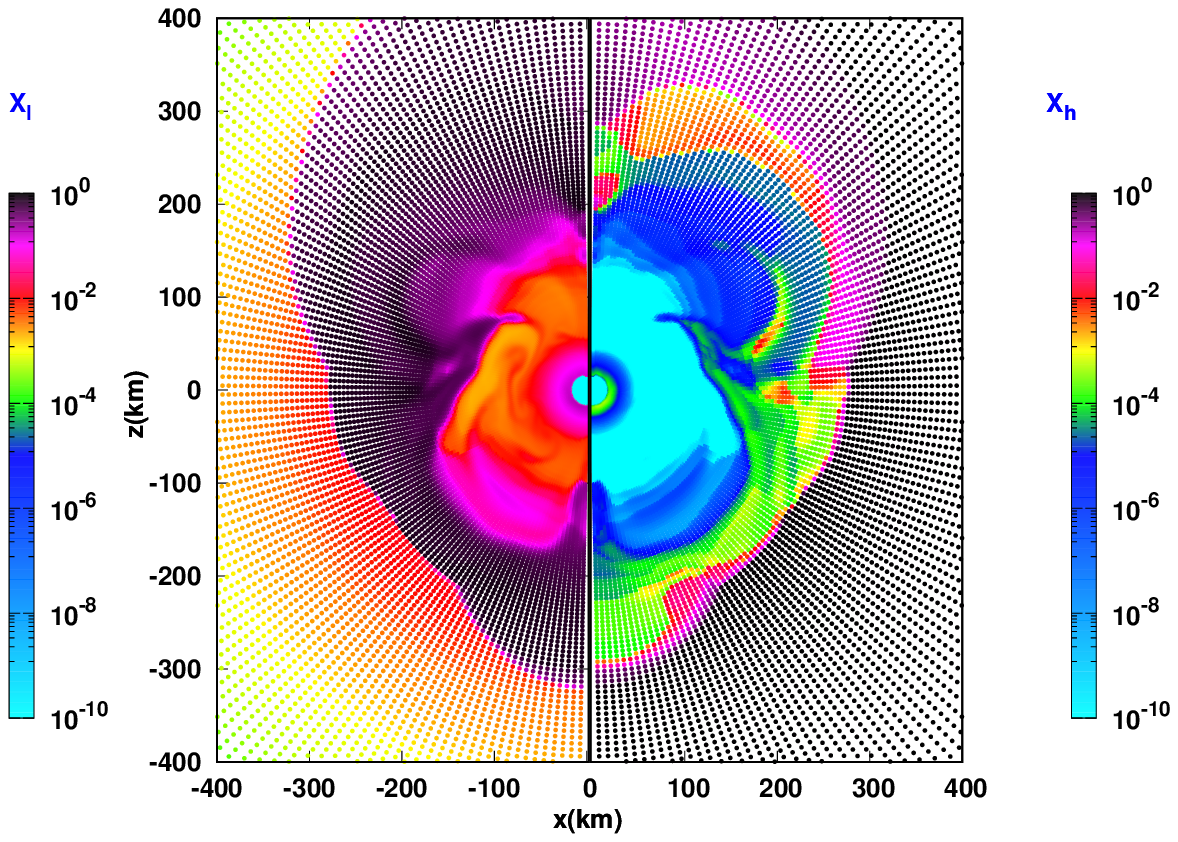}
\vspace{2cm}
\caption{Distributions of  mass fractions for light nuclei (left)  and  heavy nuclei (right) at 100 ms (top panel) and 200 ms (bottom panel)   after core bounce
in   the supernova simulation \cite{nagakura19b}  using  FYSS (VM)  EOS \cite{furusawa17d}. 
}
\label{fig_xlh}
\end{figure}

\begin{figure}
\hspace{1cm}
\includegraphics[width=14cm]{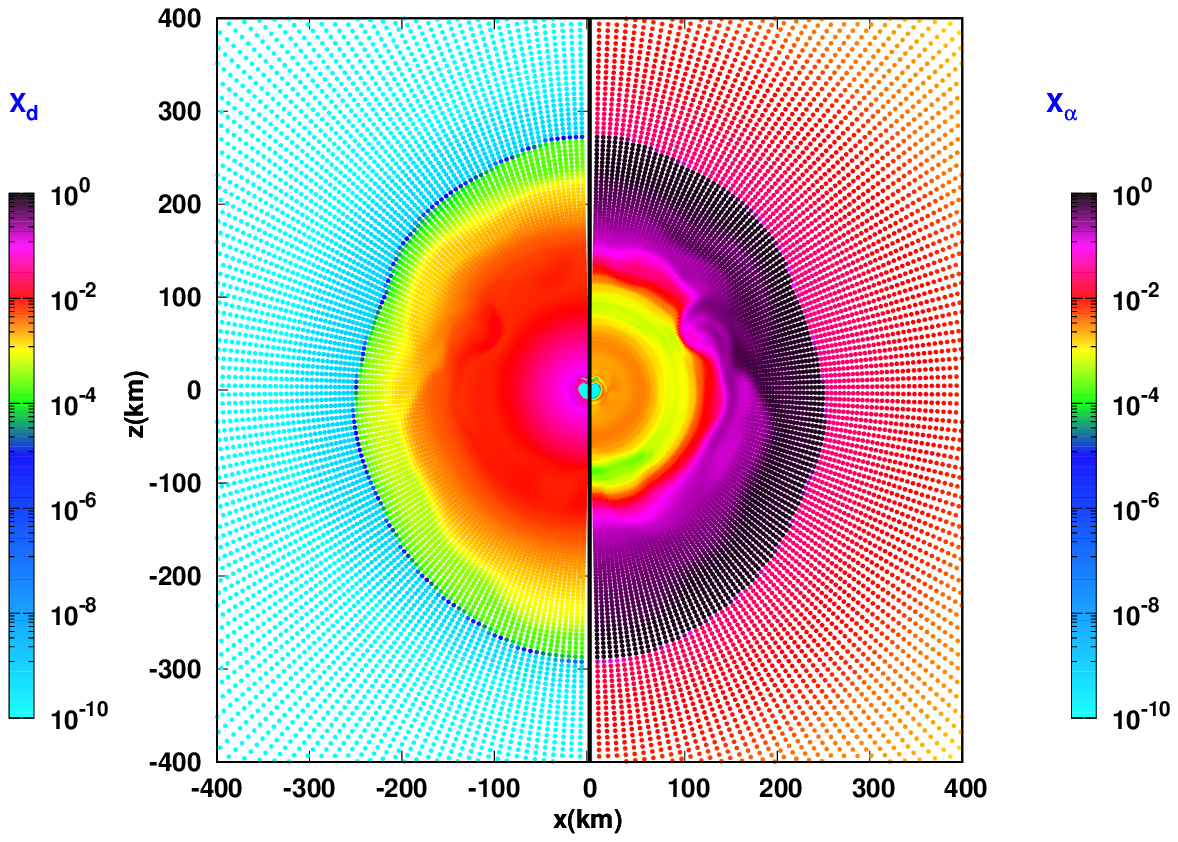}\\
\hspace{1cm}
\includegraphics[width=14cm]{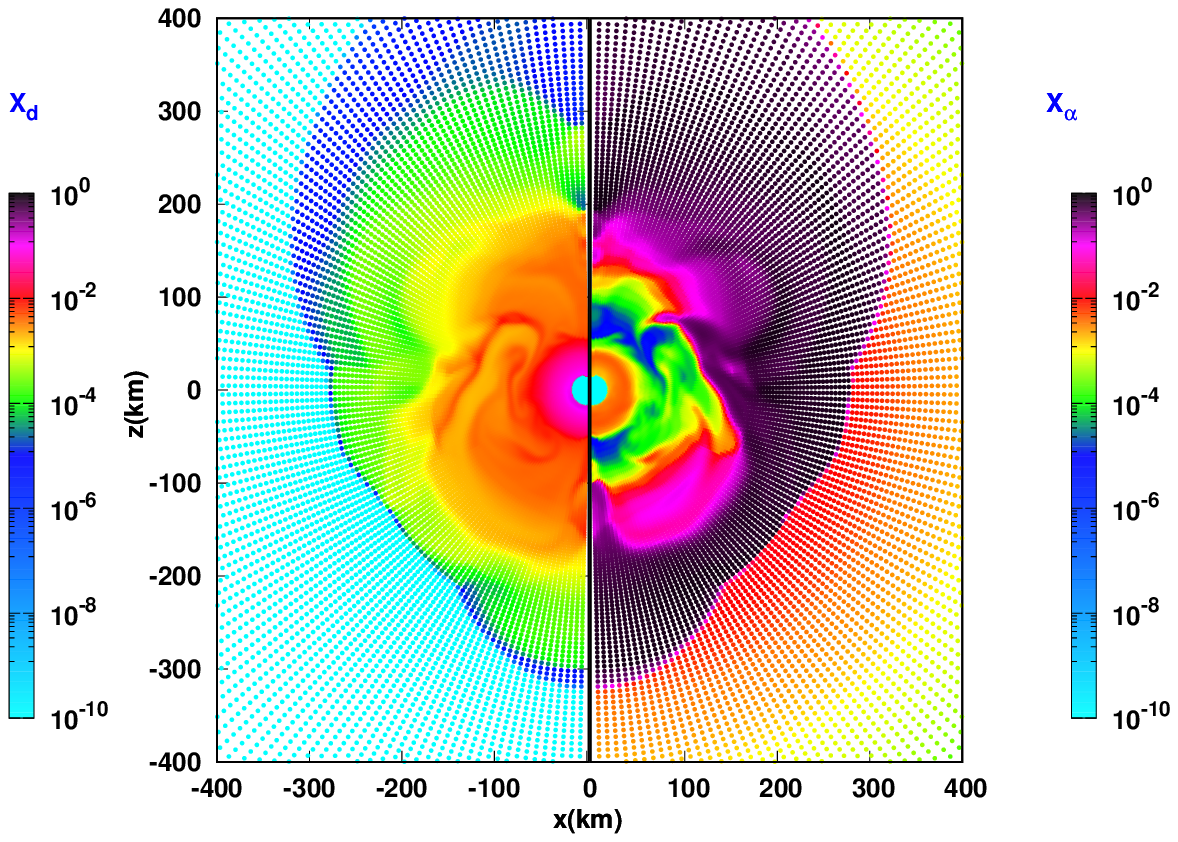}
\vspace{2cm}
\caption{Distributions of  deuteron mass fractions  (left) and $\alpha$ particle mass fractions  (right)  at 100 ms (top panel) and 200 ms (bottom panel)   after core bounce
in   the supernova simulation \cite{nagakura19b}  using  FYSS (VM)  EOS \cite{furusawa17d}.
}
\label{fig_xdal}
\end{figure}

\begin{figure}
\hspace{1cm}
\includegraphics[width=14cm]{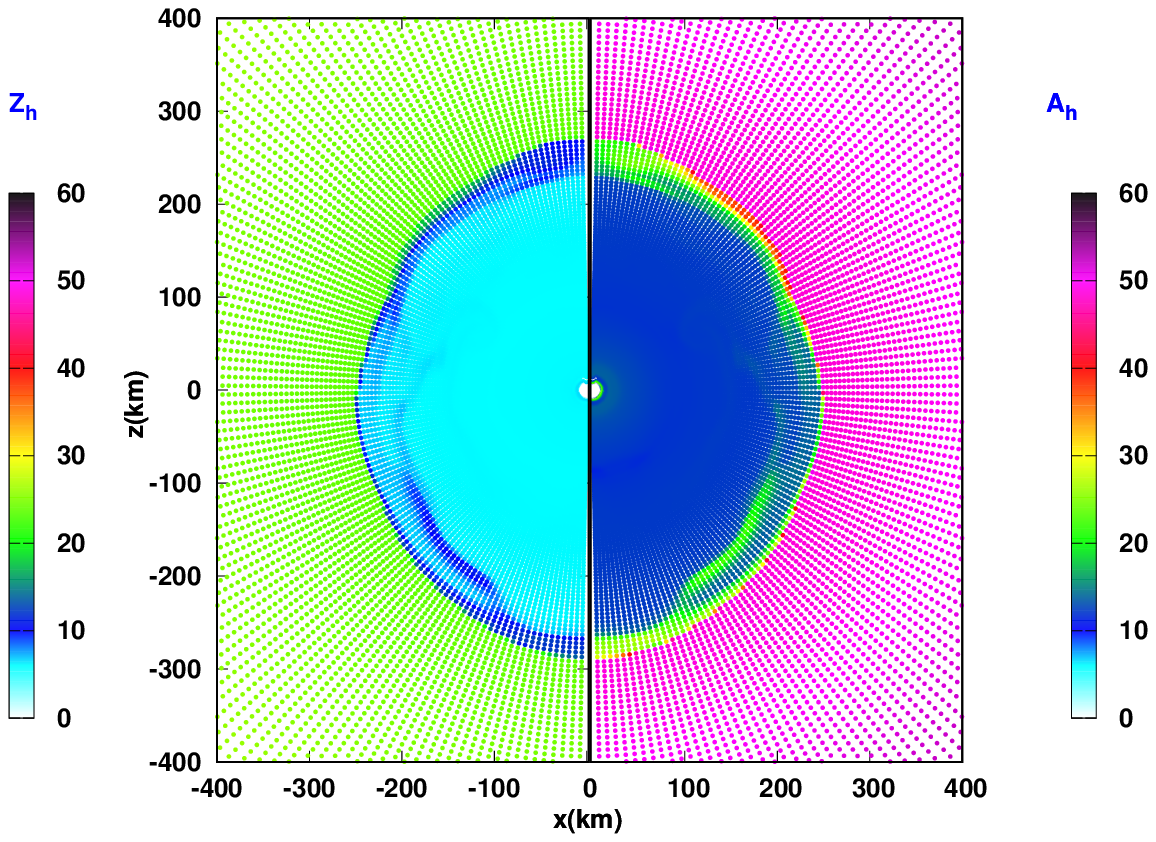}\\
\hspace{1cm}
\includegraphics[width=14cm]{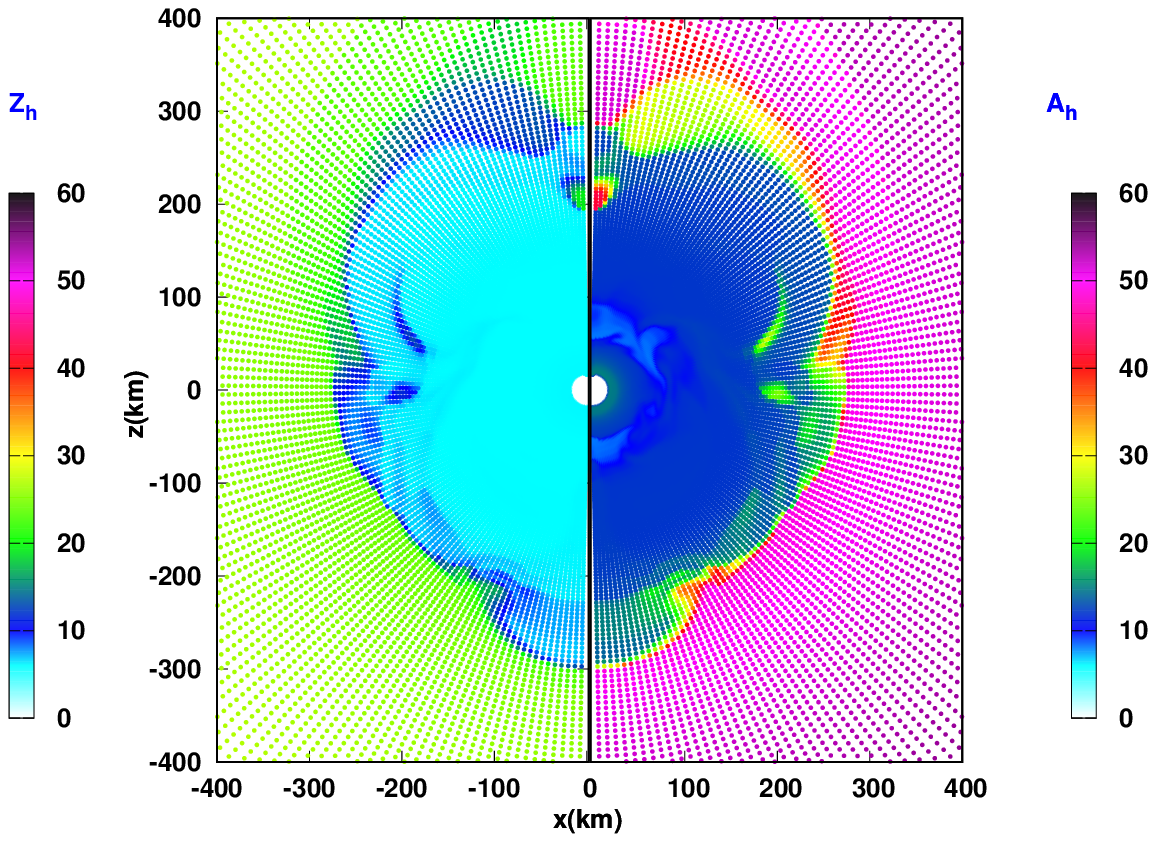}
\vspace{2cm}
\caption{Distributions of average proton numbers (left) and average mass  numbers (right)   for heavy nuclei ($Z \geq 6$)  at 100 ms (top panel) and 200 ms (bottom panel)   after core bounce
in   the supernova simulation \cite{nagakura19b}  using  FYSS (VM)  EOS \cite{furusawa17d}.
 }
\label{fig_za}
\end{figure}

\begin{figure}[h]
\hspace{2cm}
\includegraphics[width=8cm]{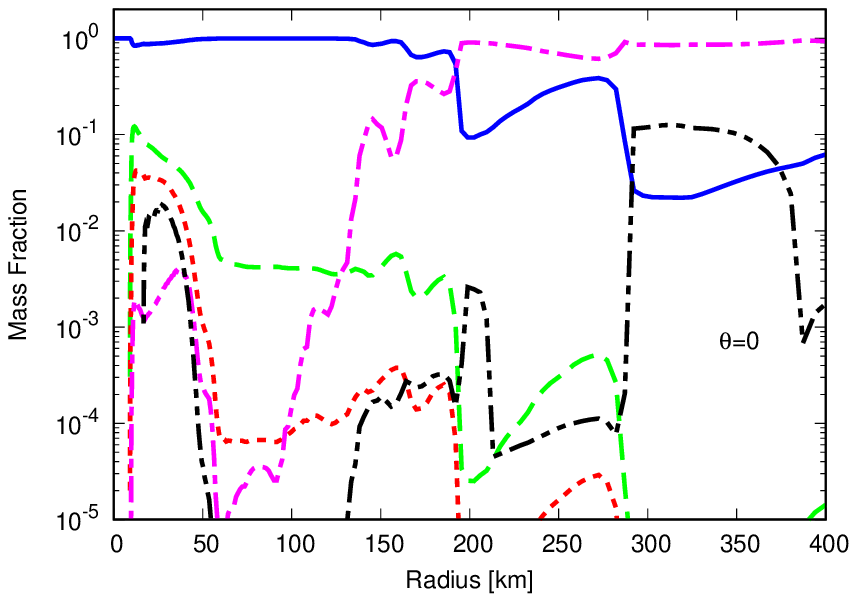}
\includegraphics[width=8cm]{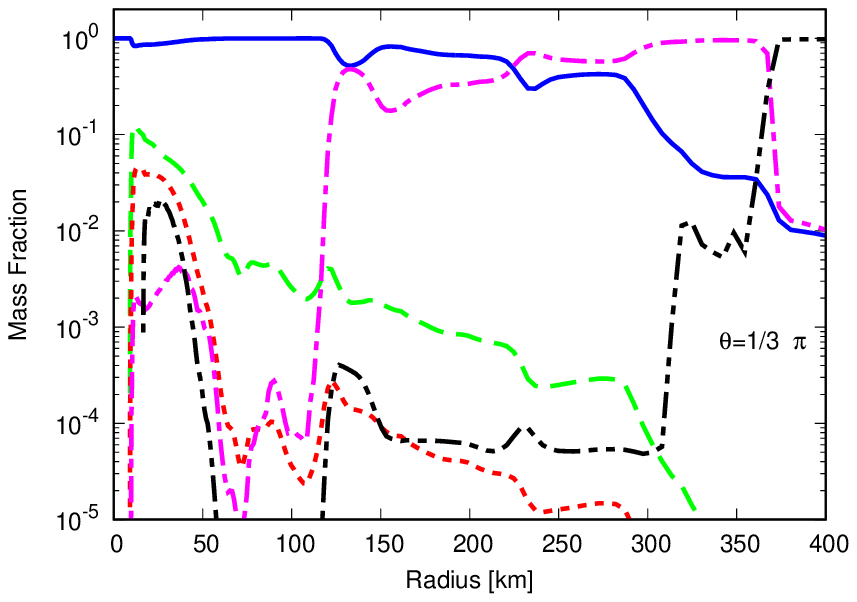}\\
\hspace{2cm}
\includegraphics[width=8cm]{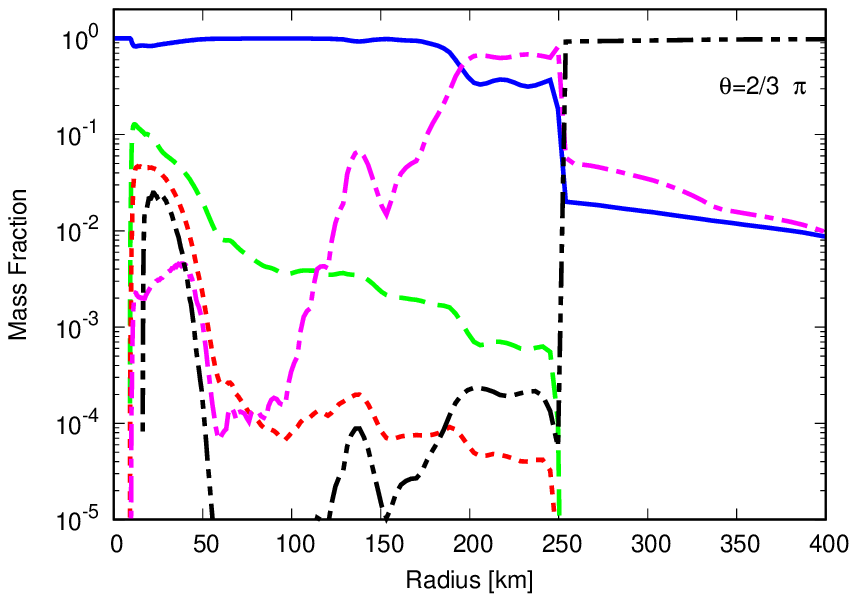}
\includegraphics[width=8cm]{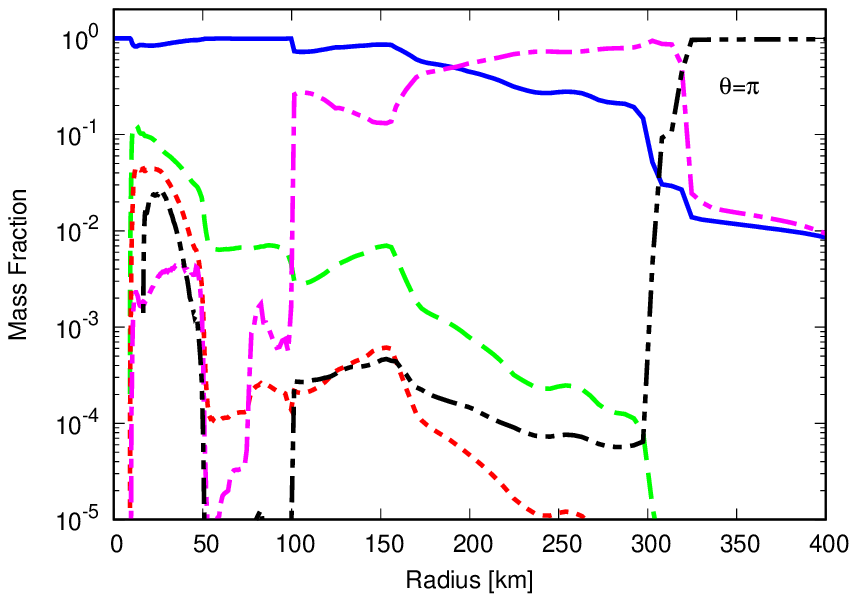}
\vspace{2cm}
\caption{
 Mass fractions for $A = 1$: protons and neutrons (blue solid lines), $A = 2$: deuterons (green dashed lines), $A = 3$:
$^3$H and $^3$He (red dotted lines), $A = 4$: $\alpha$ particles  (magenta dotted dashed lines), and $A>4$: other nuclei (black double-dotted dashed lines) 
at time after the core bounce, $t_{b}=200$ ms and in polar angles of $\theta=0$ (left upper panel),  $\theta=\frac{1}{3}\pi$ (right upper panel), $\theta=\frac{2}{3}\pi$ (left bottom panel), 
and  $\theta=\pi$ (right bottom panel)
in   the supernova simulation \cite{nagakura19b}  using  FYSS (VM)  EOS \cite{furusawa17d}.
}
\label{fig_frac}
\end{figure}

\begin{figure}[htbp]
  \begin{minipage}[b]{0.99\linewidth}
    \centering
\vspace{-1cm}
\includegraphics[width=11cm]{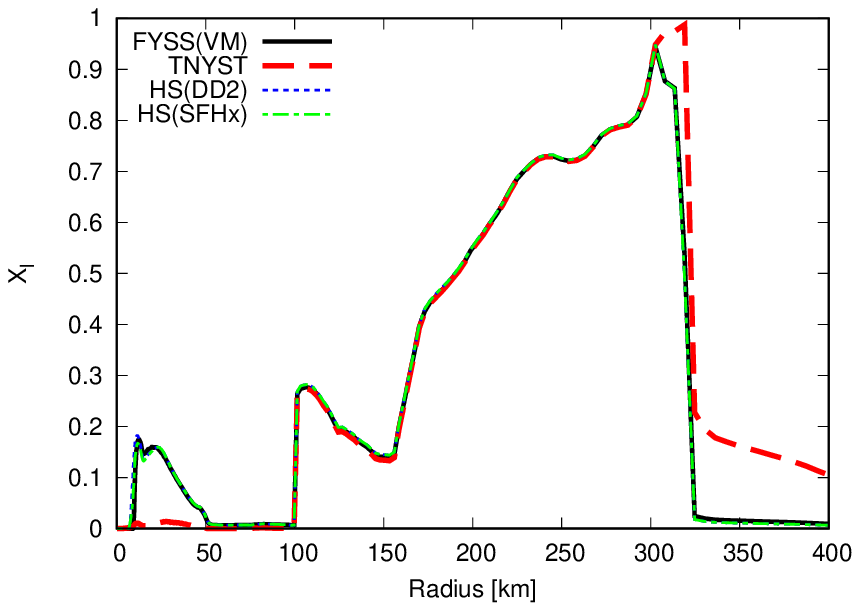}
\vspace{2cm}
\caption{
Mass fractions of light nuclei 
using FYSS (VM) (black solid lines),   TNTYST (VM) (red dashed lines), 
HS (DD2) (blue dotted lines), and HS (SHFx) (green dashed-dotted lines) EOSs
for thermodynamical conditions at polar angle  $\theta=\pi$
in  the supernova simulation \cite{nagakura19b}  using  FYSS (VM)  EOS \cite{furusawa17d}.
}
\label{fig_pbfrac}
  \end{minipage}\\
  \begin{minipage}[b]{0.99\linewidth}
    \centering
\vspace{-1cm}
\includegraphics[width=11cm]{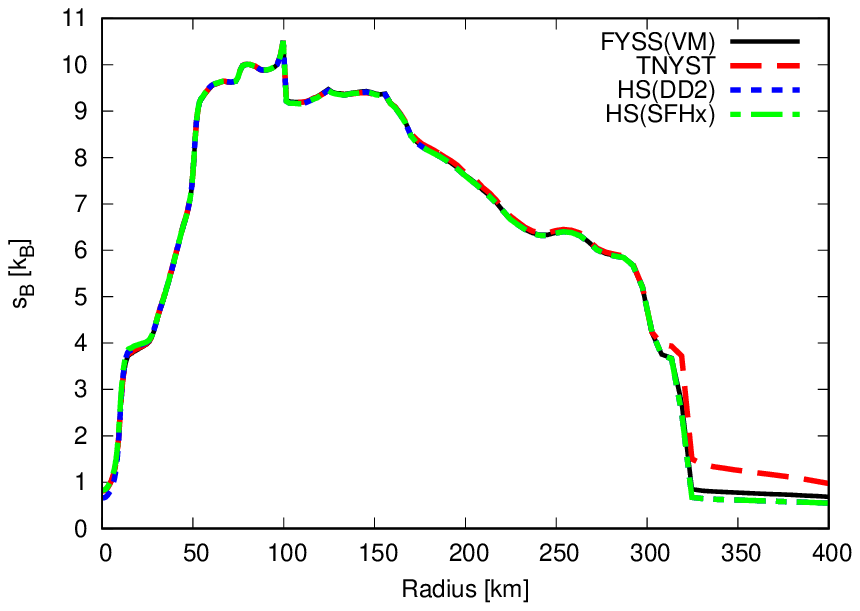}
\vspace{2cm}
\caption{
Baryonic entropy 
using FYSS (VM) (black solid lines),   TNTYST (VM) (red dashed lines), 
HS (DD2) (blue dotted lines), and HS (SHFx) (green dashed-dotted lines) EOSs
for thermodynamical conditions  at polar angle  $\theta=\pi$
in  the supernova simulation \cite{nagakura19b}  using  FYSS (VM)  EOS \cite{furusawa17d}.
}
\label{fig_pbent}  \end{minipage}
\end{figure}

\subsection{Shock revival phase}
After the core bounce, shock waves are  generated, and can be seen by jumps in some properties
 such as the temperature ($T\sim1$ MeV),
  density  ($\rho_B\sim10^9$~g/cm$^{ 3}$),  and  average nucleon number ($\braket{A}\sim10$),
which are shown in
Figs.~\ref{fig_rt} and~\ref{fig_fye}.
The average nucleon number for all baryons is defined as
  $\braket{A} =  \sum_k A_k  n_k/ n_B$, 
where  index $k$ denotes a nucleon or a nucleus, 
$A_k$ is the nucleon number of the particle (e.g., $A_n=A_p=1$,  $A_d=2$, $A_{\alpha}=4$, $A_{^{56}\rm{Fe}}=56$), and $n_k$ is the number density of the nucleons or nuclei.
The distributions of the shock waves are close to spherical  at $t=100$~ms after the core bounce. 
However, deformed shock wave are observed at $t =$~200 ms.
In some areas---$(x, z)\sim (\pm$ 100 km, 80 km$)$ at $t=100$ ms, and $(x, z)\sim$ (0 km, $\pm$200 km$)$ at $t=200$ ms---inward flows with low entropies and large values of $\braket{A}$ and $Y_p$ \cite{furusawa13b} occur.

Figures~\ref{fig_xpn}--\ref{fig_xdal} show the distributions of the mass fractions   of nucleons and
  nuclei in the shock expanding phase at $t=$~100 and 200~ms  of the supernova simulation \cite{nagakura19b}. 
The central region is composed of interacting nucleons. 
The minimum, angle-averaged, and maximum shock radii are approximately 220, 250 and  290~km  at $t=$100~ms, and approximately 250, 310 and  550~km at  $t=$200~ms, respectively. 
 In shocked matter, nucleons and light nuclei are dominant. 
 Accreting  matter consists of iron-group nuclei and $\alpha$ particles.
 
As a  temperature rise is caused by shock heating,  the entropy contribution, $S$, to the free energy, $F=U-TS$  increases,
relative to the  internal energy, $U$. 
Nucleons and light nuclei 
have  larger entropies per baryon than heavy nuclei,
and, hence, their number densities increase to lower the total free energy, 
 even if  their internal energies per baryon  is larger than those of heavy nuclei. 
Along  with inward flows,  heavy nuclei and $\alpha$ particles have large mass fractions, where nucleons and deuterons have small mass fractions.
Around shock waves, $\alpha $ particles are abundant, because of their large binding energies. 
In contrast, deuterons are populated just above the surface of the produced PNS.
The mass fractions of nuclei with $A=3$,  $^3$H, and $^3$He, are also large around the PNS surface.

As shown in  Fig.~\ref{fig_za},
various heavy nuclei  emerge around a shock wave; however their mass and proton numbers are small.
Even in shocked matter,  not all iron-group nuclei are dissolved into nucleons and light nuclei, specifically at $t=200$ ms (e.g., $x\sim0$ km and $z\sim$200 and 300 km at the  polar angle of $\theta=0$), 
because they have inward flows with low entropies.
Fig.~\ref{fig_frac} shows the radial profiles  of the mass fractions of nucleons and some nuclei
 at some polar angles $\theta=0$ ($x=0$ and $z\geq0$), $\frac{1}{3}\pi$,  $\frac{2}{3}\pi$, and $\pi$ ($x=0$ and $z\leq0$).
The radial distributions of the nuclear compositions differ with the angle, because the shock wave is deformed.
Above the shock wave, heavy nuclei are dominant, whereas below it, $\alpha$ particles  are the most abundant.
Particularly, in the low-entropy flows ($\theta=0$ or $\pi$ and $r \geq 200$~km), $\alpha$ particles dominate.
The radial profiles around the PNS ($r<50$~km) are similar to each other; however, the mass fractions of nucleons, deuterons, and  $^3 \rm H$ are large.

Figs.~\ref{fig_pbfrac} and \ref{fig_pbent} show the radial profiles  of 
the mass fractions of light nuclei and the baryonic entropies at polar angle $\theta=\pi$, respectively.
Those for the TNTYST (VM), HS (DD2), and HS (SFHx) EOSs  under the same thermodynamical
conditions $(\rho_B, T, Y_p)$ are also presented. 
The FYSS and HS EOSs present similar results, 
because the nuclear masses of light nuclei are similar. 
At  $r\sim10$~km, a slight difference is due to the in-medium effects. 
In the TNTYST EOS, the lack of deuterons and tritons decreases the light nuclei mass fractions
 and baryonic entropies at $r<50$~km,
whereas  the underestimation of the mass fractions of heavy nuclei \cite{burrows84, furusawa17b} 
increases them around and above shock radius  $r\sim330$~km.


As discussed in some  studies \cite{nagakura19a, furusawa13b, nasu15} and Sec.~\ref{sec.intro.inter}, 
light nuclei are never dominant targets in neutrino emission and absorption. 
However, their emergence  reduces the number of nucleons,
which  are dominant neutrino heating and cooling sources. 
In addition, deuteron heating and cooling reactions such as reactions~(ix) and (x) 
are non-negligible, and their contribution to the total neutrino reaction rate is up to $\sim10\%$ 
depending on the time and the location \cite{furusawa13b, nagakura19a}. 
The heating of the matter
 around an expanding shock wave
 may also be influenced by the neutrino absorption and the inelastic scattering of $\alpha$ particles  \cite{furusawa13b, ohnishi07}. 

\subsection{Effects of  EOSs on simulations} \label{sec:eoscomp}
Many EOSs have been adopted in astrophysical calculations. 
In this review, we introduce some studies related to CCSNe. 

\subsubsection{Uniform nuclear matter}
Several EOSs used in CCSN simulations have been compared,  in CCSN simulations and
  it has been found that soft EOSs are preferred for neutrino-driven supernova explosions, 
  e.g.  \cite{fischer13,fischer17}.
Sumiyoshi et al.  \cite{sumiyoshi05}  compared the  LS (220) and STOS (TM1) EOSs in 1D simulations.
The lower  $J_0$ of the LS EOS   
 allows  for more protons to be dripped than that by the STOS EOS.
The dripped protons capture electrons, and thus,
 the decrease in electron degenerate pressures causes the mass of the PNS
to be approximately 0.1 $M_\odot$ lower than that from the STOS (TM1)  EOS.
The smaller mass of a PNS leads to smaller kinetic energy of the shock waves at bounce, which is  negative for shock revival.
However, the central density of the PNS in a simulation using the LS EOS
is higher than that from the STOS EOS,  because  of the softness (low values of $K_0$, $J_0$, and $L_0$)
 in the LS EOS.
The higher luminosity and higher average energy of the neutrinos  due to the higher central densities of the PNS
 are advantages for shock revival.


Suwa et al. \cite{suwa13}  shows similar results for 2D simulations 
by comparing the LS (180, 375) and STOS (TM1) EOSs.
The simulation using the STOS (TM1) EOS did  not reproduce a shock revival.    
The electron antineutrino luminosity  in the LS EOS models was high, and 
the mass enclosed within the gain region was large.
In addition,  they also found more aspherical downward flows of
 the accretion matter on the PNS surface with the model using the LS EOS than those with the STOS EOS.
In other 2D simulations comparing the LS (220) and FYSS (TM1) EOSs,  a PNS calculated by
 the softer EOS (LS EOS) shrinks faster and emits more neutrinos, leading to
  a larger radius of the shock wave \cite{nagakura18a, harada20}. 
Fully relativistic 3D CCSN simulations were performed using the HS (TM1, DD2, and SFHx) EOSs \cite{kuroda17}.
They showed that in supernova models  with softer EOS,
the development of the SASI is more active, affecting the gravitational wave observations.


The SRO EOSs have been systematically compared in supernova simulations 
using approximately 100 parameter sets \cite{schneider19}.
Using them in 1D simulations were implemented, whereas six 3D simulations were conducted using some selected parameters.
It was found  that the effective masses of nucleons, i.e.,  $m_p^*$ and $m_n^*$, in Eq.~(\ref{eq:sky}) at $n_B \geq n_0$
are the most influential  uncertainties  in Skyrme-type EOSs  affecting  neutrino emission and dynamics.
The peak frequency of the gravitational waves is also sensitive to the effective masses, which is related to the contraction of the PNS \cite{andersen21}.
It has also been pointed out that PNS convection and supernova explosion
are dependent on the entropy of the PNS, which  is determined  by the EOS \cite{boccioli21}.

Neutrino signals from the PNS cooling after a shock revival 
are also dependent on the EOS \cite{nakazato22}.
It is known that the time between a core-bounce and
 the formation of a black hole is short for a soft EOS \cite{sumiyoshi06, schneider20}.
Gravitational wave signals and  
the possibility of a shock revival along with the collapse of the PNS to a black hole
are also  dependent on the EOS \cite{pan18}.

It is noted that the thermodynamical conditions of supernova matter (see Fig.~\ref{fig_rtyp})
are quite different of NS matter. 
NS properties are sensitive to  the EOS at $Y_p \sim 0.1$ and   $T=0$ MeV.
Supernova matter has a wider range of thermodynamic conditions than NS matter:
 $0.1<Y_p <0.5$ and $T> 0.5$~MeV, as shown in Figs.~\ref{fig_rtyp}, \ref{fig_rt}, and  \ref{fig_fye}.
In addition, the effective mass, which determines the dynamics of CCSN simulations, does not play a key role in NS properties \cite{schneider19}. 
Therefore, a general-purpose EOS has many aspect as functions of $\rho_B$, $T$, and $Y_p$. 
Regarding the constraints on NS matter,  simulations and observations of neutron star mergers
\cite{abbott17,shibata17} are more promising than those of CCSNe introduced in Sec.~\ref{sec_pro}.

\subsubsection{ Non-uniform nuclear matter}
Hempel et al. \cite{hempel12}   performed 1D supernova simulations using the STOS (TM1), LS (180, 220), 
and HS (TMA, FSUgold).
The differences between the nuclear interactions (180 and 220, or TMA and FSUgold) 
are smaller than the differences between the nuclear models (STOS, LS and HS).
In 2D simulations \cite{suwa13}
comparing the LS (180, 375) and STOS (TM1) EOSs, similar results were obtained.
Despite the large difference in the incompressibility,
both 2D simulations  of a15 $M_\odot$ progenitor  using the LS (180) and  LS (375) EOSs
 show shock revival after a shock stall.
In a simple analysis \cite{lattimer00}, the chemical potentials of protons and neutrons, 
which are directly related to the weak interaction rates around shock waves,
 are more sensitive to non-uniform nuclear matter than to uniform nuclear matter. 

Nagakura et al. \cite{nagakura19a} systematically compared not only
the  EOSs but also the weak interaction consistencies with the EOSs,
and found that the weak rate consistency  affects the  structure of 
the PNS and  the time evolution of shock-waves
to the same degree
as  the  EOS difference. 
The weak rate difference originating from the nuclear composition of the EOSs is quite large. 
For example, in an early phase of the core collapse,
electron capture rates of heavy nuclei in the SNA EOS may be underestimated  by approximately 80$\%$ true rates \cite{furusawa17c}.  
In addition to the nuclear composition given by the  EOS, weak interaction rates themselves, particularly electron capture rates of heavy nuclei, have major influences on the dynamics \cite{sullivan16}.

\subsubsection{Heavy leptons and exotic hadrons} \label{exo}
Additional degrees of freedom due to heavy leptons and exotic hadrons
at high densities and high temperatures affect the dynamics of supernovae  and  the final fate of PNSs.
The muons in the central part of a PNS, soften the EOS, and, hence, 
the luminosity and mean energy of neutrinos increase \cite{bollig17}.
It is also pointed out that the second collapse of the core
 with  a phase transition from hadronic
  to quark matter may induce shock revivals 
for massive progenitors \cite{fischer11,kuroda22}.

The time from the core-bounce of a PNS to the black hole formation is dependent on the maximum mass of  the PNS.
The transition to quark matter reduces the maximum mass,  life time of  the PNS, and duration of the  neutrino emission    \cite{nakazato08,nakazato10}.
The emergence of hyperons also decreases the time for a PNS to become a black hole
\cite{sumiyoshi08,nakazato12,peres13}.
Central densities of the PNS for the typical progenitors of massive stars (e.g., 11.2 $M_\odot$ \cite{woosely02}) 
are up to a few times nuclear saturation density. 
Populations of hyperons and quark phases in the EOS are more likely
 to be realized in the core collapse of highly massive stars (e.g., 40 $M_\odot$). 

\section{Summary \label{sec:conc}}
We reviewed the EOSs for the hot and dense stellar matter and nuclei in the central engines of CCSNe.
Dripped neutrons and dense electrons allow populations of neutron-rich heavy nuclei to form
during the core-collapse stage, owing  to the reductions in the Coulomb energy and the total proton fraction.
The nuclei with $Z=$20--50 and $N =$40--90 
 specifically determine the core-deleptonization and the subsequent core dynamics  \cite{sullivan16,furusawa18b}. 
Nucleons and light nuclei dominate shock matter, and  
their mass fractions and weak interactions have  significant effects on shock wave dynamics. 
$\alpha$ particles are available around shock waves and in inward flows with low entropies, 
 whereas deuterons exist immediately above the surface of the  PNS.

Recently, numerous  EOSs have been developed.
Numerous constrains on the EOS for uniform nuclear matter are derived from 
ab-initio calculations of neutron matter, NS observations, and terrestrial experiments.
Consequently, some phenomenological EOSs have been updated, .e.g., 
the SLy4 parameter set for Skyrme-type interactions and the TM1e parameter sets for the RMF theory \cite{schneider17, raduta19, shen20}.  
Recent experiments  on PREX-II \cite{adhikari21, reed21},
in which  the neutron skin thickness of heavy nuclei is measured,  also help to constrain and improve the calculations of asymmetric nuclear matter.
We should also study the EOSs based on microscopic calculations \cite{togashi17, furusawa20a}.

There are three major ambiguities in modeling the EOS of non-uniform nuclear matter:  (1) the temperature dependence of the free energies  of neutron-rich heavy nuclei;
 (2)  calculations of  mixtures of nucleons and light nuclei at high entropies; and 
 (3) the transition from a mixture of various nuclei (non-uniform nuclear matter) to uniform
  nuclear matter below nuclear saturation density, $\rho_B\sim 10^{14}$ g/cm$^3$.
The first one should be addressed with  nuclear experiments and up-to-date calculations of heavy nuclei \cite{sakurai18,otsuka20}.
Investigation on  few-nucleon systems  and experiments would help to improve the second task \cite{typel10,roepke09, hempel15}. 
For the third task, molecular dynamics \cite{schneider13} and precise calculations of nuclear pastas \cite{pais14, horowitz16} would be helpful. 

\section*{Acknowledgments}
We thank S. Yamada, K. Sumiyoshi, I. Mishustin, and M. Hempel for productive discussions. 
This study was supported by JSPS KAKENHI (Grant Number JP 19K14723, 20H01905)
and  HPCI Strategic Program of Japanese MEXT  (Project ID: hp170304, 180111, 180179, 180239, 190100, 190160, 200102, 200124).
A part of the numerical calculations was conducted  on PC cluster at Center for Computational Astrophysics, National Astronomical Observatory of Japan.

\bibliographystyle{elsarticle-num} 
\newcommand{\gguide}{{\it Preparing graphics for IOP Publishing journals}}
\newcommand{\apj}{ Astrophys. J. }
\newcommand{\apjl}{ Astrophys. J. lett.}
\newcommand{\apjs}{ Astrophys. J. Suppl}
\newcommand{\aap}{ Astronomy \& Astrophysics}
\newcommand{\prc}{ Phys. Rev. C}
\newcommand{\prd}{ Phys. Rev. D}
\newcommand{\physrep}{Physics Reports}
\newcommand{\infin}{$\inifty$}

\newpage

\bibliography{reference221101}

\begin{thebibliography}{100}
\expandafter\ifx\csname url\endcsname\relax
  \def\url#1{\texttt{#1}}\fi
\expandafter\ifx\csname urlprefix\endcsname\relax\def\urlprefix{URL }\fi
\expandafter\ifx\csname href\endcsname\relax
  \def\href#1#2{#2} \def\path#1{#1}\fi

\bibitem{janka12}
H.-T. {Janka}, {Explosion Mechanisms of Core-Collapse Supernovae}, Annual
  Review of Nuclear and Particle Science 62 (2012) 407--451.
\newblock \href {http://arxiv.org/abs/1206.2503} {\path{arXiv:1206.2503}},
  \href {http://dx.doi.org/10.1146/annurev-nucl-102711-094901}
  {\path{doi:10.1146/annurev-nucl-102711-094901}}.

\bibitem{kotake12}
K.~{Kotake}, T.~{Takiwaki}, Y.~{Suwa}, W.~{Iwakami Nakano}, S.~{Kawagoe},
  Y.~{Masada}, S.-i. {Fujimoto}, {Multimessengers from Core-Collapse
  Supernovae: Multidimensionality as a Key to Bridge Theory and Observation},
  Advances in Astronomy 2012 (2012) 428757.
\newblock \href {http://arxiv.org/abs/1204.2330} {\path{arXiv:1204.2330}},
  \href {http://dx.doi.org/10.1155/2012/428757}
  {\path{doi:10.1155/2012/428757}}.

\bibitem{lattimer91}
J.~M. {Lattimer}, F.~D. {Swesty}, {A generalized equation of state for hot,
  dense matter}, Nuclear Physics A 535 (1991) 331--376.
\newblock \href {http://dx.doi.org/10.1016/0375-9474(91)90452-C}
  {\path{doi:10.1016/0375-9474(91)90452-C}}.

\bibitem{schneider17}
A.~S. Schneider, L.~F. Roberts, C.~D. Ott,
  \href{https://link.aps.org/doi/10.1103/PhysRevC.96.065802}{Open-source
  nuclear equation of state framework based on the liquid-drop model with
  skyrme interaction}, Phys. Rev. C 96 (2017) 065802.
\newblock \href {http://dx.doi.org/10.1103/PhysRevC.96.065802}
  {\path{doi:10.1103/PhysRevC.96.065802}}.
\newline\urlprefix\url{https://link.aps.org/doi/10.1103/PhysRevC.96.065802}

\bibitem{schneider19}
A.~S. Schneider, C.~Constantinou, B.~Muccioli, M.~Prakash,
  \href{https://link.aps.org/doi/10.1103/PhysRevC.100.025803}{Akmal-pandharipande-ravenhall
  equation of state for simulations of supernovae, neutron stars, and binary
  mergers}, Phys. Rev. C 100 (2019) 025803.
\newblock \href {http://dx.doi.org/10.1103/PhysRevC.100.025803}
  {\path{doi:10.1103/PhysRevC.100.025803}}.
\newline\urlprefix\url{https://link.aps.org/doi/10.1103/PhysRevC.100.025803}

\bibitem{raduta19}
A.~Raduta, F.~Gulminelli,
  \href{http://www.sciencedirect.com/science/article/pii/S0375947418303816}{Nuclear
  statistical equilibrium equation of state for core collapse}, Nuclear Physics
  A 983 (2019) 252 -- 275.
\newblock \href
  {http://dx.doi.org/https://doi.org/10.1016/j.nuclphysa.2018.11.003}
  {\path{doi:https://doi.org/10.1016/j.nuclphysa.2018.11.003}}.
\newline\urlprefix\url{http://www.sciencedirect.com/science/article/pii/S0375947418303816}

\bibitem{shen98a}
H.~{Shen}, H.~{Toki}, K.~{Oyamatsu}, K.~{Sumiyoshi}, {Relativistic equation of
  state of nuclear matter for supernova and neutron star}, Nuclear Physics A
  637 (1998) 435--450.
\newblock \href {http://dx.doi.org/10.1016/S0375-9474(98)00236-X}
  {\path{doi:10.1016/S0375-9474(98)00236-X}}.

\bibitem{shen98b}
H.~{Shen}, H.~{Toki}, K.~{Oyamatsu}, K.~{Sumiyoshi}, {Relativistic Equation of
  State of Nuclear Matter for Supernova Explosion}, Progress of Theoretical
  Physics 100 (1998) 1013--1031.
\newblock \href {http://dx.doi.org/10.1143/PTP.100.1013}
  {\path{doi:10.1143/PTP.100.1013}}.

\bibitem{shen11}
H.~Shen, H.~Toki, K.~Oyamatsu, K.~Sumiyoshi, {Relativistic Equation of State
  for Core-Collapse Supernova Simulations}, Astrophys.J.Suppl. 197 (2011) 20.
\newblock \href {http://arxiv.org/abs/1105.1666} {\path{arXiv:1105.1666}},
  \href {http://dx.doi.org/10.1088/0067-0049/197/2/20}
  {\path{doi:10.1088/0067-0049/197/2/20}}.

\bibitem{hempel10}
M.~{Hempel}, J.~{Schaffner-Bielich}, {A statistical model for a complete
  supernova equation of state}, Nuclear Physics A 837 (2010) 210--254.
\newblock \href {http://arxiv.org/abs/0911.4073} {\path{arXiv:0911.4073}},
  \href {http://dx.doi.org/10.1016/j.nuclphysa.2010.02.010}
  {\path{doi:10.1016/j.nuclphysa.2010.02.010}}.

\bibitem{steiner13}
A.~W. {Steiner}, M.~{Hempel}, T.~{Fischer}, {Core-collapse Supernova Equations
  of State Based on Neutron Star Observations}, \apj 774 (2013) 17.
\newblock \href {http://arxiv.org/abs/1207.2184} {\path{arXiv:1207.2184}},
  \href {http://dx.doi.org/10.1088/0004-637X/774/1/17}
  {\path{doi:10.1088/0004-637X/774/1/17}}.

\bibitem{sheng11a}
G.~{Shen}, C.~J. {Horowitz}, S.~{Teige}, {New equation of state for
  astrophysical simulations}, \prc 83~(3) (2011) 035802.
\newblock \href {http://arxiv.org/abs/1101.3715} {\path{arXiv:1101.3715}},
  \href {http://dx.doi.org/10.1103/PhysRevC.83.035802}
  {\path{doi:10.1103/PhysRevC.83.035802}}.

\bibitem{sheng11b}
G.~{Shen}, C.~J. {Horowitz}, E.~{O'Connor},
  \href{https://link.aps.org/doi/10.1103/PhysRevC.83.065808}{Second
  relativistic mean field and virial equation of state for astrophysical
  simulations}, Phys. Rev. C 83 (2011) 065808.
\newblock \href {http://dx.doi.org/10.1103/PhysRevC.83.065808}
  {\path{doi:10.1103/PhysRevC.83.065808}}.
\newline\urlprefix\url{https://link.aps.org/doi/10.1103/PhysRevC.83.065808}

\bibitem{furusawa13a}
S.~{Furusawa}, K.~{Sumiyoshi}, S.~{Yamada}, H.~{Suzuki}, {New Equations of
  State Based on the Liquid Drop Model of Heavy Nuclei and Quantum Approach to
  Light Nuclei for Core-collapse Supernova Simulations}, \apj 772 (2013) 95.
\newblock \href {http://arxiv.org/abs/1305.1508} {\path{arXiv:1305.1508}},
  \href {http://dx.doi.org/10.1088/0004-637X/772/2/95}
  {\path{doi:10.1088/0004-637X/772/2/95}}.

\bibitem{furusawa17a}
S.~Furusawa, K.~Sumiyoshi, S.~Yamada, H.~Suzuki, Supernova equations of state
  including full nuclear ensemble with in-medium effects, Nuclear Physics A 957
  (2017) 188 -- 207.
\newblock \href
  {http://dx.doi.org/http://dx.doi.org/10.1016/j.nuclphysa.2016.09.002}
  {\path{doi:http://dx.doi.org/10.1016/j.nuclphysa.2016.09.002}}.

\bibitem{typel18}
S.~Typel, \href{https://doi.org/10.1088{\%}2F1361-6471{\%}2Faadea5}{Equations
  of state for astrophysical simulations from generalized relativistic density
  functionals}, Journal of Physics G: Nuclear and Particle Physics 45~(11)
  (2018) 114001.
\newblock \href {http://dx.doi.org/10.1088/1361-6471/aadea5}
  {\path{doi:10.1088/1361-6471/aadea5}}.
\newline\urlprefix\url{https://doi.org/10.1088{\%}2F1361-6471{\%}2Faadea5}

\bibitem{togashi17}
H.~Togashi, K.~Nakazato, Y.~Takehara, S.~Yamamuro, H.~Suzuki, M.~Takano,
  \href{http://www.sciencedirect.com/science/article/pii/S0375947417300350}{Nuclear
  equation of state for core-collapse supernova simulations with realistic
  nuclear forces}, Nuclear Physics A 961 (2017) 78 -- 105.
\newblock \href
  {http://dx.doi.org/https://doi.org/10.1016/j.nuclphysa.2017.02.010}
  {\path{doi:https://doi.org/10.1016/j.nuclphysa.2017.02.010}}.
\newline\urlprefix\url{http://www.sciencedirect.com/science/article/pii/S0375947417300350}

\bibitem{furusawa17d}
S.~Furusawa, H.~Togashi, H.~Nagakura, K.~Sumiyoshi, S.~Yamada, H.~Suzuki,
  M.~Takano, \href{http://stacks.iop.org/0954-3899/44/i=9/a=094001}{A new
  equation of state for core-collapse supernovae based on realistic nuclear
  forces and including a full nuclear ensemble}, Journal of Physics G: Nuclear
  and Particle Physics 44~(9) (2017) 094001.
\newline\urlprefix\url{http://stacks.iop.org/0954-3899/44/i=9/a=094001}

\bibitem{furusawa20a}
S.~Furusawa, H.~Togashi, K.~Sumiyoshi, K.~Saito, S.~Yamada, H.~Suzuki,
  \href{https://doi.org/10.1093/ptep/ptz135}{{Nuclear statistical equilibrium
  equation of state with a parametrized Dirac-Br{\" u}ckner Hartree-Fock
  calculation}}, Progress of Theoretical and Experimental Physics 2020~(1),
  013D05.
\newblock \href
  {http://arxiv.org/abs/https://academic.oup.com/ptep/article-pdf/2020/1/013D05/32290084/ptz135.pdf}
  {\path{arXiv:https://academic.oup.com/ptep/article-pdf/2020/1/013D05/32290084/ptz135.pdf}},
  \href {http://dx.doi.org/10.1093/ptep/ptz135}
  {\path{doi:10.1093/ptep/ptz135}}.
\newline\urlprefix\url{https://doi.org/10.1093/ptep/ptz135}

\bibitem{burrows84}
A.~{Burrows}, J.~M. {Lattimer}, {On the accuracy of the single-nucleus
  approximation in the equation of state of hot, dense matter}, \apj 285 (1984)
  294--303.
\newblock \href {http://dx.doi.org/10.1086/162505} {\path{doi:10.1086/162505}}.

\bibitem{furusawa17b}
S.~Furusawa, H.~Nagakura, K.~Sumiyoshi, C.~Kato, S.~Yamada,
  \href{http://link.aps.org/doi/10.1103/PhysRevC.95.025809}{Dependence of weak
  interaction rates on the nuclear composition during stellar core collapse},
  Phys. Rev. C 95 (2017) 025809.
\newblock \href {http://dx.doi.org/10.1103/PhysRevC.95.025809}
  {\path{doi:10.1103/PhysRevC.95.025809}}.
\newline\urlprefix\url{http://link.aps.org/doi/10.1103/PhysRevC.95.025809}

\bibitem{blinnikov11}
S.~I. {Blinnikov}, I.~V. {Panov}, M.~A. {Rudzsky}, K.~{Sumiyoshi}, {The
  equation of state and composition of hot, dense matter in core-collapse
  supernovae}, \aap 535 (2011) A37.
\newblock \href {http://arxiv.org/abs/0904.3849} {\path{arXiv:0904.3849}},
  \href {http://dx.doi.org/10.1051/0004-6361/201117225}
  {\path{doi:10.1051/0004-6361/201117225}}.

\bibitem{furusawa11}
S.~{Furusawa}, S.~{Yamada}, K.~{Sumiyoshi}, H.~{Suzuki}, {A New Baryonic
  Equation of State at Sub-nuclear Densities for Core-collapse Simulations},
  \apj 738 (2011) 178.
\newblock \href {http://arxiv.org/abs/1103.6129} {\path{arXiv:1103.6129}},
  \href {http://dx.doi.org/10.1088/0004-637X/738/2/178}
  {\path{doi:10.1088/0004-637X/738/2/178}}.

\bibitem{sumiyoshi05}
K.~{Sumiyoshi}, S.~{Yamada}, H.~{Suzuki}, H.~{Shen}, S.~{Chiba}, H.~{Toki},
  {Postbounce Evolution of Core-Collapse Supernovae: Long-Term Effects of the
  Equation of State}, \apj 629 (2005) 922--932.
\newblock \href {http://arxiv.org/abs/astro-ph/0506620}
  {\path{arXiv:astro-ph/0506620}}, \href {http://dx.doi.org/10.1086/431788}
  {\path{doi:10.1086/431788}}.

\bibitem{nagakura19a}
H.~Nagakura, S.~Furusawa, H.~Togashi, S.~Richers, K.~Sumiyoshi, S.~Yamada,
  \apjs 240~(2) (2019) 38.
\newblock \href {http://dx.doi.org/10.3847/1538-4365/aafac9}
  {\path{doi:10.3847/1538-4365/aafac9}},
  \href{https://doi.org/10.3847{\%}2F1538-4365{\%}2Faafac9}{[link]}.
\newline\urlprefix\url{https://doi.org/10.3847{\%}2F1538-4365{\%}2Faafac9}

\bibitem{oertel17}
M.~{Oertel}, M.~{Hempel}, T.~{Kl{\"a}hn}, S.~{Typel}, {Equations of state for
  supernovae and compact stars}, Reviews of Modern Physics 89~(1) (2017)
  015007.
\newblock \href {http://arxiv.org/abs/1610.03361} {\path{arXiv:1610.03361}},
  \href {http://dx.doi.org/10.1103/RevModPhys.89.015007}
  {\path{doi:10.1103/RevModPhys.89.015007}}.

\bibitem{raduta21}
A.~R. {Raduta}, F.~{Nacu}, M.~{Oertel}, {EoS for hot neutron stars}, arXiv
  e-prints (2021) arXiv:2109.00251\href {http://arxiv.org/abs/2109.00251}
  {\path{arXiv:2109.00251}}.

\bibitem{typel15}
S.~Typel, M.~Oertel, T.~Kl\"ahn, {CompOSE CompStar online supernova equations
  of state harmonising the concert of nuclear physics and astrophysics
  compose.obspm.fr}, Phys. Part. Nucl. 46~(4) (2015) 633--664.
\newblock \href {http://arxiv.org/abs/1307.5715} {\path{arXiv:1307.5715}},
  \href {http://dx.doi.org/10.1134/S1063779615040061}
  {\path{doi:10.1134/S1063779615040061}}.

\bibitem{janka16}
H.-T. Janka, T.~Melson, A.~Summa,
  \href{https://doi.org/10.1146/annurev-nucl-102115-044747}{Physics of
  core-collapse supernovae in three dimensions: A sneak preview}, Annual Review
  of Nuclear and Particle Science 66~(1) (2016) 341--375.
\newblock \href
  {http://arxiv.org/abs/https://doi.org/10.1146/annurev-nucl-102115-044747}
  {\path{arXiv:https://doi.org/10.1146/annurev-nucl-102115-044747}}, \href
  {http://dx.doi.org/10.1146/annurev-nucl-102115-044747}
  {\path{doi:10.1146/annurev-nucl-102115-044747}}.
\newline\urlprefix\url{https://doi.org/10.1146/annurev-nucl-102115-044747}

\bibitem{burrows21}
A.~{Burrows}, D.~{Vartanyan}, {Core-collapse supernova explosion theory},
  Nature 589~(7840) (2021) 29--39.
\newblock \href {http://arxiv.org/abs/2009.14157} {\path{arXiv:2009.14157}},
  \href {http://dx.doi.org/10.1038/s41586-020-03059-w}
  {\path{doi:10.1038/s41586-020-03059-w}}.

\bibitem{woosely02}
S.~E. Woosley, A.~Heger, T.~A. Weaver,
  \href{https://link.aps.org/doi/10.1103/RevModPhys.74.1015}{The evolution and
  explosion of massive stars}, Rev. Mod. Phys. 74 (2002) 1015--1071.
\newblock \href {http://dx.doi.org/10.1103/RevModPhys.74.1015}
  {\path{doi:10.1103/RevModPhys.74.1015}}.
\newline\urlprefix\url{https://link.aps.org/doi/10.1103/RevModPhys.74.1015}

\bibitem{iwakami08}
W.~Iwakami, K.~Kotake, N.~Ohnishi, S.~Yamada, K.~Sawada,
  \href{https://doi.org/10.1086/533582}{Three-dimensional simulations of
  standing accretion shock instability in core-collapse supernovae}, The
  Astrophysical Journal 678~(2) (2008) 1207--1222.
\newblock \href {http://dx.doi.org/10.1086/533582} {\path{doi:10.1086/533582}}.
\newline\urlprefix\url{https://doi.org/10.1086/533582}

\bibitem{takiwaki14}
T.~Takiwaki, K.~Kotake, Y.~Suwa,
  \href{https://doi.org/10.1088/0004-637x/786/2/83}{A {COMPARISON} {OF} {TWO}-
  {AND} {THREE}-{DIMENSIONAL} {NEUTRINO}-{HYDRODYNAMICS} {SIMULATIONS} {OF}
  {CORE}-{COLLAPSE} {SUPERNOVAE}}, The Astrophysical Journal 786~(2) (2014) 83.
\newblock \href {http://dx.doi.org/10.1088/0004-637x/786/2/83}
  {\path{doi:10.1088/0004-637x/786/2/83}}.
\newline\urlprefix\url{https://doi.org/10.1088/0004-637x/786/2/83}

\bibitem{lentz15}
E.~J. {Lentz}, S.~W. {Bruenn}, W.~R. {Hix}, A.~{Mezzacappa}, O.~E.~B. {Messer},
  E.~{Endeve}, J.~M. {Blondin}, J.~A. {Harris}, P.~{Marronetti}, K.~N.
  {Yakunin}, {Three-dimensional Core-collapse Supernova Simulated Using a 15 M
  $_{{\ensuremath{\odot}}}$ Progenitor}, \apjl 807~(2) (2015) L31.
\newblock \href {http://arxiv.org/abs/1505.05110} {\path{arXiv:1505.05110}},
  \href {http://dx.doi.org/10.1088/2041-8205/807/2/L31}
  {\path{doi:10.1088/2041-8205/807/2/L31}}.

\bibitem{roberts16}
L.~F. {Roberts}, C.~D. {Ott}, R.~{Haas}, E.~P. {O'Connor}, P.~{Diener},
  E.~{Schnetter}, {General-Relativistic Three-Dimensional Multi-group Neutrino
  Radiation-Hydrodynamics Simulations of Core-Collapse Supernovae}, \apj
  831~(1) (2016) 98.
\newblock \href {http://arxiv.org/abs/1604.07848} {\path{arXiv:1604.07848}},
  \href {http://dx.doi.org/10.3847/0004-637X/831/1/98}
  {\path{doi:10.3847/0004-637X/831/1/98}}.

\bibitem{oconnor18}
E.~P. {O'Connor}, S.~M. {Couch}, {Exploring Fundamentally Three-dimensional
  Phenomena in High-fidelity Simulations of Core-collapse Supernovae}, \apj
  865~(2) (2018) 81.
\newblock \href {http://arxiv.org/abs/1807.07579} {\path{arXiv:1807.07579}},
  \href {http://dx.doi.org/10.3847/1538-4357/aadcf7}
  {\path{doi:10.3847/1538-4357/aadcf7}}.

\bibitem{bernhard19}
B.~{M{\"u}ller}, T.~M. {Tauris}, A.~{Heger}, P.~{Banerjee}, Y.-Z. {Qian},
  J.~{Powell}, C.~{Chan}, D.~W. {Gay}, N.~{Langer}, {Three-dimensional
  simulations of neutrino-driven core-collapse supernovae from low-mass single
  and binary star progenitors}, \mnras 484~(3) (2019) 3307--3324.
\newblock \href {http://arxiv.org/abs/1811.05483} {\path{arXiv:1811.05483}},
  \href {http://dx.doi.org/10.1093/mnras/stz216}
  {\path{doi:10.1093/mnras/stz216}}.

\bibitem{nagakura19b}
H.~{Nagakura}, K.~{Sumiyoshi}, S.~{Yamada}, {Possible Early Linear Acceleration
  of Proto-neutron Stars via Asymmetric Neutrino Emission in Core-collapse
  Supernovae}, \apjl 880~(2) (2019) L28.
\newblock \href {http://arxiv.org/abs/1907.04863} {\path{arXiv:1907.04863}},
  \href {http://dx.doi.org/10.3847/2041-8213/ab30ca}
  {\path{doi:10.3847/2041-8213/ab30ca}}.

\bibitem{robert19}
R.~{Glas}, H.~T. {Janka}, T.~{Melson}, G.~{Stockinger}, O.~{Just}, {Effects of
  LESA in Three-dimensional Supernova Simulations with Multidimensional and
  Ray-by-ray-plus Neutrino Transport}, \apj 881~(1) (2019) 36.
\newblock \href {http://arxiv.org/abs/1809.10150} {\path{arXiv:1809.10150}},
  \href {http://dx.doi.org/10.3847/1538-4357/ab275c}
  {\path{doi:10.3847/1538-4357/ab275c}}.

\bibitem{vartanyan19}
D.~{Vartanyan}, A.~{Burrows}, D.~{Radice}, {Temporal and angular variations of
  3D core-collapse supernova emissions and their physical correlations}, \mnras
  489~(2) (2019) 2227--2246.
\newblock \href {http://arxiv.org/abs/1906.08787} {\path{arXiv:1906.08787}},
  \href {http://dx.doi.org/10.1093/mnras/stz2307}
  {\path{doi:10.1093/mnras/stz2307}}.

\bibitem{nagakura19c}
H.~{Nagakura}, A.~{Burrows}, D.~{Radice}, D.~{Vartanyan}, {Towards an
  understanding of the resolution dependence of Core-Collapse Supernova
  simulations}, \mnras 490~(4) (2019) 4622--4637.
\newblock \href {http://arxiv.org/abs/1905.03786} {\path{arXiv:1905.03786}},
  \href {http://dx.doi.org/10.1093/mnras/stz2730}
  {\path{doi:10.1093/mnras/stz2730}}.

\bibitem{burrows20}
A.~{Burrows}, D.~{Radice}, D.~{Vartanyan}, H.~{Nagakura}, M.~A. {Skinner},
  J.~C. {Dolence}, {The overarching framework of core-collapse supernova
  explosions as revealed by 3D FORNAX simulations}, \mnras 491~(2) (2020)
  2715--2735.
\newblock \href {http://arxiv.org/abs/1909.04152} {\path{arXiv:1909.04152}},
  \href {http://dx.doi.org/10.1093/mnras/stz3223}
  {\path{doi:10.1093/mnras/stz3223}}.

\bibitem{nagakura20}
H.~{Nagakura}, A.~{Burrows}, D.~{Radice}, D.~{Vartanyan}, {A systematic study
  of proto-neutron star convection in three-dimensional core-collapse supernova
  simulations}, \mnras 492~(4) (2020) 5764--5779.
\newblock \href {http://arxiv.org/abs/1912.07615} {\path{arXiv:1912.07615}},
  \href {http://dx.doi.org/10.1093/mnras/staa261}
  {\path{doi:10.1093/mnras/staa261}}.

\bibitem{jade20}
J.~{Powell}, B.~{M{\"u}ller}, {Three-dimensional core-collapse supernova
  simulations of massive and rotating progenitors}, \mnras 494~(4) (2020)
  4665--4675.
\newblock \href {http://arxiv.org/abs/2002.10115} {\path{arXiv:2002.10115}},
  \href {http://dx.doi.org/10.1093/mnras/staa1048}
  {\path{doi:10.1093/mnras/staa1048}}.

\bibitem{kuroda20}
T.~{Kuroda}, A.~{Arcones}, T.~{Takiwaki}, K.~{Kotake}, {Magnetorotational
  Explosion of a Massive Star Supported by Neutrino Heating in General
  Relativistic Three-dimensional Simulations}, \apj 896~(2) (2020) 102.
\newblock \href {http://arxiv.org/abs/2003.02004} {\path{arXiv:2003.02004}},
  \href {http://dx.doi.org/10.3847/1538-4357/ab9308}
  {\path{doi:10.3847/1538-4357/ab9308}}.

\bibitem{takiwaki21}
T.~{Takiwaki}, K.~{Kotake}, T.~{Foglizzo}, {Insights into non-axisymmetric
  instabilities in three-dimensional rotating supernova models with neutrino
  and gravitational-wave signatures}, \mnras 508~(1) (2021) 966--985.
\newblock \href {http://arxiv.org/abs/2107.02933} {\path{arXiv:2107.02933}},
  \href {http://dx.doi.org/10.1093/mnras/stab2607}
  {\path{doi:10.1093/mnras/stab2607}}.

\bibitem{yamamoto13}
Y.~Yamamoto, S.~ichiro Fujimoto, H.~Nagakura, S.~Yamada,
  \href{https://doi.org/10.1088/0004-637x/771/1/27}{{POST}-{SHOCK}-{REVIVAL}
  {EVOLUTION} {IN} {THE} {NEUTRINO}-{HEATING} {MECHANISM} {OF}
  {CORE}-{COLLAPSE} {SUPERNOVAE}}, The Astrophysical Journal 771~(1) (2013) 27.
\newblock \href {http://dx.doi.org/10.1088/0004-637x/771/1/27}
  {\path{doi:10.1088/0004-637x/771/1/27}}.
\newline\urlprefix\url{https://doi.org/10.1088/0004-637x/771/1/27}

\bibitem{nakamura14}
K.~Nakamura, T.~Takiwaki, K.~Kotake, N.~Nishimura,
  \href{https://doi.org/10.1088/0004-637x/782/2/91}{{REVISITING} {IMPACTS} {OF}
  {NUCLEAR} {BURNING} {FOR} {REVIVING} {WEAK} {SHOCKS} {IN} {NEUTRINO}-{DRIVEN}
  {SUPERNOVAE}}, The Astrophysical Journal 782~(2) (2014) 91.
\newblock \href {http://dx.doi.org/10.1088/0004-637x/782/2/91}
  {\path{doi:10.1088/0004-637x/782/2/91}}.
\newline\urlprefix\url{https://doi.org/10.1088/0004-637x/782/2/91}

\bibitem{nakamura19}
K.~Nakamura, T.~Takiwaki, K.~Kotake,
  \href{https://doi.org/10.1093/pasj/psz080}{{Long-term simulations of
  multi-dimensional core-collapse supernovae: Implications for neutron star
  kicks}}, Publications of the Astronomical Society of Japan 71~(5), 98.
\newblock \href
  {http://arxiv.org/abs/https://academic.oup.com/pasj/article-pdf/71/5/98/30161428/psz080.pdf}
  {\path{arXiv:https://academic.oup.com/pasj/article-pdf/71/5/98/30161428/psz080.pdf}},
  \href {http://dx.doi.org/10.1093/pasj/psz080}
  {\path{doi:10.1093/pasj/psz080}}.
\newline\urlprefix\url{https://doi.org/10.1093/pasj/psz080}

\bibitem{suwa15}
Y.~Suwa, S.~Yamada, T.~Takiwaki, K.~Kotake,
  \href{https://doi.org/10.3847/0004-637x/816/1/43}{{THE} {CRITERION} {OF}
  {SUPERNOVA} {EXPLOSION} {REVISITED}: {THE} {MASS} {ACCRETION} {HISTORY}}, The
  Astrophysical Journal 816~(1) (2015) 43.
\newblock \href {http://dx.doi.org/10.3847/0004-637x/816/1/43}
  {\path{doi:10.3847/0004-637x/816/1/43}}.
\newline\urlprefix\url{https://doi.org/10.3847/0004-637x/816/1/43}

\bibitem{suwa16}
Y.~Suwa, E.~M{\"u}ller,
  \href{http://dx.doi.org/10.1093/mnras/stw1150}{Parametric initial conditions
  for core-collapse supernova simulations}, Monthly Notices of the Royal
  Astronomical Society 460~(3) (2016) 2664--2674.
\newblock \href {http://dx.doi.org/10.1093/mnras/stw1150}
  {\path{doi:10.1093/mnras/stw1150}}.
\newline\urlprefix\url{http://dx.doi.org/10.1093/mnras/stw1150}

\bibitem{nagakura18b}
H.~Nagakura, K.~Takahashi, Y.~Yamamoto,
  \href{https://doi.org/10.1093/mnras/sty3114}{{On the importance of progenitor
  asymmetry to shock revival in core-collapse supernovae}}, Monthly Notices of
  the Royal Astronomical Society 483~(1) (2018) 208--222.
\newblock \href
  {http://arxiv.org/abs/https://academic.oup.com/mnras/article-pdf/483/1/208/26996558/sty3114.pdf}
  {\path{arXiv:https://academic.oup.com/mnras/article-pdf/483/1/208/26996558/sty3114.pdf}},
  \href {http://dx.doi.org/10.1093/mnras/sty3114}
  {\path{doi:10.1093/mnras/sty3114}}.
\newline\urlprefix\url{https://doi.org/10.1093/mnras/sty3114}

\bibitem{yoshida19}
T.~Yoshida, T.~Takiwaki, K.~Kotake, K.~Takahashi, K.~Nakamura, H.~Umeda,
  \href{https://doi.org/10.3847/1538-4357/ab2b9d}{One-, two-, and
  three-dimensional simulations of oxygen-shell burning just before the core
  collapse of massive stars}, The Astrophysical Journal 881~(1) (2019) 16.
\newblock \href {http://dx.doi.org/10.3847/1538-4357/ab2b9d}
  {\path{doi:10.3847/1538-4357/ab2b9d}}.
\newline\urlprefix\url{https://doi.org/10.3847/1538-4357/ab2b9d}

\bibitem{yoshida21}
T.~Yoshida, T.~Takiwaki, D.~R. Aguilera-Dena, K.~Kotake, K.~Takahashi,
  K.~Nakamura, H.~Umeda, N.~Langer, {A three-dimensional hydrodynamics
  simulation of oxygen-shell burning in the final evolution of a fast-rotating
  massive star}, Mon. Not. Roy. Astron. Soc. 506~(1) (2021) L20--L25.
\newblock \href {http://arxiv.org/abs/2106.09909} {\path{arXiv:2106.09909}},
  \href {http://dx.doi.org/10.1093/mnrasl/slab067}
  {\path{doi:10.1093/mnrasl/slab067}}.

\bibitem{nagakura14}
H.~Nagakura, K.~Sumiyoshi, S.~Yamada,
  \href{https://doi.org/10.1088/0067-0049/214/2/16}{{THREE}-{DIMENSIONAL}
  {BOLTZMANN} {HYDRO} {CODE} {FOR} {CORE} {COLLAPSE} {IN} {MASSIVE} {STARS}. i.
  {SPECIAL} {RELATIVISTIC} {TREATMENTS}}, The Astrophysical Journal Supplement
  Series 214~(2) (2014) 16.
\newblock \href {http://dx.doi.org/10.1088/0067-0049/214/2/16}
  {\path{doi:10.1088/0067-0049/214/2/16}}.
\newline\urlprefix\url{https://doi.org/10.1088/0067-0049/214/2/16}

\bibitem{akaho21}
R.~Akaho, A.~Harada, H.~Nagakura, K.~Sumiyoshi, W.~Iwakami, H.~Okawa,
  S.~Furusawa, H.~Matsufuru, S.~Yamada,
  \href{https://doi.org/10.3847/1538-4357/abe1bf}{Multidimensional boltzmann
  neutrino transport code in full general relativity for core-collapse
  simulations}, The Astrophysical Journal 909~(2) (2021) 210.
\newblock \href {http://dx.doi.org/10.3847/1538-4357/abe1bf}
  {\path{doi:10.3847/1538-4357/abe1bf}}.
\newline\urlprefix\url{https://doi.org/10.3847/1538-4357/abe1bf}

\bibitem{bollig17}
R.~Bollig, H.-T. Janka, A.~Lohs, G.~Mart\'{\i}nez-Pinedo, C.~J. Horowitz,
  T.~Melson,
  \href{https://link.aps.org/doi/10.1103/PhysRevLett.119.242702}{Muon creation
  in supernova matter facilitates neutrino-driven explosions}, Phys. Rev. Lett.
  119 (2017) 242702.
\newblock \href {http://dx.doi.org/10.1103/PhysRevLett.119.242702}
  {\path{doi:10.1103/PhysRevLett.119.242702}}.
\newline\urlprefix\url{https://link.aps.org/doi/10.1103/PhysRevLett.119.242702}

\bibitem{fischer11}
T.~Fischer, I.~Sagert, G.~Pagliara, M.~Hempel, J.~Schaffner-Bielich,
  T.~Rauscher, F.-K. Thielemann, R.~Kappeli, G.~Martinez-Pinedo,
  M.~Liebendorfer,
  \href{https://doi.org/10.1088/0067-0049/194/2/39}{{CORE}-{COLLAPSE}
  {SUPERNOVA} {EXPLOSIONS} {TRIGGERED} {BY} a {QUARK}-{HADRON} {PHASE}
  {TRANSITION} {DURING} {THE} {EARLY} {POST}-{BOUNCE} {PHASE}}, The
  Astrophysical Journal Supplement Series 194~(2) (2011) 39.
\newblock \href {http://dx.doi.org/10.1088/0067-0049/194/2/39}
  {\path{doi:10.1088/0067-0049/194/2/39}}.
\newline\urlprefix\url{https://doi.org/10.1088/0067-0049/194/2/39}

\bibitem{newton09}
W.~G. {Newton}, J.~R. {Stone}, {Modeling nuclear ``pasta'' and the transition
  to uniform nuclear matter with the 3D Skyrme-Hartree-Fock method at finite
  temperature: Core-collapse supernovae}, \prc 79~(5) (2009) 055801.
\newblock \href {http://dx.doi.org/10.1103/PhysRevC.79.055801}
  {\path{doi:10.1103/PhysRevC.79.055801}}.

\bibitem{watanabe11}
G.~{Watanabe}, T.~{Maruyama}, {Nuclear pasta in supernovae and neutron stars},
  ArXiv e-prints\href {http://arxiv.org/abs/1109.3511}
  {\path{arXiv:1109.3511}}.

\bibitem{roggero17}
A.~{Roggero}, J.~{Margueron}, L.~F. {Roberts}, S.~{Reddy}, {Nuclear pasta in
  hot dense matter and its implications for neutrino scattering}, ArXiv
  e-prints\href {http://arxiv.org/abs/1710.10206} {\path{arXiv:1710.10206}}.

\bibitem{shapiro83}
S.~L. {Shapiro}, S.~A. {Teukolsky}, {Black holes, white dwarfs, and neutron
  stars : the physics of compact objects}, 1983.

\bibitem{burrows06}
A.~{Burrows}, S.~{Reddy}, T.~A. {Thompson}, {Neutrino opacities in nuclear
  matter}, Nuclear Physics A 777 (2006) 356--394.
\newblock \href {http://arxiv.org/abs/astro-ph/0404432}
  {\path{arXiv:astro-ph/0404432}}, \href
  {http://dx.doi.org/10.1016/j.nuclphysa.2004.06.012}
  {\path{doi:10.1016/j.nuclphysa.2004.06.012}}.

\bibitem{sullivan16}
C.~{Sullivan}, E.~{O'Connor}, R.~G.~T. {Zegers}, T.~{Grubb}, S.~M. {Austin},
  {The Sensitivity of Core-collapse Supernovae to Nuclear Electron Capture},
  \apj 816 (2016) 44.
\newblock \href {http://arxiv.org/abs/1508.07348} {\path{arXiv:1508.07348}},
  \href {http://dx.doi.org/10.3847/0004-637X/816/1/44}
  {\path{doi:10.3847/0004-637X/816/1/44}}.

\bibitem{burrows98}
A.~{Burrows}, R.~F. {Sawyer}, {Effects of correlations on neutrino opacities in
  nuclear matter}, \prc 58~(1) (1998) 554--571.
\newblock \href {http://arxiv.org/abs/astro-ph/9801082}
  {\path{arXiv:astro-ph/9801082}}, \href
  {http://dx.doi.org/10.1103/PhysRevC.58.554}
  {\path{doi:10.1103/PhysRevC.58.554}}.

\bibitem{burrows99}
A.~{Burrows}, R.~F. {Sawyer}, {Many-body corrections to charged-current
  neutrino absorption rates in nuclear matter}, \prc 59~(1) (1999) 510--514.
\newblock \href {http://arxiv.org/abs/astro-ph/9804264}
  {\path{arXiv:astro-ph/9804264}}, \href
  {http://dx.doi.org/10.1103/PhysRevC.59.510}
  {\path{doi:10.1103/PhysRevC.59.510}}.

\bibitem{horowitz17}
C.~J. Horowitz, O.~L. Caballero, Z.~Lin, E.~O'Connor, A.~Schwenk,
  \href{https://link.aps.org/doi/10.1103/PhysRevC.95.025801}{Neutrino-nucleon
  scattering in supernova matter from the virial expansion}, Phys. Rev. C 95
  (2017) 025801.
\newblock \href {http://dx.doi.org/10.1103/PhysRevC.95.025801}
  {\path{doi:10.1103/PhysRevC.95.025801}}.
\newline\urlprefix\url{https://link.aps.org/doi/10.1103/PhysRevC.95.025801}

\bibitem{fuller82}
G.~M. {Fuller}, W.~A. {Fowler}, M.~J. {Newman}, {Stellar weak interaction rates
  for intermediate-mass nuclei. II - A = 21 to A = 60}, \apj 252 (1982)
  715--740.
\newblock \href {http://dx.doi.org/10.1086/159597} {\path{doi:10.1086/159597}}.

\bibitem{oda94}
T.~{Oda}, M.~{Hino}, K.~{Muto}, M.~{Takahara}, K.~{Sato}, {Rate Tables for the
  Weak Processes of sd-Shell Nuclei in Stellar Matter}, Atomic Data and Nuclear
  Data Tables 56 (1994) 231--403.
\newblock \href {http://dx.doi.org/10.1006/adnd.1994.1007}
  {\path{doi:10.1006/adnd.1994.1007}}.

\bibitem{langanke00}
K.~{Langanke}, G.~{Mart{\'{\i}}nez-Pinedo}, {Shell-model calculations of
  stellar weak interaction rates: II. Weak rates for nuclei in the mass range
  /A=45-65 in supernovae environments}, Nuclear Physics A 673 (2000) 481--508.
\newblock \href {http://arxiv.org/abs/nucl-th/0001018}
  {\path{arXiv:nucl-th/0001018}}, \href
  {http://dx.doi.org/10.1016/S0375-9474(00)00131-7}
  {\path{doi:10.1016/S0375-9474(00)00131-7}}.

\bibitem{langanke03}
K.~Langanke, G.~Mart\'{\i}nez-Pinedo, J.~M. Sampaio, D.~J. Dean, W.~R. Hix,
  O.~E.~B. Messer, A.~Mezzacappa, M.~Liebend\"orfer, H.-T. Janka, M.~Rampp,
  \href{http://link.aps.org/doi/10.1103/PhysRevLett.90.241102}{Electron capture
  rates on nuclei and implications for stellar core collapse}, Phys. Rev. Lett.
  90 (2003) 241102.
\newblock \href {http://dx.doi.org/10.1103/PhysRevLett.90.241102}
  {\path{doi:10.1103/PhysRevLett.90.241102}}.
\newline\urlprefix\url{http://link.aps.org/doi/10.1103/PhysRevLett.90.241102}

\bibitem{fuller85}
G.~M. {Fuller}, W.~A. {Fowler}, M.~J. {Newman}, {Stellar weak interaction rates
  for intermediate-mass nuclei. IV - Interpolation procedures for rapidly
  varying lepton capture rates using effective log (ft)-values}, \apj 293
  (1985) 1--16.
\newblock \href {http://dx.doi.org/10.1086/163208} {\path{doi:10.1086/163208}}.

\bibitem{kato21}
M.~Kato, S.~Furusawa, K.~Suzuki,
  \href{https://doi.org/10.7566/JPSJ.90.105001}{Electron in-medium effects on
  electron captures of neutron-rich nuclei in stellar core collapse}, Journal
  of the Physical Society of Japan 90~(10) (2021) 105001.
\newblock \href {http://arxiv.org/abs/https://doi.org/10.7566/JPSJ.90.105001}
  {\path{arXiv:https://doi.org/10.7566/JPSJ.90.105001}}, \href
  {http://dx.doi.org/10.7566/JPSJ.90.105001}
  {\path{doi:10.7566/JPSJ.90.105001}}.
\newline\urlprefix\url{https://doi.org/10.7566/JPSJ.90.105001}

\bibitem{langanke21}
K.~Langanke, G.~{Mart{\'{\i}}nez-Pinedo}, R.~G.~T. Zegers,
  \href{https://dx.doi.org/10.1088/1361-6633/abf207}{Electron capture in
  stars}, Reports on Progress in Physics 84~(6) (2021) 066301.
\newblock \href {http://dx.doi.org/10.1088/1361-6633/abf207}
  {\path{doi:10.1088/1361-6633/abf207}}.
\newline\urlprefix\url{https://dx.doi.org/10.1088/1361-6633/abf207}

\bibitem{titus18}
R.~Titus, C.~Sullivan, R.~G.~T. Zegers, B.~A. Brown, B.~Gao,
  \href{http://stacks.iop.org/0954-3899/45/i=1/a=014004}{Impact of
  electron-captures on nuclei near n = 50 on core-collapse supernovae}, Journal
  of Physics G: Nuclear and Particle Physics 45~(1) (2018) 014004.
\newline\urlprefix\url{http://stacks.iop.org/0954-3899/45/i=1/a=014004}

\bibitem{raduta17}
A.~R. Raduta, F.~Gulminelli, M.~Oertel,
  \href{http://link.aps.org/doi/10.1103/PhysRevC.95.025805}{Stellar electron
  capture rates on neutron-rich nuclei and their impact on stellar core
  collapse}, Phys. Rev. C 95 (2017) 025805.
\newblock \href {http://dx.doi.org/10.1103/PhysRevC.95.025805}
  {\path{doi:10.1103/PhysRevC.95.025805}}.
\newline\urlprefix\url{http://link.aps.org/doi/10.1103/PhysRevC.95.025805}

\bibitem{dzhioev20}
A.~A. Dzhioev, K.~Langanke, G.~Mart{\'{\i}}nez-Pinedo, A.~I. Vdovin,
  C.~Stoyanov,
  \href{https://link.aps.org/doi/10.1103/PhysRevC.101.025805}{Unblocking of
  stellar electron capture for neutron-rich $n=50$ nuclei at finite
  temperature}, Phys. Rev. C 101 (2020) 025805.
\newblock \href {http://dx.doi.org/10.1103/PhysRevC.101.025805}
  {\path{doi:10.1103/PhysRevC.101.025805}}.
\newline\urlprefix\url{https://link.aps.org/doi/10.1103/PhysRevC.101.025805}

\bibitem{litvinova21}
E.~Litvinova, C.~Robin,
  \href{https://link.aps.org/doi/10.1103/PhysRevC.103.024326}{Impact of complex
  many-body correlations on electron capture in thermally excited nuclei around
  $^{78}\mathrm{Ni}$}, Phys. Rev. C 103 (2021) 024326.
\newblock \href {http://dx.doi.org/10.1103/PhysRevC.103.024326}
  {\path{doi:10.1103/PhysRevC.103.024326}}.
\newline\urlprefix\url{https://link.aps.org/doi/10.1103/PhysRevC.103.024326}

\bibitem{giraud22}
S.~Giraud, R.~G.~T. Zegers, B.~A. Brown, J.-M. Gabler, J.~Lesniak,
  J.~Rebenstock, E.~M. Ney, J.~Engel, A.~Ravli\ifmmode~\acute{c}\else
  \'{c}\fi{}, N.~Paar,
  \href{https://link.aps.org/doi/10.1103/PhysRevC.105.055801}{Finite-temperature
  electron-capture rates for neutron-rich nuclei near $n=50$ and effects on
  core-collapse supernova simulations}, Phys. Rev. C 105 (2022) 055801.
\newblock \href {http://dx.doi.org/10.1103/PhysRevC.105.055801}
  {\path{doi:10.1103/PhysRevC.105.055801}}.
\newline\urlprefix\url{https://link.aps.org/doi/10.1103/PhysRevC.105.055801}

\bibitem{bruenn85}
S.~W. {Bruenn}, {Stellar core collapse - Numerical model and infall epoch},
  \apjs 58 (1985) 771--841.
\newblock \href {http://dx.doi.org/10.1086/191056} {\path{doi:10.1086/191056}}.

\bibitem{fischer16}
T.~{Fischer}, G.~{Mart{\'{\i}}nez-Pinedo}, M.~{Hempel}, L.~{Huther},
  G.~{R{\"o}pke}, S.~{Typel}, A.~{Lohs}, {Expected impact from weak reactions
  with light nuclei in corecollapse supernova simulations}, in: European
  Physical Journal Web of Conferences, Vol. 109 of European Physical Journal
  Web of Conferences, 2016, p. 06002.
\newblock \href {http://arxiv.org/abs/1512.00193} {\path{arXiv:1512.00193}},
  \href {http://dx.doi.org/10.1051/epjconf/201610906002}
  {\path{doi:10.1051/epjconf/201610906002}}.

\bibitem{nakamura01}
S.~Nakamura, T.~Sato, V.~Gudkov, K.~Kubodera,
  \href{https://link.aps.org/doi/10.1103/PhysRevC.63.034617}{Neutrino reactions
  on the deuteron}, Phys. Rev. C 63 (2001) 034617.
\newblock \href {http://dx.doi.org/10.1103/PhysRevC.63.034617}
  {\path{doi:10.1103/PhysRevC.63.034617}}.
\newline\urlprefix\url{https://link.aps.org/doi/10.1103/PhysRevC.63.034617}

\bibitem{Haxton88}
W.~C. {Haxton}, {Neutrino heating in supernovae}, Physical Review Letters 60
  (1988) 1999--2002.
\newblock \href {http://dx.doi.org/10.1103/PhysRevLett.60.1999}
  {\path{doi:10.1103/PhysRevLett.60.1999}}.

\bibitem{langanke08}
K.~{Langanke}, G.~{Mart{\'{\i}}nez-Pinedo}, B.~{M{\"u}ller}, H.-T. {Janka},
  A.~{Marek}, W.~R. {Hix}, A.~{Juodagalvis}, J.~M. {Sampaio}, {Effects of
  Inelastic Neutrino-Nucleus Scattering on Supernova Dynamics and Radiated
  Neutrino Spectra}, Physical Review Letters 100~(1) (2008) 011101.
\newblock \href {http://arxiv.org/abs/0706.1687} {\path{arXiv:0706.1687}},
  \href {http://dx.doi.org/10.1103/PhysRevLett.100.011101}
  {\path{doi:10.1103/PhysRevLett.100.011101}}.

\bibitem{juodagalvis05}
A.~{Juodagalvis}, K.~{Langanke}, G.~{Mart{\'\i}nez-Pinedo}, W.~R. {Hix}, D.~J.
  {Dean}, J.~M. {Sampaio}, {Neutral-current neutrino-nucleus cross sections for
  A{\ensuremath{\sim}}50-65 nuclei}, Nuclear Physics A 747~(1) (2005) 87--108.
\newblock \href {http://arxiv.org/abs/nucl-th/0404078}
  {\path{arXiv:nucl-th/0404078}}, \href
  {http://dx.doi.org/10.1016/j.nuclphysa.2004.09.005}
  {\path{doi:10.1016/j.nuclphysa.2004.09.005}}.

\bibitem{furusawa13b}
S.~{Furusawa}, H.~{Nagakura}, K.~{Sumiyoshi}, S.~{Yamada}, {The Influence of
  Inelastic Neutrino Reactions with Light Nuclei on the Standing Accretion
  Shock Instability in Core-collapse Supernovae}, \apj 774 (2013) 78.
\newblock \href {http://arxiv.org/abs/1305.1510} {\path{arXiv:1305.1510}},
  \href {http://dx.doi.org/10.1088/0004-637X/774/1/78}
  {\path{doi:10.1088/0004-637X/774/1/78}}.

\bibitem{ohnishi07}
N.~{Ohnishi}, K.~{Kotake}, S.~{Yamada}, {Inelastic Neutrino-Helium Scatterings
  and Standing Accretion Shock Instability in Core-Collapse Supernovae}, \apj
  667 (2007) 375--381.
\newblock \href {http://arxiv.org/abs/astro-ph/0606187}
  {\path{arXiv:astro-ph/0606187}}, \href {http://dx.doi.org/10.1086/520755}
  {\path{doi:10.1086/520755}}.

\bibitem{dutra12}
M.~Dutra, O.~Lourenco, J.~S. Sa~Martins, A.~Delfino, J.~R. Stone, P.~D.
  Stevenson, {Skyrme Interaction and Nuclear Matter Constraints}, Phys. Rev. C
  85 (2012) 035201.
\newblock \href {http://arxiv.org/abs/1202.3902} {\path{arXiv:1202.3902}},
  \href {http://dx.doi.org/10.1103/PhysRevC.85.035201}
  {\path{doi:10.1103/PhysRevC.85.035201}}.

\bibitem{chabant98}
E.~Chabanat, P.~Bonche, P.~Haensel, J.~Meyer, R.~Schaeffer,
  \href{https://www.sciencedirect.com/science/article/pii/S0375947498001808}{A
  skyrme parametrization from subnuclear to neutron star densities part ii.
  nuclei far from stabilities}, Nuclear Physics A 635~(1) (1998) 231--256.
\newblock \href
  {http://dx.doi.org/https://doi.org/10.1016/S0375-9474(98)00180-8}
  {\path{doi:https://doi.org/10.1016/S0375-9474(98)00180-8}}.
\newline\urlprefix\url{https://www.sciencedirect.com/science/article/pii/S0375947498001808}

\bibitem{akmal98}
A.~Akmal, V.~R. Pandharipande, D.~G. Ravenhall,
  \href{https://link.aps.org/doi/10.1103/PhysRevC.58.1804}{Equation of state of
  nucleon matter and neutron star structure}, Phys. Rev. C 58 (1998)
  1804--1828.
\newblock \href {http://dx.doi.org/10.1103/PhysRevC.58.1804}
  {\path{doi:10.1103/PhysRevC.58.1804}}.
\newline\urlprefix\url{https://link.aps.org/doi/10.1103/PhysRevC.58.1804}

\bibitem{sugahara94}
Y.~Sugahara, H.~Toki, {Relativistic mean field theory for unstable nuclei with
  nonlinear sigma and omega terms}, Nucl.Phys. A579 (1994) 557--572.
\newblock \href {http://dx.doi.org/10.1016/0375-9474(94)90923-7}
  {\path{doi:10.1016/0375-9474(94)90923-7}}.

\bibitem{dalen10}
E.~Dalen, H.~Ml{{\"u}}ther, Relativistic effects in nuclear matter and nuclei,
  International Journal of Modern Physics E-nuclear Physics - IJMPE 19.
\newblock \href {http://dx.doi.org/10.1142/S0218301310016533}
  {\path{doi:10.1142/S0218301310016533}}.

\bibitem{li18}
B.-A. Li, B.-J. Cai, L.-W. Chen, J.~Xu,
  \href{https://www.sciencedirect.com/science/article/pii/S0146641018300012}{Nucleon
  effective masses in neutron-rich matter}, Progress in Particle and Nuclear
  Physics 99 (2018) 29--119.
\newblock \href {http://dx.doi.org/https://doi.org/10.1016/j.ppnp.2018.01.001}
  {\path{doi:https://doi.org/10.1016/j.ppnp.2018.01.001}}.
\newline\urlprefix\url{https://www.sciencedirect.com/science/article/pii/S0146641018300012}

\bibitem{shen20}
H.~Shen, F.~Ji, J.~Hu, K.~Sumiyoshi,
  \href{https://doi.org/10.3847/1538-4357/ab72fd}{Effects of symmetry energy on
  the equation of state for simulations of core-collapse supernovae and
  neutron-star mergers}, The Astrophysical Journal 891~(2) (2020) 148.
\newblock \href {http://dx.doi.org/10.3847/1538-4357/ab72fd}
  {\path{doi:10.3847/1538-4357/ab72fd}}.
\newline\urlprefix\url{https://doi.org/10.3847/1538-4357/ab72fd}

\bibitem{wiringa95}
R.~B. Wiringa, V.~G.~J. Stoks, R.~Schiavilla,
  \href{http://link.aps.org/doi/10.1103/PhysRevC.51.38}{Accurate
  nucleon-nucleon potential with charge-independence breaking}, Phys. Rev. C 51
  (1995) 38--51.
\newblock \href {http://dx.doi.org/10.1103/PhysRevC.51.38}
  {\path{doi:10.1103/PhysRevC.51.38}}.
\newline\urlprefix\url{http://link.aps.org/doi/10.1103/PhysRevC.51.38}

\bibitem{carlson83}
J.~{Carlson}, V.~R. {Pandharipande}, R.~B. {Wiringa}, {Three-nucleon
  interaction in 3-, 4- and {$\infty$}-body systems}, Nuclear Physics A 401
  (1983) 59--85.
\newblock \href {http://dx.doi.org/10.1016/0375-9474(83)90336-6}
  {\path{doi:10.1016/0375-9474(83)90336-6}}.

\bibitem{pudliner95}
B.~S. Pudliner, V.~R. Pandharipande, J.~Carlson, R.~B. Wiringa,
  \href{http://link.aps.org/doi/10.1103/PhysRevLett.74.4396}{Quantum monte
  carlo calculations of $\mathit{A }\ensuremath{\le}6$ nuclei}, Phys. Rev.
  Lett. 74 (1995) 4396--4399.
\newblock \href {http://dx.doi.org/10.1103/PhysRevLett.74.4396}
  {\path{doi:10.1103/PhysRevLett.74.4396}}.
\newline\urlprefix\url{http://link.aps.org/doi/10.1103/PhysRevLett.74.4396}

\bibitem{togashi13}
H.~{Togashi}, M.~{Takano}, {Variational study for the equation of state of
  asymmetric nuclear matter at finite temperatures}, Nuclear Physics A 902
  (2013) 53--73.
\newblock \href {http://arxiv.org/abs/1302.4261} {\path{arXiv:1302.4261}},
  \href {http://dx.doi.org/10.1016/j.nuclphysa.2013.02.014}
  {\path{doi:10.1016/j.nuclphysa.2013.02.014}}.

\bibitem{kanzawa07}
H.~{Kanzawa}, K.~{Oyamatsu}, K.~{Sumiyoshi}, M.~{Takano}, {Variational
  calculation for the equation of state of nuclear matter at finite
  temperatures}, Nuclear Physics A 791 (2007) 232--250.
\newblock \href {http://arxiv.org/abs/nucl-th/0701069}
  {\path{arXiv:nucl-th/0701069}}, \href
  {http://dx.doi.org/10.1016/j.nuclphysa.2007.01.098}
  {\path{doi:10.1016/j.nuclphysa.2007.01.098}}.

\bibitem{kanzawa09}
H.~{Kanzawa}, M.~{Takano}, K.~{Oyamatsu}, K.~{Sumiyoshi}, Prog. Theor. Phys.
  122 (2009) 673.

\bibitem{schmidt79}
K.~E. {Schmidt}, V.~R. {Pandharipande}, {Variational theory of nuclear matter
  at finite temperatures}, Physics Letters B 87 (1979) 11--14.
\newblock \href {http://dx.doi.org/10.1016/0370-2693(79)90004-2}
  {\path{doi:10.1016/0370-2693(79)90004-2}}.

\bibitem{mukherjee07}
A.~{Mukherjee}, V.~R. {Pandharipande}, {Variational theory of hot nucleon
  matter}, \prc 75~(3) (2007) 035802.
\newblock \href {http://arxiv.org/abs/nucl-th/0609058}
  {\path{arXiv:nucl-th/0609058}}, \href
  {http://dx.doi.org/10.1103/PhysRevC.75.035802}
  {\path{doi:10.1103/PhysRevC.75.035802}}.

\bibitem{mukherjee09}
A.~Mukherjee,
  \href{http://link.aps.org/doi/10.1103/PhysRevC.79.045811}{Variational theory
  of hot nucleon matter. ii. spin-isospin correlations and equation of state of
  nuclear and neutron matter}, Phys. Rev. C 79 (2009) 045811.
\newblock \href {http://dx.doi.org/10.1103/PhysRevC.79.045811}
  {\path{doi:10.1103/PhysRevC.79.045811}}.
\newline\urlprefix\url{http://link.aps.org/doi/10.1103/PhysRevC.79.045811}

\bibitem{katayama13}
T.~{Katayama}, K.~{Saito}, {Properties of dense, asymmetric nuclear matter in
  Dirac-Brueckner-Hartree-Fock approach}, \prc 88~(3) (2013) 035805.
\newblock \href {http://arxiv.org/abs/1307.2067} {\path{arXiv:1307.2067}},
  \href {http://dx.doi.org/10.1103/PhysRevC.88.035805}
  {\path{doi:10.1103/PhysRevC.88.035805}}.

\bibitem{brockmann90}
R.~Brockmann, R.~Machleidt,
  \href{https://link.aps.org/doi/10.1103/PhysRevC.42.1965}{Relativistic nuclear
  structure. i. nuclear matter}, Phys. Rev. C 42 (1990) 1965--1980.
\newblock \href {http://dx.doi.org/10.1103/PhysRevC.42.1965}
  {\path{doi:10.1103/PhysRevC.42.1965}}.
\newline\urlprefix\url{https://link.aps.org/doi/10.1103/PhysRevC.42.1965}

\bibitem{gross99}
T.~Gross-Boelting, C.~Fuchs, A.~Faessler,
  \href{http://www.sciencedirect.com/science/article/pii/S0375947499000226}{Covariant
  representations of the relativistic brueckner t-matrix and the nuclear matter
  problem}, Nuclear Physics A 648~(1) (1999) 105 -- 137.
\newblock \href
  {http://dx.doi.org/https://doi.org/10.1016/S0375-9474(99)00022-6}
  {\path{doi:https://doi.org/10.1016/S0375-9474(99)00022-6}}.
\newline\urlprefix\url{http://www.sciencedirect.com/science/article/pii/S0375947499000226}

\bibitem{horowitz87}
C.~Horowitz, B.~D. Serot,
  \href{http://www.sciencedirect.com/science/article/pii/0375947487903708}{The
  relativistic two-nucleon problem in nuclear matter}, Nuclear Physics A
  464~(4) (1987) 613 -- 699.
\newblock \href
  {http://dx.doi.org/https://doi.org/10.1016/0375-9474(87)90370-8}
  {\path{doi:https://doi.org/10.1016/0375-9474(87)90370-8}}.
\newline\urlprefix\url{http://www.sciencedirect.com/science/article/pii/0375947487903708}

\bibitem{sumiyoshi94}
K.~{Sumiyoshi}, H.~{Toki}, {Relativistic equation of state of nuclear matter
  for the supernova explosion and the birth of neutron stars}, \apj 422 (1994)
  700--718.
\newblock \href {http://dx.doi.org/10.1086/173763} {\path{doi:10.1086/173763}}.

\bibitem{aoki12}
S.~Aoki, T.~Doi, T.~Hatsuda, Y.~Ikeda, T.~Inoue, N.~Ishii, K.~Murano,
  H.~Nemura, K.~Sasaki, H.~Q. Collaboration),
  \href{https://doi.org/10.1093/ptep/pts010}{{Lattice quantum chromodynamical
  approach to nuclear physics}}, Progress of Theoretical and Experimental
  Physics 2012~(1), 01A105.
\newblock \href
  {http://arxiv.org/abs/https://academic.oup.com/ptep/article-pdf/2012/1/01A105/11584959/pts010.pdf}
  {\path{arXiv:https://academic.oup.com/ptep/article-pdf/2012/1/01A105/11584959/pts010.pdf}},
  \href {http://dx.doi.org/10.1093/ptep/pts010}
  {\path{doi:10.1093/ptep/pts010}}.
\newline\urlprefix\url{https://doi.org/10.1093/ptep/pts010}

\bibitem{holt17}
J.~W. Holt, N.~Kaiser,
  \href{https://link.aps.org/doi/10.1103/PhysRevC.95.034326}{Equation of state
  of nuclear and neutron matter at third-order in perturbation theory from
  chiral effective field theory}, Phys. Rev. C 95 (2017) 034326.
\newblock \href {http://dx.doi.org/10.1103/PhysRevC.95.034326}
  {\path{doi:10.1103/PhysRevC.95.034326}}.
\newline\urlprefix\url{https://link.aps.org/doi/10.1103/PhysRevC.95.034326}

\bibitem{hebeler10}
K.~Hebeler, J.~M. Lattimer, C.~J. Pethick, A.~Schwenk,
  \href{https://link.aps.org/doi/10.1103/PhysRevLett.105.161102}{Constraints on
  neutron star radii based on chiral effective field theory interactions},
  Phys. Rev. Lett. 105 (2010) 161102.
\newblock \href {http://dx.doi.org/10.1103/PhysRevLett.105.161102}
  {\path{doi:10.1103/PhysRevLett.105.161102}}.
\newline\urlprefix\url{https://link.aps.org/doi/10.1103/PhysRevLett.105.161102}

\bibitem{hebeler14}
K.~Hebeler, A.~Schwenk,
  \href{http://dx.doi.org/10.1140/epja/i2014-14011-4}{Symmetry energy, neutron
  skin, and neutron star radius from chiral effective field theory
  interactions}, The European Physical Journal A 50~(2) (2014) 1--7.
\newblock \href {http://dx.doi.org/10.1140/epja/i2014-14011-4}
  {\path{doi:10.1140/epja/i2014-14011-4}}.
\newline\urlprefix\url{http://dx.doi.org/10.1140/epja/i2014-14011-4}

\bibitem{drischler21}
C.~Drischler, J.~Holt, C.~Wellenhofer,
  \href{https://doi.org/10.1146/annurev-nucl-102419-041903}{Chiral effective
  field theory and the high-density nuclear equation of state}, Annual Review
  of Nuclear and Particle Science 71~(1) (2021) 403--432.
\newblock \href
  {http://arxiv.org/abs/https://doi.org/10.1146/annurev-nucl-102419-041903}
  {\path{arXiv:https://doi.org/10.1146/annurev-nucl-102419-041903}}, \href
  {http://dx.doi.org/10.1146/annurev-nucl-102419-041903}
  {\path{doi:10.1146/annurev-nucl-102419-041903}}.
\newline\urlprefix\url{https://doi.org/10.1146/annurev-nucl-102419-041903}

\bibitem{vries87}
H.~{De Vries}, C.~{De Jager}, C.~{De Vries},
  \href{https://www.sciencedirect.com/science/article/pii/0092640X87900131}{Nuclear
  charge-density-distribution parameters from elastic electron scattering},
  Atomic Data and Nuclear Data Tables 36~(3) (1987) 495--536.
\newblock \href
  {http://dx.doi.org/https://doi.org/10.1016/0092-640X(87)90013-1}
  {\path{doi:https://doi.org/10.1016/0092-640X(87)90013-1}}.
\newline\urlprefix\url{https://www.sciencedirect.com/science/article/pii/0092640X87900131}

\bibitem{audi03}
G.~{Audi}, A.~H. {Wapstra}, C.~{Thibault}, {The AME2003 atomic mass evaluation
  . (II). Tables, graphs and references}, Nuclear Physics A 729 (2003)
  337--676.
\newblock \href {http://dx.doi.org/10.1016/j.nuclphysa.2003.11.003}
  {\path{doi:10.1016/j.nuclphysa.2003.11.003}}.

\bibitem{audi12}
G.~Audi, M.~Wang, A.~Wapstra, F.~Kondev, M.~MacCormick, X.~Xu, B.~Pfeiffer,
  \href{http://stacks.iop.org/1674-1137/36/i=12/a=002}{The ame2012 atomic mass
  evaluation}, Chinese Physics C 36~(12) (2012) 1287.
\newblock \href {http://dx.doi.org/10.1088/1674-1137/36/12/002}
  {\path{doi:10.1088/1674-1137/36/12/002}}.
\newline\urlprefix\url{http://stacks.iop.org/1674-1137/36/i=12/a=002}

\bibitem{wang14}
N.~Wang, M.~Liu, X.~Wu, J.~Meng,
  \href{http://www.sciencedirect.com/science/article/pii/S037026931400358X}{Surface
  diffuseness correction in global mass formula}, Physics Letters B 734 (2014)
  215 -- 219.
\newblock \href
  {http://dx.doi.org/https://doi.org/10.1016/j.physletb.2014.05.049}
  {\path{doi:https://doi.org/10.1016/j.physletb.2014.05.049}}.
\newline\urlprefix\url{http://www.sciencedirect.com/science/article/pii/S037026931400358X}

\bibitem{colo04}
G.~Col\`o, N.~Van~Giai, J.~Meyer, K.~Bennaceur, P.~Bonche,
  \href{https://link.aps.org/doi/10.1103/PhysRevC.70.024307}{Microscopic
  determination of the nuclear incompressibility within the nonrelativistic
  framework}, Phys. Rev. C 70 (2004) 024307.
\newblock \href {http://dx.doi.org/10.1103/PhysRevC.70.024307}
  {\path{doi:10.1103/PhysRevC.70.024307}}.
\newline\urlprefix\url{https://link.aps.org/doi/10.1103/PhysRevC.70.024307}

\bibitem{piekarewicz04}
J.~Piekarewicz,
  \href{https://link.aps.org/doi/10.1103/PhysRevC.69.041301}{Unmasking the
  nuclear matter equation of state}, Phys. Rev. C 69 (2004) 041301.
\newblock \href {http://dx.doi.org/10.1103/PhysRevC.69.041301}
  {\path{doi:10.1103/PhysRevC.69.041301}}.
\newline\urlprefix\url{https://link.aps.org/doi/10.1103/PhysRevC.69.041301}

\bibitem{burgio21}
G.~Burgio, H.-J. Schulze, I.~Vidana, J.-B. Wei,
  \href{https://www.sciencedirect.com/science/article/pii/S0146641021000338}{Neutron
  stars and the nuclear equation of state}, Progress in Particle and Nuclear
  Physics 120 (2021) 103879.
\newblock \href {http://dx.doi.org/https://doi.org/10.1016/j.ppnp.2021.103879}
  {\path{doi:https://doi.org/10.1016/j.ppnp.2021.103879}}.
\newline\urlprefix\url{https://www.sciencedirect.com/science/article/pii/S0146641021000338}

\bibitem{lattimer13}
J.~M. {Lattimer}, Y.~{Lim}, {Constraining the Symmetry Parameters of the
  Nuclear Interaction}, \apj 771 (2013) 51.
\newblock \href {http://arxiv.org/abs/1203.4286} {\path{arXiv:1203.4286}},
  \href {http://dx.doi.org/10.1088/0004-637X/771/1/51}
  {\path{doi:10.1088/0004-637X/771/1/51}}.

\bibitem{danielewicz14}
P.~Danielewicz, J.~Lee,
  \href{http://www.sciencedirect.com/science/article/pii/S0375947413007872}{Symmetry
  energy ii: Isobaric analog states}, Nuclear Physics A 922 (2014) 1 -- 70.
\newblock \href
  {http://dx.doi.org/http://dx.doi.org/10.1016/j.nuclphysa.2013.11.005}
  {\path{doi:http://dx.doi.org/10.1016/j.nuclphysa.2013.11.005}}.
\newline\urlprefix\url{http://www.sciencedirect.com/science/article/pii/S0375947413007872}

\bibitem{zenihiro21}
J.~Zenihiro, T.~Uesaka, H.~Sagawa, S.~Yoshida,
  \href{https://doi.org/10.1093/ptep/ptab001}{{Proton density polarization of
  the doubly magic 40Ca core in 48Ca and EoS parameters}}, Progress of
  Theoretical and Experimental Physics 2021~(2), 023D05.
\newblock \href
  {http://arxiv.org/abs/https://academic.oup.com/ptep/article-pdf/2021/2/023D05/36644942/ptab001.pdf}
  {\path{arXiv:https://academic.oup.com/ptep/article-pdf/2021/2/023D05/36644942/ptab001.pdf}},
  \href {http://dx.doi.org/10.1093/ptep/ptab001}
  {\path{doi:10.1093/ptep/ptab001}}.
\newline\urlprefix\url{https://doi.org/10.1093/ptep/ptab001}

\bibitem{estee21}
J.~Estee, W.~G. Lynch, C.~Y. Tsang, J.~Barney, G.~Jhang, M.~B. Tsang, R.~Wang,
  M.~Kaneko, J.~W. Lee, T.~Isobe, M.~Kurata-Nishimura, T.~Murakami, D.~S. Ahn,
  L.~Atar, T.~Aumann, H.~Baba, K.~Boretzky, J.~Brzychczyk, G.~Cerizza,
  N.~Chiga, N.~Fukuda, I.~Gasparic, B.~Hong, A.~Horvat, K.~Ieki, N.~Inabe,
  Y.~J. Kim, T.~Kobayashi, Y.~Kondo, P.~Lasko, H.~S. Lee, Y.~Leifels,
  J.~\L{}ukasik, J.~Manfredi, A.~B. McIntosh, P.~Morfouace, T.~Nakamura,
  N.~Nakatsuka, S.~Nishimura, H.~Otsu, P.~Paw\l{}owski, K.~Pelczar, D.~Rossi,
  H.~Sakurai, C.~Santamaria, H.~Sato, H.~Scheit, R.~Shane, Y.~Shimizu,
  H.~Simon, A.~Snoch, A.~Sochocka, T.~Sumikama, H.~Suzuki, D.~Suzuki,
  H.~Takeda, S.~Tangwancharoen, H.~Toernqvist, Y.~Togano, Z.~G. Xiao, S.~J.
  Yennello, Y.~Zhang, M.~D. Cozma,
  \href{https://link.aps.org/doi/10.1103/PhysRevLett.126.162701}{Probing the
  symmetry energy with the spectral pion ratio}, Phys. Rev. Lett. 126 (2021)
  162701.
\newblock \href {http://dx.doi.org/10.1103/PhysRevLett.126.162701}
  {\path{doi:10.1103/PhysRevLett.126.162701}}.
\newline\urlprefix\url{https://link.aps.org/doi/10.1103/PhysRevLett.126.162701}

\bibitem{reed21}
B.~T. Reed, F.~J. Fattoyev, C.~J. Horowitz, J.~Piekarewicz,
  \href{https://link.aps.org/doi/10.1103/PhysRevLett.126.172503}{Implications
  of prex-2 on the equation of state of neutron-rich matter}, Phys. Rev. Lett.
  126 (2021) 172503.
\newblock \href {http://dx.doi.org/10.1103/PhysRevLett.126.172503}
  {\path{doi:10.1103/PhysRevLett.126.172503}}.
\newline\urlprefix\url{https://link.aps.org/doi/10.1103/PhysRevLett.126.172503}

\bibitem{adhikari21}
D.~Adhikari, H.~Albataineh, D.~Androic, K.~Aniol, D.~S. Armstrong, T.~Averett,
  C.~Ayerbe~Gayoso, S.~Barcus, V.~Bellini, R.~S. Beminiwattha, J.~F. Benesch,
  H.~Bhatt, D.~Bhatta~Pathak, D.~Bhetuwal, B.~Blaikie, Q.~Campagna,
  A.~Camsonne, G.~D. Cates, Y.~Chen, C.~Clarke, J.~C. Cornejo, S.~Covrig~Dusa,
  P.~Datta, A.~Deshpande, D.~Dutta, C.~Feldman, E.~Fuchey, C.~Gal, D.~Gaskell,
  T.~Gautam, M.~Gericke, C.~Ghosh, I.~Halilovic, J.-O. Hansen, F.~Hauenstein,
  W.~Henry, C.~J. Horowitz, C.~Jantzi, S.~Jian, S.~Johnston, D.~C. Jones,
  B.~Karki, S.~Katugampola, C.~Keppel, P.~M. King, D.~E. King, M.~Knauss, K.~S.
  Kumar, T.~Kutz, N.~Lashley-Colthirst, G.~Leverick, H.~Liu, N.~Liyange,
  S.~Malace, R.~Mammei, J.~Mammei, M.~McCaughan, D.~McNulty, D.~Meekins,
  C.~Metts, R.~Michaels, M.~M. Mondal, J.~Napolitano, A.~Narayan, D.~Nikolaev,
  M.~N.~H. Rashad, V.~Owen, C.~Palatchi, J.~Pan, B.~Pandey, S.~Park, K.~D.
  Paschke, M.~Petrusky, M.~L. Pitt, S.~Premathilake, A.~J.~R. Puckett,
  B.~Quinn, R.~Radloff, S.~Rahman, A.~Rathnayake, B.~T. Reed, P.~E. Reimer,
  R.~Richards, S.~Riordan, Y.~Roblin, S.~Seeds, A.~Shahinyan, P.~Souder,
  L.~Tang, M.~Thiel, Y.~Tian, G.~M. Urciuoli, E.~W. Wertz, B.~Wojtsekhowski,
  B.~Yale, T.~Ye, A.~Yoon, A.~Zec, W.~Zhang, J.~Zhang, X.~Zheng,
  \href{https://link.aps.org/doi/10.1103/PhysRevLett.126.172502}{Accurate
  determination of the neutron skin thickness of $^{208}\mathrm{Pb}$ through
  parity-violation in electron scattering}, Phys. Rev. Lett. 126 (2021) 172502.
\newblock \href {http://dx.doi.org/10.1103/PhysRevLett.126.172502}
  {\path{doi:10.1103/PhysRevLett.126.172502}}.
\newline\urlprefix\url{https://link.aps.org/doi/10.1103/PhysRevLett.126.172502}

\bibitem{cromartie19}
H.~T. {Cromartie}, E.~{Fonseca}, S.~M. {Ransom}, P.~B. {Demorest},
  Z.~{Arzoumanian}, H.~{Blumer}, P.~R. {Brook}, M.~E. {DeCesar}, T.~{Dolch},
  J.~A. {Ellis}, R.~D. {Ferdman}, E.~C. {Ferrara}, N.~{Garver-Daniels}, P.~A.
  {Gentile}, M.~L. {Jones}, M.~T. {Lam}, D.~R. {Lorimer}, R.~S. {Lynch}, M.~A.
  {McLaughlin}, C.~{Ng}, D.~J. {Nice}, T.~T. {Pennucci}, R.~{Spiewak}, I.~H.
  {Stairs}, K.~{Stovall}, J.~K. {Swiggum}, W.~W. {Zhu}, {Relativistic Shapiro
  delay measurements of an extremely massive millisecond pulsar}, Nature
  Astronomy\href {http://dx.doi.org/10.1038/s41550-019-0880-2}
  {\path{doi:10.1038/s41550-019-0880-2}}.

\bibitem{abbott17}
B.~{Abbott et al.},
  \href{https://link.aps.org/doi/10.1103/PhysRevLett.119.161101}{Gw170817:
  Observation of gravitational waves from a binary neutron star inspiral},
  Phys. Rev. Lett. 119 (2017) 161101.
\newblock \href {http://dx.doi.org/10.1103/PhysRevLett.119.161101}
  {\path{doi:10.1103/PhysRevLett.119.161101}}.
\newline\urlprefix\url{https://link.aps.org/doi/10.1103/PhysRevLett.119.161101}

\bibitem{shibata17}
M.~{Shibata}, S.~{Fujibayashi}, K.~{Hotokezaka}, K.~{Kiuchi}, K.~{Kyutoku},
  Y.~{Sekiguchi}, M.~{Tanaka}, {Modeling GW170817 based on numerical relativity
  and its implications}, \prd 96~(12) (2017) 123012.
\newblock \href {http://arxiv.org/abs/1710.07579} {\path{arXiv:1710.07579}},
  \href {http://dx.doi.org/10.1103/PhysRevD.96.123012}
  {\path{doi:10.1103/PhysRevD.96.123012}}.

\bibitem{abbott18}
B.~{Abbott et al.},
  \href{https://link.aps.org/doi/10.1103/PhysRevLett.121.161101}{Gw170817:
  Measurements of neutron star radii and equation of state}, Phys. Rev. Lett.
  121 (2018) 161101.
\newblock \href {http://dx.doi.org/10.1103/PhysRevLett.121.161101}
  {\path{doi:10.1103/PhysRevLett.121.161101}}.
\newline\urlprefix\url{https://link.aps.org/doi/10.1103/PhysRevLett.121.161101}

\bibitem{annala18}
E.~Annala, T.~Gorda, A.~Kurkela, A.~Vuorinen,
  \href{https://link.aps.org/doi/10.1103/PhysRevLett.120.172703}{Gravitational-wave
  constraints on the neutron-star-matter equation of state}, Phys. Rev. Lett.
  120 (2018) 172703.
\newblock \href {http://dx.doi.org/10.1103/PhysRevLett.120.172703}
  {\path{doi:10.1103/PhysRevLett.120.172703}}.
\newline\urlprefix\url{https://link.aps.org/doi/10.1103/PhysRevLett.120.172703}

\bibitem{typel10}
S.~{Typel}, G.~{R{\"o}pke}, T.~{Kl{\"a}hn}, D.~{Blaschke}, H.~H. {Wolter},
  Composition and thermodynamics of nuclear matter with light clusters, \prc
  81~(1) (2010) 015803.
\newblock \href {http://arxiv.org/abs/0908.2344} {\path{arXiv:0908.2344}},
  \href {http://dx.doi.org/10.1103/PhysRevC.81.015803}
  {\path{doi:10.1103/PhysRevC.81.015803}}.

\bibitem{oyamatsu93}
K.~{Oyamatsu}, {Nuclear shapes in the inner crust of a neutron star}, Nuclear
  Physics A 561 (1993) 431--452.
\newblock \href {http://dx.doi.org/10.1016/0375-9474(93)90020-X}
  {\path{doi:10.1016/0375-9474(93)90020-X}}.

\bibitem{juodagalvis10}
A.~{Juodagalvis}, K.~{Langanke}, W.~R. {Hix}, G.~{Mart{\'{\i}}nez-Pinedo},
  J.~M. {Sampaio}, {Improved estimate of electron capture rates on nuclei
  during stellar core collapse}, Nuclear Physics A 848 (2010) 454--478.
\newblock \href {http://arxiv.org/abs/0909.0179} {\path{arXiv:0909.0179}},
  \href {http://dx.doi.org/10.1016/j.nuclphysa.2010.09.012}
  {\path{doi:10.1016/j.nuclphysa.2010.09.012}}.

\bibitem{furusawa20c}
S.~Furusawa, I.~Mishustin,
  \href{https://www.sciencedirect.com/science/article/pii/S0375947420303018}{Degeneracy
  effects and bose condensation in warm nuclear matter with light and heavy
  clusters}, Nuclear Physics A 1002 (2020) 121991.
\newblock \href
  {http://dx.doi.org/https://doi.org/10.1016/j.nuclphysa.2020.121991}
  {\path{doi:https://doi.org/10.1016/j.nuclphysa.2020.121991}}.
\newline\urlprefix\url{https://www.sciencedirect.com/science/article/pii/S0375947420303018}

\bibitem{huang21}
W.~Huang, M.~Wang, F.~Kondev, G.~Audi, S.~Naimi,
  \href{https://doi.org/10.1088/1674-1137/abddb0}{The {AME} 2020 atomic mass
  evaluation (i). evaluation of input data, and adjustment procedures{$\ast$}},
  Chinese Physics C 45~(3) (2021) 030002.
\newblock \href {http://dx.doi.org/10.1088/1674-1137/abddb0}
  {\path{doi:10.1088/1674-1137/abddb0}}.
\newline\urlprefix\url{https://doi.org/10.1088/1674-1137/abddb0}

\bibitem{wang21}
M.~Wang, W.~Huang, F.~Kondev, G.~Audi, S.~Naimi,
  \href{https://doi.org/10.1088/1674-1137/abddaf}{The {AME} 2020 atomic mass
  evaluation ({II}). tables, graphs and references{$\ast$}}, Chinese Physics C
  45~(3) (2021) 030003.
\newblock \href {http://dx.doi.org/10.1088/1674-1137/abddaf}
  {\path{doi:10.1088/1674-1137/abddaf}}.
\newline\urlprefix\url{https://doi.org/10.1088/1674-1137/abddaf}

\bibitem{koura05}
H.~{Koura}, T.~{Tachibana}, M.~{Uno}, M.~{Yamada}, {Nuclidic Mass Formula on a
  Spherical Basis with an Improved Even-Odd Term}, Progress of Theoretical
  Physics 113 (2005) 305--325.
\newblock \href {http://dx.doi.org/10.1143/PTP.113.305}
  {\path{doi:10.1143/PTP.113.305}}.

\bibitem{nishimura14}
S.~{Nishimura}, M.~{Takano}, {Shell effects in hot nuclei and their influence
  on nuclear composition in supernova matter}, Vol. 1594, 2014, pp. 239--244.
\newblock \href {http://dx.doi.org/10.1063/1.4874076}
  {\path{doi:10.1063/1.4874076}}.

\bibitem{ravenhall83}
D.~G. {Ravenhall}, C.~J. {Pethick}, J.~R. {Wilson}, {Structure of Matter below
  Nuclear Saturation Density}, Physical Review Letters 50 (1983) 2066--2069.
\newblock \href {http://dx.doi.org/10.1103/PhysRevLett.50.2066}
  {\path{doi:10.1103/PhysRevLett.50.2066}}.

\bibitem{furusawa18a}
S.~Furusawa, I.~Mishustin,
  \href{https://link.aps.org/doi/10.1103/PhysRevC.97.025804}{Equilibrium
  nuclear ensembles taking into account vaporization of hot nuclei in dense
  stellar matter}, Phys. Rev. C 97 (2018) 025804.
\newblock \href {http://dx.doi.org/10.1103/PhysRevC.97.025804}
  {\path{doi:10.1103/PhysRevC.97.025804}}.
\newline\urlprefix\url{https://link.aps.org/doi/10.1103/PhysRevC.97.025804}

\bibitem{furusawa20b}
S.~Furusawa,
  \href{https://doi.org/10.1088{$\%$}2F1402-4896{$\%$}2Fab8c15}{Nuclei in
  central engine of core-collapse supernovae}, Physica Scripta 95~(7) (2020)
  074002.
\newblock \href {http://dx.doi.org/10.1088/1402-4896/ab8c15}
  {\path{doi:10.1088/1402-4896/ab8c15}}.
\newline\urlprefix\url{https://doi.org/10.1088{$\%$}2F1402-4896{$\%$}2Fab8c15}

\bibitem{roepke09}
G.~{R{\"o}pke}, {Light nuclei quasiparticle energy shifts in hot and dense
  nuclear matter}, \prc 79~(1) (2009) 014002.
\newblock \href {http://arxiv.org/abs/0810.4645} {\path{arXiv:0810.4645}},
  \href {http://dx.doi.org/10.1103/PhysRevC.79.014002}
  {\path{doi:10.1103/PhysRevC.79.014002}}.

\bibitem{fai82}
G.~{F{\'a}i}, J.~{Randrup}, {Explosion-evaporation model for fragment
  production in medium-energy nuclear collisions}, Nuclear Physics A 381 (1982)
  557--576.
\newblock \href {http://dx.doi.org/10.1016/0375-9474(82)90376-1}
  {\path{doi:10.1016/0375-9474(82)90376-1}}.

\bibitem{egidy05}
T.~v. Egidy, D.~Bucurescu,
  \href{https://link.aps.org/doi/10.1103/PhysRevC.72.044311}{Systematics of
  nuclear level density parameters}, Phys. Rev. C 72 (2005) 044311.
\newblock \href {http://dx.doi.org/10.1103/PhysRevC.72.044311}
  {\path{doi:10.1103/PhysRevC.72.044311}}.
\newline\urlprefix\url{https://link.aps.org/doi/10.1103/PhysRevC.72.044311}

\bibitem{rauscher00}
T.~Rauscher, F.-K. Thielemann,
  \href{http://www.sciencedirect.com/science/article/pii/S0092640X00908349}{Astrophysical
  reaction rates from statistical model calculations}, Atomic Data and Nuclear
  Data Tables 75~(1) (2000) 1 -- 351.
\newblock \href {http://dx.doi.org/https://doi.org/10.1006/adnd.2000.0834}
  {\path{doi:https://doi.org/10.1006/adnd.2000.0834}}.
\newline\urlprefix\url{http://www.sciencedirect.com/science/article/pii/S0092640X00908349}

\bibitem{furusawa18b}
S.~Furusawa,
  \href{https://link.aps.org/doi/10.1103/PhysRevC.98.065802}{Sensitivity of
  nuclear statistical equilibrium to nuclear uncertainties during stellar core
  collapse}, Phys. Rev. C 98 (2018) 065802.
\newblock \href {http://dx.doi.org/10.1103/PhysRevC.98.065802}
  {\path{doi:10.1103/PhysRevC.98.065802}}.
\newline\urlprefix\url{https://link.aps.org/doi/10.1103/PhysRevC.98.065802}

\bibitem{pais16}
H.~{Pais}, S.~{Typel}, {Comparison of equation of state models with different
  cluster dissolution mechanisms}, ArXiv e-prints\href
  {http://arxiv.org/abs/1612.07022} {\path{arXiv:1612.07022}}.

\bibitem{furusawa17c}
S.~Furusawa, I.~Mishustin,
  \href{http://link.aps.org/doi/10.1103/PhysRevC.95.035802}{Self-consistent
  calculation of the nuclear composition in hot and dense stellar matter},
  Phys. Rev. C 95 (2017) 035802.
\newblock \href {http://dx.doi.org/10.1103/PhysRevC.95.035802}
  {\path{doi:10.1103/PhysRevC.95.035802}}.
\newline\urlprefix\url{http://link.aps.org/doi/10.1103/PhysRevC.95.035802}

\bibitem{schneider13}
A.~S. Schneider, C.~J. Horowitz, J.~Hughto, D.~K. Berry,
  \href{http://link.aps.org/doi/10.1103/PhysRevC.88.065807}{Nuclear ``pasta''
  formation}, Phys. Rev. C 88 (2013) 065807.
\newblock \href {http://dx.doi.org/10.1103/PhysRevC.88.065807}
  {\path{doi:10.1103/PhysRevC.88.065807}}.
\newline\urlprefix\url{http://link.aps.org/doi/10.1103/PhysRevC.88.065807}

\bibitem{pais14}
H.~Pais, W.~G. Newton, J.~R. Stone,
  \href{https://link.aps.org/doi/10.1103/PhysRevC.90.065802}{Phase transitions
  in core-collapse supernova matter at sub-saturation densities}, Phys. Rev. C
  90 (2014) 065802.
\newblock \href {http://dx.doi.org/10.1103/PhysRevC.90.065802}
  {\path{doi:10.1103/PhysRevC.90.065802}}.
\newline\urlprefix\url{https://link.aps.org/doi/10.1103/PhysRevC.90.065802}

\bibitem{gulminelli15}
F.~Gulminelli, A.~R. Raduta,
  \href{http://link.aps.org/doi/10.1103/PhysRevC.92.055803}{Unified treatment
  of subsaturation stellar matter at zero and finite temperature}, Phys. Rev. C
  92 (2015) 055803.
\newblock \href {http://dx.doi.org/10.1103/PhysRevC.92.055803}
  {\path{doi:10.1103/PhysRevC.92.055803}}.
\newline\urlprefix\url{http://link.aps.org/doi/10.1103/PhysRevC.92.055803}

\bibitem{buyukcizmeci13}
N.~{Buyukcizmeci}, A.~S. {Botvina}, I.~N. {Mishustin}, R.~{Ogul}, M.~{Hempel},
  J.~{Schaffner-Bielich}, F.-K. {Thielemann}, S.~{Furusawa}, K.~{Sumiyoshi},
  S.~{Yamada}, H.~{Suzuki}, {A comparative study of statistical models for
  nuclear equation of state of stellar matter}, Nuclear Physics A 907 (2013)
  13--54.
\newblock \href {http://arxiv.org/abs/1211.5990} {\path{arXiv:1211.5990}},
  \href {http://dx.doi.org/10.1016/j.nuclphysa.2013.03.010}
  {\path{doi:10.1016/j.nuclphysa.2013.03.010}}.

\bibitem{hempel15}
M.~Hempel, K.~Hagel, J.~Natowitz, G.~R\"opke, S.~Typel,
  \href{https://link.aps.org/doi/10.1103/PhysRevC.91.045805}{Constraining
  supernova equations of state with equilibrium constants from heavy-ion
  collisions}, Phys. Rev. C 91 (2015) 045805.
\newblock \href {http://dx.doi.org/10.1103/PhysRevC.91.045805}
  {\path{doi:10.1103/PhysRevC.91.045805}}.
\newline\urlprefix\url{https://link.aps.org/doi/10.1103/PhysRevC.91.045805}

\bibitem{hempel10d}
M.~{Hempel}, {Hot and Dense Matter in Compact Stars: From Nuclei to Quarks },
  Ph.D. thesis (2010).

\bibitem{nasu15}
S.~{Nasu}, S.~X. {Nakamura}, K.~{Sumiyoshi}, T.~{Sato}, F.~{Myhrer},
  K.~{Kubodera}, {Neutrino Emissivities from Deuteron Breakup and Formation in
  Supernovae}, \apj 801 (2015) 78.
\newblock \href {http://arxiv.org/abs/1402.0959} {\path{arXiv:1402.0959}},
  \href {http://dx.doi.org/10.1088/0004-637X/801/2/78}
  {\path{doi:10.1088/0004-637X/801/2/78}}.

\bibitem{fischer13}
T.~Fischer, M.~Hempel, I.~Sagert, Y.~Suwa, J.~Schaffner-Bielich, {Symmetry
  energy impact in simulations of core-collapse supernovae}, Eur. Phys. J. A50
  (2014) 46.
\newblock \href {http://arxiv.org/abs/1307.6190} {\path{arXiv:1307.6190}},
  \href {http://dx.doi.org/10.1140/epja/i2014-14046-5}
  {\path{doi:10.1140/epja/i2014-14046-5}}.

\bibitem{fischer17}
T.~Fischer, N.-U. Bastian, D.~Blaschke, M.~Cierniak, M.~Hempel, T.~Kl{{\"a}}hn,
  G.~Mart{\'{\i}}nez-Pinedo, W.~G. Newton, G.~Rl{{\"o}}pke, S.~Typel, et~al.,
  The state of matter in simulations of core-collapse supernovae―reflections
  and recent developments, Publications of the Astronomical Society of
  Australia 34 (2017) e067.
\newblock \href {http://dx.doi.org/10.1017/pasa.2017.63}
  {\path{doi:10.1017/pasa.2017.63}}.

\bibitem{suwa13}
Y.~{Suwa}, T.~{Takiwaki}, K.~{Kotake}, T.~{Fischer}, M.~{Liebendorfer},
  K.~{Sato}, {On the Importance of the Equation of State for the
  Neutrino-driven Supernova Explosion Mechanism}, \apj 764 (2013) 99.
\newblock \href {http://arxiv.org/abs/1206.6101} {\path{arXiv:1206.6101}},
  \href {http://dx.doi.org/10.1088/0004-637X/764/1/99}
  {\path{doi:10.1088/0004-637X/764/1/99}}.

\bibitem{nagakura18a}
H.~{Nagakura}, W.~{Iwakami}, S.~{Furusawa}, H.~{Okawa}, A.~{Harada},
  K.~{Sumiyoshi}, S.~{Yamada}, H.~{Matsufuru}, A.~{Imakura}, {Simulations of
  Core-collapse Supernovae in Spatial Axisymmetry with Full Boltzmann Neutrino
  Transport}, \apj 854 (2018) 136.
\newblock \href {http://arxiv.org/abs/1702.01752} {\path{arXiv:1702.01752}},
  \href {http://dx.doi.org/10.3847/1538-4357/aaac29}
  {\path{doi:10.3847/1538-4357/aaac29}}.

\bibitem{harada20}
A.~Harada, H.~Nagakura, W.~Iwakami, H.~Okawa, S.~Furusawa, K.~Sumiyoshi,
  H.~Matsufuru, S.~Yamada, The boltzmann-radiation-hydrodynamics simulations of
  the core-collapse supernova with the different equations of state: the role
  of nuclear composition and the behavior of neutrinos (2020).
\newblock \href {http://arxiv.org/abs/2003.08630} {\path{arXiv:2003.08630}}.

\bibitem{kuroda17}
T.~Kuroda, K.~Kotake, K.~Hayama, T.~Takiwaki,
  \href{https://doi.org/10.3847/1538-4357/aa988d}{Correlated signatures of
  gravitational-wave and neutrino emission in three-dimensional
  general-relativistic core-collapse supernova simulations}, The Astrophysical
  Journal 851~(1) (2017) 62.
\newblock \href {http://dx.doi.org/10.3847/1538-4357/aa988d}
  {\path{doi:10.3847/1538-4357/aa988d}}.
\newline\urlprefix\url{https://doi.org/10.3847/1538-4357/aa988d}

\bibitem{andersen21}
O.~{Eggenberger Andersen}, S.~{Zha}, A.~{da Silva Schneider}, A.~{Betranhandy},
  S.~M. {Couch}, E.~P. {O'Connor}, {Equation-of-state Dependence of
  Gravitational Waves in Core-collapse Supernovae}, \apj 923~(2) (2021) 201.
\newblock \href {http://arxiv.org/abs/2106.09734} {\path{arXiv:2106.09734}},
  \href {http://dx.doi.org/10.3847/1538-4357/ac294c}
  {\path{doi:10.3847/1538-4357/ac294c}}.

\bibitem{boccioli21}
L.~{Boccioli}, G.~J. {Mathews}, I.-S. {Suh}, E.~P. {O'Connor}, {Effect of the
  Nuclear Equation of State on Relativistic-Turbulence Induced Core-Collapse
  Supernovae}, arXiv e-prints (2021) arXiv:2110.05544\href
  {http://arxiv.org/abs/2110.05544} {\path{arXiv:2110.05544}}.

\bibitem{nakazato22}
K.~Nakazato, F.~Nakanishi, M.~Harada, Y.~Koshio, Y.~Suwa, K.~Sumiyoshi,
  A.~Harada, M.~Mori, R.~A. Wendell,
  \href{https://doi.org/10.3847/1538-4357/ac3ae2}{Observing supernova neutrino
  light curves with super-kamiokande. {II}. impact of the nuclear equation of
  state}, The Astrophysical Journal 925~(1) (2022) 98.
\newblock \href {http://dx.doi.org/10.3847/1538-4357/ac3ae2}
  {\path{doi:10.3847/1538-4357/ac3ae2}}.
\newline\urlprefix\url{https://doi.org/10.3847/1538-4357/ac3ae2}

\bibitem{sumiyoshi06}
K.~Sumiyoshi, S.~Yamada, H.~Suzuki, S.~Chiba,
  \href{https://link.aps.org/doi/10.1103/PhysRevLett.97.091101}{Neutrino
  signals from the formation of a black hole: A probe of the equation of state
  of dense matter}, Phys. Rev. Lett. 97 (2006) 091101.
\newblock \href {http://dx.doi.org/10.1103/PhysRevLett.97.091101}
  {\path{doi:10.1103/PhysRevLett.97.091101}}.
\newline\urlprefix\url{https://link.aps.org/doi/10.1103/PhysRevLett.97.091101}

\bibitem{schneider20}
A.~{da Silva Schneider}, E.~{O'Connor}, E.~{Granqvist}, A.~{Betranhandy}, S.~M.
  {Couch}, {Equation of State and Progenitor Dependence of Stellar-mass Black
  Hole Formation}, \apj 894~(1) (2020) 4.
\newblock \href {http://arxiv.org/abs/2001.10434} {\path{arXiv:2001.10434}},
  \href {http://dx.doi.org/10.3847/1538-4357/ab8308}
  {\path{doi:10.3847/1538-4357/ab8308}}.

\bibitem{pan18}
K.-C. {Pan}, M.~{Liebend{\"o}rfer}, S.~M. {Couch}, F.-K. {Thielemann},
  {Equation of State Dependent Dynamics and Multi-messenger Signals from
  Stellar-mass Black Hole Formation}, \apj 857~(1) (2018) 13.
\newblock \href {http://arxiv.org/abs/1710.01690} {\path{arXiv:1710.01690}},
  \href {http://dx.doi.org/10.3847/1538-4357/aab71d}
  {\path{doi:10.3847/1538-4357/aab71d}}.

\bibitem{hempel12}
M.~Hempel, T.~Fischer, J.~Schaffner-Bielich, M.~Liebendorfer,
  \href{https://doi.org/10.1088/0004-637x/748/1/70}{{NEW} {EQUATIONS} {OF}
  {STATE} {IN} {SIMULATIONS} {OF} {CORE}-{COLLAPSE} {SUPERNOVAE}}, The
  Astrophysical Journal 748~(1) (2012) 70.
\newblock \href {http://dx.doi.org/10.1088/0004-637x/748/1/70}
  {\path{doi:10.1088/0004-637x/748/1/70}}.
\newline\urlprefix\url{https://doi.org/10.1088/0004-637x/748/1/70}

\bibitem{lattimer00}
J.~M. Lattimer, M.~Prakash,
  \href{https://www.sciencedirect.com/science/article/pii/S0370157300000193}{Nuclear
  matter and its role in supernovae, neutron stars and compact object binary
  mergers}, Physics Reports 333-334 (2000) 121--146.
\newblock \href
  {http://dx.doi.org/https://doi.org/10.1016/S0370-1573(00)00019-3}
  {\path{doi:https://doi.org/10.1016/S0370-1573(00)00019-3}}.
\newline\urlprefix\url{https://www.sciencedirect.com/science/article/pii/S0370157300000193}

\bibitem{kuroda22}
T.~{Kuroda}, T.~{Fischer}, T.~{Takiwaki}, K.~{Kotake}, {Core-collapse Supernova
  Simulations and the Formation of Neutron Stars, Hybrid Stars, and Black
  Holes}, \apj 924~(1) (2022) 38.
\newblock \href {http://arxiv.org/abs/2109.01508} {\path{arXiv:2109.01508}},
  \href {http://dx.doi.org/10.3847/1538-4357/ac31a8}
  {\path{doi:10.3847/1538-4357/ac31a8}}.

\bibitem{nakazato08}
K.~Nakazato, K.~Sumiyoshi, S.~Yamada,
  \href{https://link.aps.org/doi/10.1103/PhysRevD.77.103006}{Astrophysical
  implications of equation of state for hadron-quark mixed phase: Compact stars
  and stellar collapses}, Phys. Rev. D 77 (2008) 103006.
\newblock \href {http://dx.doi.org/10.1103/PhysRevD.77.103006}
  {\path{doi:10.1103/PhysRevD.77.103006}}.
\newline\urlprefix\url{https://link.aps.org/doi/10.1103/PhysRevD.77.103006}

\bibitem{nakazato10}
K.~Nakazato, K.~Sumiyoshi, S.~Yamada,
  \href{https://doi.org/10.1088/0004-637x/721/2/1284}{{IMPACT} {OF} {QUARKS}
  {AND} {PIONS} {ON} {DYNAMICS} {AND} {NEUTRINO} {SIGNAL} {OF} {BLACK} {HOLE}
  {FORMATION} {IN} {NON}-{ROTATING} {STELLAR} {CORE} {COLLAPSE}}, The
  Astrophysical Journal 721~(2) (2010) 1284--1294.
\newblock \href {http://dx.doi.org/10.1088/0004-637x/721/2/1284}
  {\path{doi:10.1088/0004-637x/721/2/1284}}.
\newline\urlprefix\url{https://doi.org/10.1088/0004-637x/721/2/1284}

\bibitem{sumiyoshi08}
K.~Sumiyoshi, G.~R\"opke,
  \href{http://link.aps.org/doi/10.1103/PhysRevC.77.055804}{Appearance of light
  clusters in post-bounce evolution of core-collapse supernovae}, Phys. Rev. C
  77 (2008) 055804.
\newblock \href {http://dx.doi.org/10.1103/PhysRevC.77.055804}
  {\path{doi:10.1103/PhysRevC.77.055804}}.
\newline\urlprefix\url{http://link.aps.org/doi/10.1103/PhysRevC.77.055804}

\bibitem{nakazato12}
K.~Nakazato, S.~Furusawa, K.~Sumiyoshi, A.~Ohnishi, S.~Yamada, H.~Suzuki,
  \href{https://doi.org/10.1088/0004-637x/745/2/197}{{HYPERON} {MATTER} {AND}
  {BLACK} {HOLE} {FORMATION} {IN} {FAILED} {SUPERNOVAE}}, The Astrophysical
  Journal 745~(2) (2012) 197.
\newblock \href {http://dx.doi.org/10.1088/0004-637x/745/2/197}
  {\path{doi:10.1088/0004-637x/745/2/197}}.
\newline\urlprefix\url{https://doi.org/10.1088/0004-637x/745/2/197}

\bibitem{peres13}
B.~Peres, M.~Oertel, J.~Novak,
  \href{https://link.aps.org/doi/10.1103/PhysRevD.87.043006}{Influence of pions
  and hyperons on stellar black hole formation}, Phys. Rev. D 87 (2013) 043006.
\newblock \href {http://dx.doi.org/10.1103/PhysRevD.87.043006}
  {\path{doi:10.1103/PhysRevD.87.043006}}.
\newline\urlprefix\url{https://link.aps.org/doi/10.1103/PhysRevD.87.043006}

\bibitem{sakurai18}
H.~Sakurai, Nuclear physics with ri beam factory, Frontiers of Physics 13~(6)
  (2018) 132111.

\bibitem{otsuka20}
T.~Otsuka, A.~Gade, O.~Sorlin, T.~Suzuki, Y.~Utsuno,
  \href{https://link.aps.org/doi/10.1103/RevModPhys.92.015002}{Evolution of
  shell structure in exotic nuclei}, Rev. Mod. Phys. 92 (2020) 015002.
\newblock \href {http://dx.doi.org/10.1103/RevModPhys.92.015002}
  {\path{doi:10.1103/RevModPhys.92.015002}}.
\newline\urlprefix\url{https://link.aps.org/doi/10.1103/RevModPhys.92.015002}

\bibitem{horowitz16}
C.~J. {Horowitz}, D.~K. {Berry}, M.~E. {Caplan}, T.~{Fischer}, Z.~{Lin}, W.~G.
  {Newton}, E.~{O'Connor}, L.~F. {Roberts}, {Nuclear pasta and supernova
  neutrinos at late times}, ArXiv e-prints\href
  {http://arxiv.org/abs/1611.10226} {\path{arXiv:1611.10226}}.

\end{thebibliography}


\end{document}